\newcommand{\logrhk}{$\rm log\,R^{\prime}_\mathrm{HK}$}
\newcommand{\kms}{\,km\,s$^{-1}$} 
\newcommand{\ms}{\,m\,s$^{-1}$} 
\newcommand{\gcm}{\,g\,cm$^{-3}$} 

\newcommand{\vsini}{$v$\,sin\,$i_\star$}

\newcommand{\teff}{$T_{\rm eff}$}

\newcommand{\mearth}{$M_{\oplus}$}
\newcommand{\rearth}{$R_{\oplus}$}

\newcommand{\mpl}{M_{\rm p}}

\documentclass[longauth]{aa}  

\usepackage{graphicx}
\usepackage{txfonts}
\usepackage{subfigure}
\usepackage{hyperref,threeparttable,lscape,supertabular}
\usepackage[rightcaption]{sidecap}
\def\ms{\hbox{\,m\,s$^{-1}$}}         

\begin{document}

   \title{The GAPS Programme at TNG}

   \subtitle{LIX. A characterisation study of the $\sim$300 Myr old multi-planetary system orbiting the star
BD+40\,2790 (TOI-2076)\thanks{Based on observations made with i) the Italian Telescopio Nazionale
\textit{Galileo} (TNG) operated by the Fundaci\'on Galileo Galilei (FGG) of the
Istituto Nazionale di Astrofisica (INAF) at the
Observatorio del Roque de los Muchachos (La Palma, Canary Islands, Spain); ii) the Spanish 3.5m
telescope at Calar Alto Observatory (Almería, Spain); iii) the US WIYN 3.5m telescope at Kitt Peak National Observatory (Tucson, Arizona). }}

   \author{M.~Damasso  
          \inst{1},
          D.~Locci \inst{2},
          S.~Benatti \inst{2},
          A.~Maggio \inst{2},
          M.~Baratella \inst{3},
          S.~Desidera \inst{4},
          K.~Biazzo \inst{5},
          E.~Palle \inst{6,7}, 
          S.~Wang \inst{8}, 
          D.~Nardiello \inst{14,4},
          L.~Borsato \inst{4},
          A.~S.~Bonomo \inst{1},
          S.~Messina \inst{9},
          G.~Nowak \inst{10,6,7},  
          A.~Goyal \inst{8}, 
          V.\,J.\,S.\,B\'ejar \inst{6,7},
          A.~Bignamini \inst{11},
          L.~Cabona \inst{12},
          I.~Carleo \inst{6,7}, 
          R.~Claudi \inst{4},
          R.~Cosentino \inst{13},
          S.~Filomeno \inst{5,15,16},
          C.~Knapic \inst{11},
          N.~Lodieu \inst{6,7},
          V.~Lorenzi \inst{13,6},
          L.~Malavolta \inst{14,4},
          M.~Mallorqu\'in \inst{6,7},
          L.~Mancini \inst{15,1,17},
          G.~Mantovan \inst{14},
          G.~Micela \inst{2},
          F.~Murgas \inst{6,7},  
          J.~Orell-Miquel \inst{6,7}, 
          M.~Pedani \inst{13},
          M.~Pinamonti \inst{1},
          A.~Sozzetti \inst{1},
          R.~Spinelli \inst{2},
          M.\,R.\,Zapatero Osorio \inst{18},
          T.~Zingales \inst{14,4}
    }

   \institute{INAF - Osservatorio Astrofisico di Torino, Via Osservatorio 20, I-10025 Pino Torinese, Italy\\
           	\email{mario.damasso@inaf.it}
    \and INAF -- Osservatorio Astronomico di Palermo, Piazza del Parlamento 1, I-90134, Palermo, Italy
    \and ESO-European Southern Observatory, Alonso de Cordova 3107, Vitacura, Santiago, Chile
    \and INAF -- Osservatorio Astronomico di Padova, Vicolo dell'Osservatorio 5, I-35122, Padova, Italy
    \and INAF -- Osservatorio Astronomico di Roma, Via Frascati 33, 00078 -- Monte Porzio Catone (Roma), Italy
    \and Instituto de Astrof\'isica de Canarias (IAC), Calle V\'ia L\'actea s/n, E-38200 La Laguna, Tenerife, Spain 
    \and Departamento de Astrof\'isica, Universidad de La Laguna (ULL), E-38206 La Laguna, Tenerife, Spain
    \and Department of Astronomy, Indiana University, Bloomington, IN 47405, USA  
    \and INAF -- Osservatorio Astrofisico di Catania, Via S.~Sofia,78 - 95123 Catania, Italy
    \and Institute of Astronomy, Faculty of Physics, Astronomy and Informatics, Nicolaus Copernicus University, Grudzi\c{a}dzka 5, 87-100 Toru\'n, Poland
    \and INAF -- Osservatorio Astronomico di Trieste, via Tiepolo 11, 34143 Trieste
    \and INAF -- Osservatorio Astronomico di Brera, Via E. Bianchi 46, 23807 -- Merate (LC), Italy 
    \and Fundación Galileo Galilei-INAF, Rambla José Ana Fernández Pérez 7, E-38712, Breña Baja, Spain
    \and Dipartimento di Fisica e Astronomia "G. Galilei"-- Universt\`a degli Studi di Padova, Vicolo dell'Osservatorio 3, I-35122 Padova
    \and Dipartimento di Fisica, Università di Roma ``Tor Vergata'', Via della Ricerca Scientifica 1, I-00133, Rome, Italy 
   \and Dipartimento di Fisica, Sapienza Università di Roma, Piazzale Aldo Moro 5, 00185 Roma, Italy
   \and Max Planck Institute for Astronomy, K\"{o}nigstuhl 17, D-69117, Heidelberg, Germany
   \and Centro de Astrobiolog\'{\i}a CSIC-INTA, Carretera de Ajalvir km 4, E-28850 Torrej\'on de Ardoz, Madrid, Spain
          }

   \date{}

 
  \abstract
   {The long-term Global Architecture of Planetary Systems (GAPS) programme has been characterising a sample of young systems with transiting planets through a spectroscopic and photometric follow-up. One of the main goals of GAPS is measuring planets' dynamical masses and bulk densities. This will help to build a picture of how planets evolve within the early stages of their formation, through a comparison between the fundamental physical properties of young and mature exoplanets.} 
   {We collected more than 300 high-resolution spectra of the $\sim$300 Myr old star BD+40~2790 (TOI-2076) over $\sim$3 years. This star hosts three transiting planets discovered by TESS, with orbital periods $\sim$10, 21, and 35 days. From our determined fundamental planetary physical properties, we investigate the temporal evolution of the planetary atmospheres by calculating the expected mass loss rate due to photo-evaporation up to a system age of 5 Gyr.}
   {BD+40~2790 shows an activity-induced scatter larger than 30 $\ms$ in the radial velocities. We employed different methods to measure the stellar radial velocities and several models to filter out the dominant stellar activity signal, in order to bring to light the planet-induced signals which are expected to have semi-amplitudes one order of magnitude lower. We evaluated the mass loss rate of the planetary atmospheres using photoionization hydrodynamic modeling, accounting for the temporal evolution of the stellar high-energy flux through the adoption of different models for X-rays and EUV irradiation.}
   {The dynamical analysis confirms that the three sub-Neptune-sized companions (our radius measurements are $R_b$=2.54$\pm$0.04, $R_c$=3.35$\pm$0.05, and $R_d$=3.29$\pm$0.06 \rearth) have masses in the planetary regime. We derive 3$\sigma$ upper limits below or close to the mass of Neptune for all the planets: 11--12, 12--13.5, and 14--19 \mearth for planet $b$, $c$, and $d$ respectively. In the case of planet $d$, we found promising clues that the mass could be between $\sim$7 and 8 \mearth, with a significance level between 2.3--2.5$\sigma$ (at best). This result must be further investigated using other analysis methods or using high-precision near-IR spectrographs to collect new radial velocities, which could be less affected by stellar activity. Atmospheric photo-evaporation simulations predict that BD+40~2790\,b is currently losing its H-He gaseous envelope, which will be completely lost at an age within 0.5--3\,Gyr if its current mass is lower than 12\,M$_\oplus$. BD+40~2790\,c could have a lower bulk density than $b$, and it could retain its atmosphere up to an age of 5 Gyr. For the outermost planet $d$, we predict almost negligible evolution of its mass and radius induced by photo-evaporation.}
   {}

    \titlerunning{Characterisation of the multi-planet system orbiting BD+40\,2790}
    \authorrunning{Damasso et al.}
   \maketitle
%
\section{Introduction}
For more than five years, the Global Architecture of Planetary Systems (GAPS) collaboration \citep{2013A&A...554A..28C} has carried out a radial velocity (RV) survey to search for and characterise planets around young stars (i.e with an age less than $\sim$800 Myr) within the Young Objects (YO) sub-program (e.g. \citealt{carleo2020A&A...638A...5C,BenattiSAIt}). The main goal of the survey is measuring the fundamental orbital and physical properties of planetary systems during their early life stages, when most of the processes that shape their final architectures occur. The typical mechanisms that leave some imprints on the systems' properties include planetary migration (e.g. \citealt{2008ApJ...686..621F,2014A&A...569A..56C,2014prpl.conf..667B}), tidal circularisation (e.g. \citealt{2008ApJ...678.1396J,2008ApJ...686..580C}) and photo-evaporation (e.g. \citealt{2013ApJ...775..105O}). Along with GAPS, other RV surveys like ZEIT \citep{2016ApJ...818...46M}, THYME \citep{2019ApJ...880L..17N}, and RVSPY \citep{2022A&A...667A..63Z}, or observational campaigns focused on specific targets (e.g. \citealt{2020A&A...640A..48L,2021AJ....162..295C,2022MNRAS.514.1606B,2023A&A...671A.163M}) pursued a similar goal. 
The current observational scenario indicates that small planets with a gaseous envelope on close-in orbits around stars younger than $\sim 150$ Myr show inflated radii, as previously suggested by \cite{2015ApJ...807....3K} and \cite{2016AJ....152...61M}. At this stage, they are subject to the Kelvin-Helmholtz contraction and possibly to the photo-evaporation of their atmosphere, as shown in the simulations discussed e.g. by \cite{2022ApJ...925..172M}. This condition leads to a decrease in the radii and, as a consequence, we can expect an evolution of the location on a mass-radius with the age of the systems. This hypothesis requires the follow-up of a larger sample of young planets to be effectively tested (see e.g. \citealt{2023MmSAI..94b.195B}). The main reason to explain the current relative scarcity of young exoplanets followed-up with the RV method is the intrinsic difficulty in obtaining robust mass measurements due to the effects of the stellar activity that can impact the modeling of RV data in a severe way. 

The experience matured within the GAPS-YO survey suggests that aiming for accurate measurements of planetary masses requires the collection of a considerable number of observations per target, possibly spanning several seasons and with a cadence dense enough to guarantee a good data sampling over the stellar rotational cycles. In fact, this can be a preferred way to obtain a proper treatment and mitigation of the stellar activity with state-of-the-art data modeling tools.
That is particularly true in the cases of small-size planets or multi-planet systems. 
Despite the increased challenges in measuring the planetary masses due to the multiplicity of Doppler signals present in the RV time series, multi-planet systems represent a valuable resource for comparative planetology, for example by investigating the effects of the photo-evaporation as a function of the distance from the host stars (e.g. \citealt{2021A&A...645A..71C,damasso_2023A&A...672A.126D}), and those induced on the planetary formation and migration mechanisms by the presence of multiple bodies in a system (e.g. \citealt{2023A&A...679A..55T,2024A&A...682A.129M}).

In this paper, we present a characterisation study of the $\sim$300 Myr old multi-planetary system orbiting the K-type star BD+40\,2790, also known as TOI-2076, 
based on photometric and high-resolution spectroscopic observations. The discovery and validation of three transiting sub-Neptune-sized planets (with orbital periods $\sim$10, 21, and 35 days) by using photometric data collected by the Transiting Exoplanet Survey Satellite (TESS; \citealt{ricker2015JATIS...1a4003R}) was presented by \cite{hedges2021}. As only two non-consecutive transits were observed for the two outer planets $c$ and $d$, the TESS data were not sufficient to calculate their orbital periods and constrain their orbits. Thanks to observations with the CHaracterising ExOPlanets Satellite (CHEOPS) space telescope and a ground-based follow-up, \cite{osborn2022} were able to pin down the orbital periods of BD+402790\,c and d. They found a period ratio close to 5:3 for BD+402790\,c and d, and an anti-correlated transit timing variation (TTV) signal between planets $b$ and $c$, likely linked to the 2:1 ratio of their orbital periods. \cite{osborn2022} revised the radius measurements of the planets, revealing that all of them fall in the sub-Neptune size range. By modeling the Rossiter–McLaughlin effect of TOI-2076\,b, \cite{Frazier_2023} found that the planet has a low sky-projected obliquity, concluding that a well-aligned orbit, together with the presence of TTVs, suggests that the compact multi-planet system likely evolved through disk migration in an initially well-aligned disk. In the attempt to search for atmospheric escape from young sub-Neptunes with Keck/NIRSPEC, \cite{Zhang_2023} reported the detection of helium absorption in the atmosphere of TOI-2076\,b, whose observed properties appear consistent with the expectations from photo-evaporation. However, \cite{gaidos2023MNRAS.518.3777G}, who have observed helium excess absorption during the same transit as in \cite{Zhang_2023}, argued that this is stellar in origin.   

In this study, we employ an RV sample composed by more than 300 data points, that have been collected using three high-resolution spectrographs, with the main goal of measuring dynamical masses and bulk densities of the three planets. In Section \ref{sec:datadescription} we describe the photometric and spectroscopic dataset, which we also used to revise the fundamental stellar parameters, as discussed in Section \ref{sec:stellarparam}. In Section \ref{sec:freqcontentanalysis} we investigate which are the significant periodic signals in our spectroscopic and photometric dataset, and make use of the results to set up a thorough modeling of the data described in Section \ref{sec:datamodelling}. We investigate in Sect. \ref{sec:atmophotoeva} the planetary atmospheric evolution through models based on photo-evaporation, and summarise our results in Sect. \ref{sec:conclusions}. 


\section{Description of the dataset} \label{sec:datadescription}
\subsection{TESS photometric observations} \label{sec:tessdata}

TESS observed TOI-2076 during Cycle~2 (in Sectors~16, from 11 September to 6 October 2019, and Sector~23, from 19 March to 15 April 2020), and during Cycle~4 (in Sector~50, from 26 March to 22 April 2022). This star belongs to the list of targets of the Guest Investigator programs GO-4195 (PI: Villanueva), GO-4231 (PI: Dragomir), GO-4023 (PI: Kipping), GO-4191 (PI: Burt), GO-4242 (PI: Mayo), and GO-4039 (PI: Davenport). An analysis of the complete TESS dataset is presented in \cite{Zhang_2023}. 

\begin{figure}
    \centering
    \includegraphics[width=0.5\textwidth]{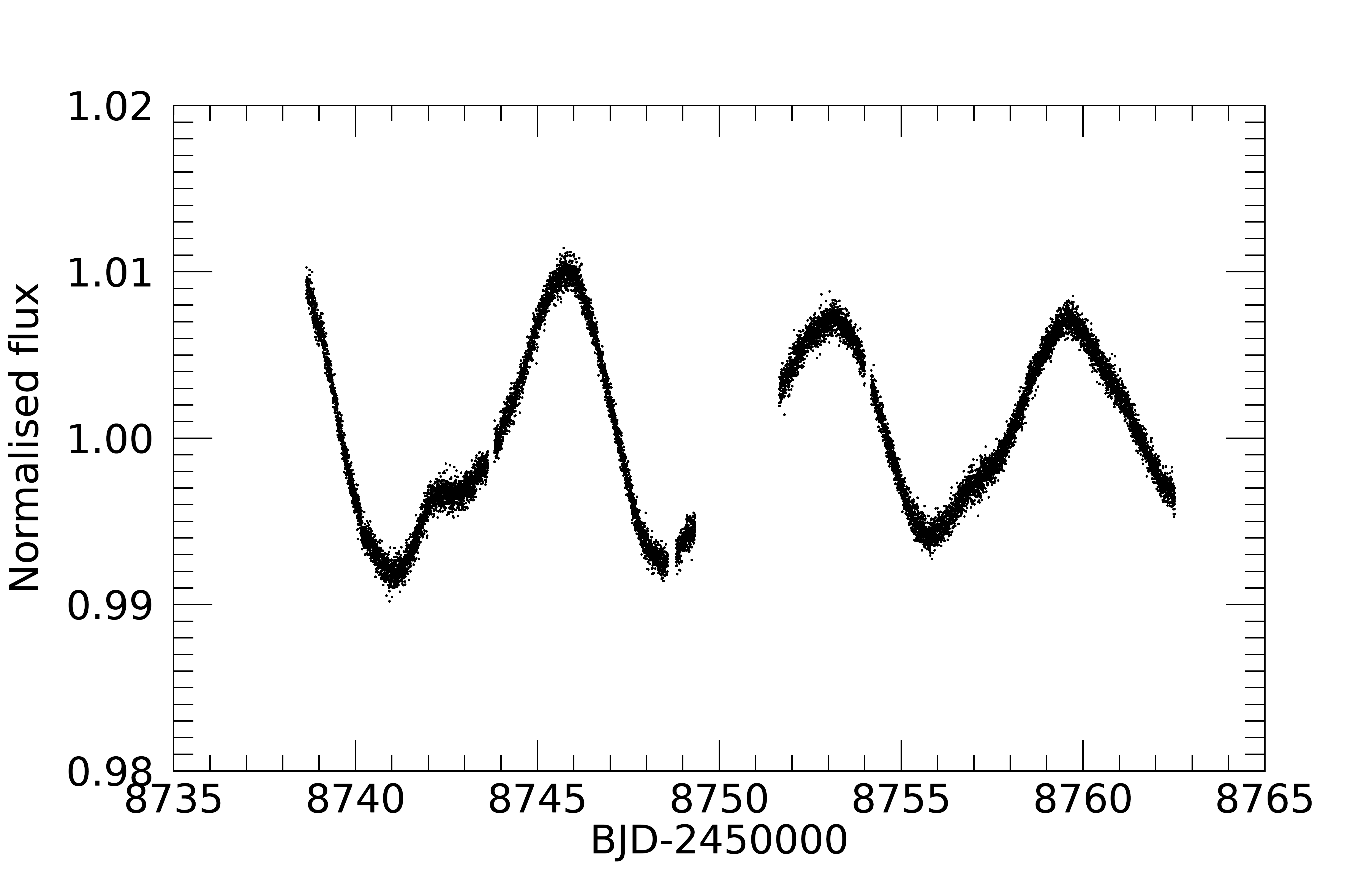}
    \includegraphics[width=0.5\textwidth]{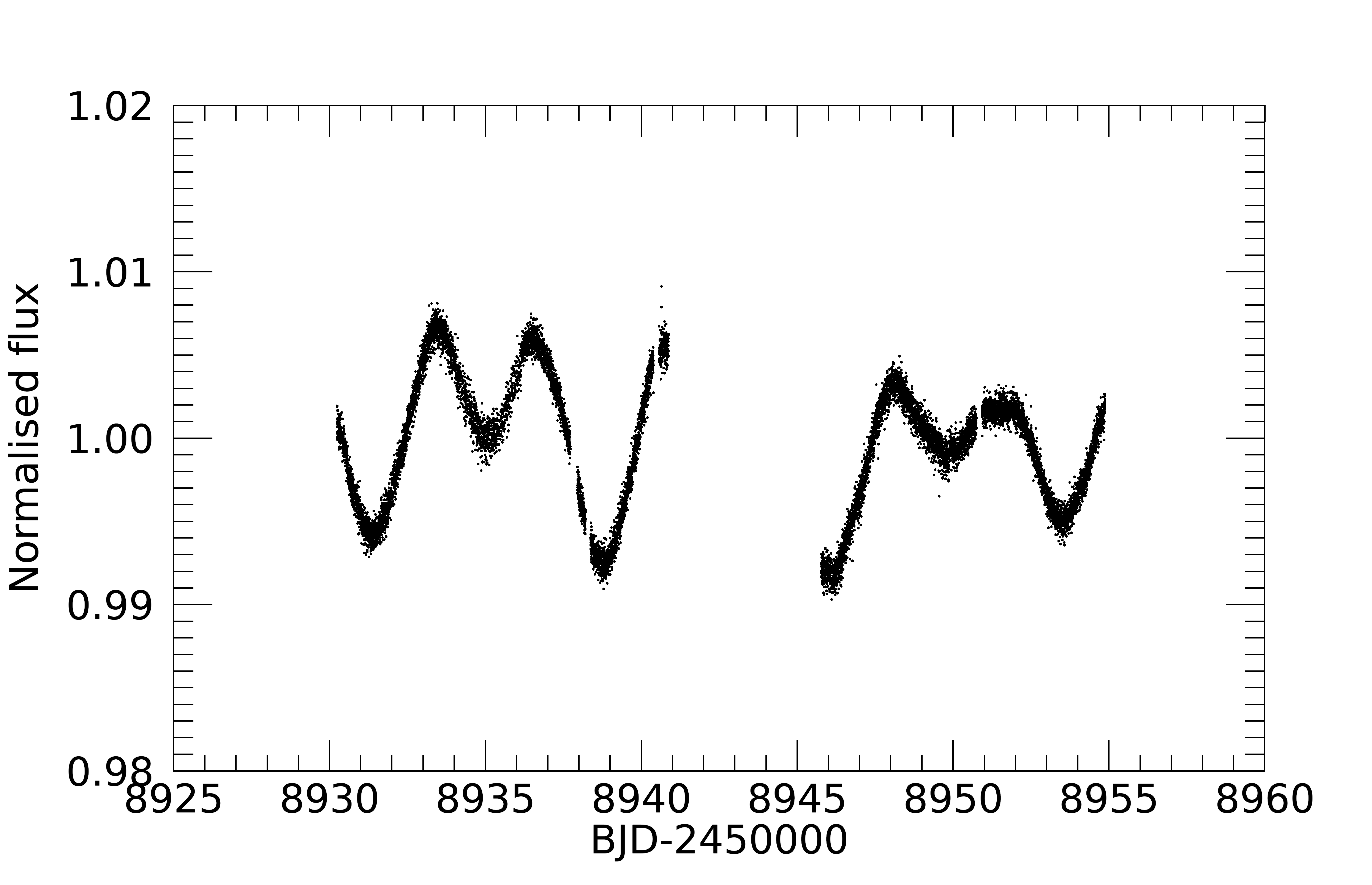}
    \includegraphics[width=0.5\textwidth]{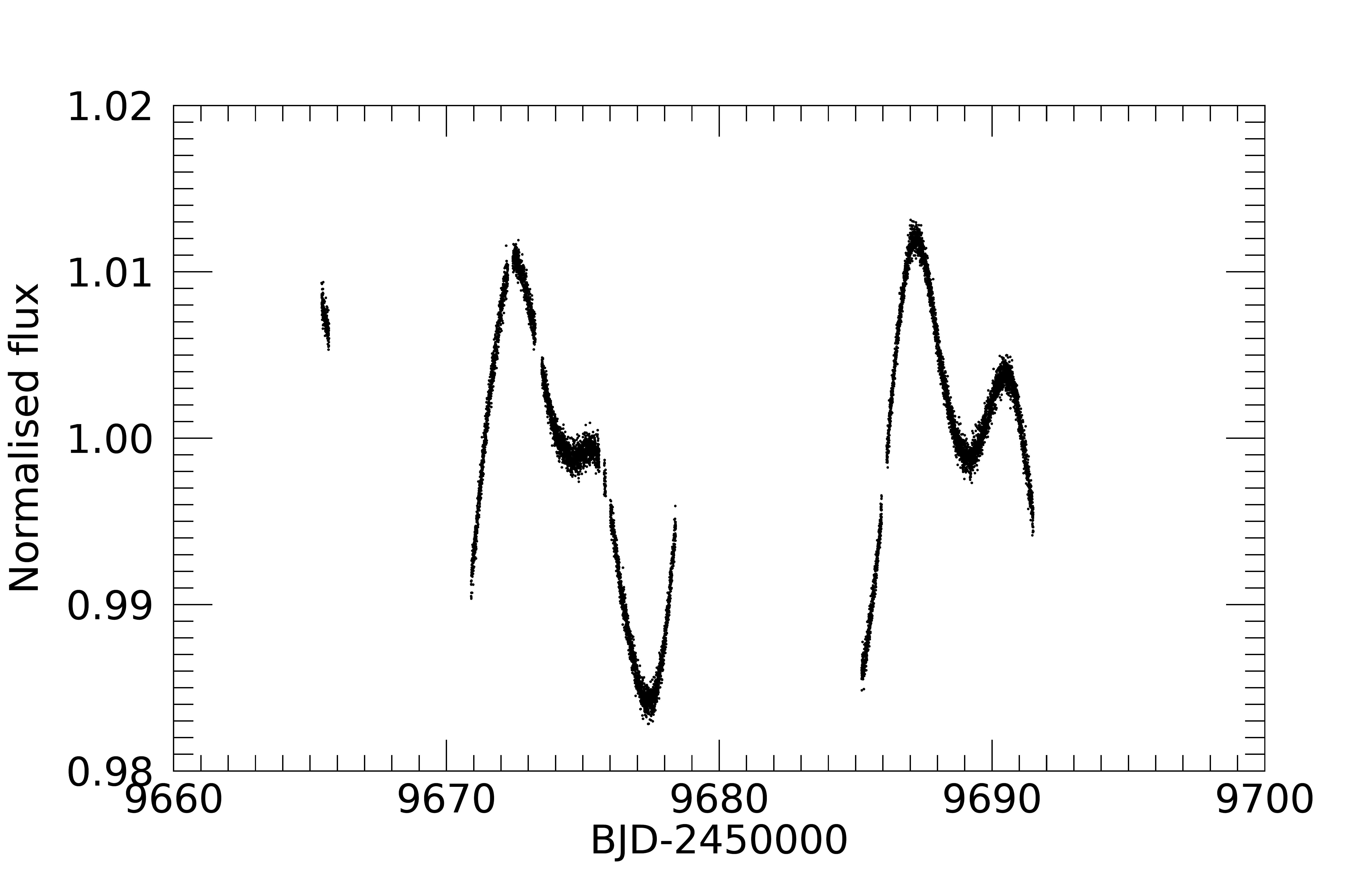}
    \caption{\textit{From top to bottom:} TESS light curves of Sectors 16, 23, and 50, extracted as described in Sect. \ref{sec:tessdata}. The transits of the three planets were masked and bad-quality points were removed. }
    \label{fig:tesslcnotrans}
\end{figure}

In our work, we used the short-cadence light curve (2-minute sampling). As already done in previous works of the GAPS series, we did not adopt the pre-search data conditioning simple aperture photometry (PDCSAP, \citealt{2012PASP..124.1000S,2012PASP..124..985S,2014PASP..126..100S}) light curves, because they could be affected by systematics due to over-corrections and/or injection of spurious signals. Instead, by using the cotrending basis vectors extracted as in \cite{2021MNRAS.505.3767N,nardiello2022}, we corrected the simple aperture photometry (SAP) light curve. The resulting light curve, with the transits removed, is shown in Fig. \ref{fig:tesslcnotrans}.

The measurement of the planetary radii from light curves of young and active stars can be influenced by the choice of the algorithm used to model and remove the variability due to the stellar activity (e.g. see \citealt{canocchi23}). To this purpose, we used the publicly available \texttt{Python} package \texttt{w{\={o}}tan}\footnote{\url{https://github.com/hippke/wotan}} \citep{wotan2019AJ....158..143H} to detrend (or flatten) the TESS light curve. 
From \texttt{w{\={o}}tan}, we selected the \texttt{cosine} (a sum of sines and cosines, with an iterative clipping of 2$\sigma$ outliers until convergence) and \texttt{hspline} (spline with a robust Huber estimator) from the available algorithms. The transits of the three planets were masked before detrending the light curve using the ephemeris reported by \cite{osborn2022}, then the derived model of stellar activity was interpolated to the in-transit data points, and the original light curve flattened. In Sect. \ref{sec:transitmodel} we compare the radius measurements obtained from the analysis of these two versions of the light curve, to inspect whether there are significant differences due to the choice of a specific detrending algorithm over the other. 


\subsection{Analysis of additional archival photometry}
For the measurement of the stellar rotation period, $P_{\rm rot,\,\star}$, for which \cite{nardiello2022} provided a measurement based on TESS data ($P_{\rm rot,\,\star}$=7.29$\pm$0.12 d), we made use of additional archive datasets, which are of adequate photometric precision and have a longer baseline with respect to the TESS observations. We made use of the All-Sky Automated Survey for SuperNovae (ASAS-SN; \citealt{2014ApJ...788...48S,2017PASP..129j4502K}) V-band observations. We downloaded the light curve from the public archive{\footnote{\url{https://asas-sn.osu.edu/}}}, and we removed low-quality data and all the points associated with a seeing larger than $3\sigma$ the mean seeing of each photometric series. The photometric series of TOI-2076 spans from March 2013 to August 2018 (765 points). 

We also analysed the publicly available light curves obtained by the SuperWASP survey (\citealt{2006PASP..118.1407P,2010A&A...520L..10B}), after removing from the time series the data with the quality flag \texttt{FLAG=0} (not-corrected photometric points) and outliers. 


\subsection{HARPS-N spectroscopic observations} \label{sec:harpsrv}
TOI-2076 was observed with the HARPS-N spectrograph \citep{Cosentino2012}, mounted at the Telescopio Nazionale \textit{Galileo} (TNG) on the island of La Palma (Canary Islands, Spain), from 6 August 2020 to 14 September 2022, with exposure times of 900s or 1200s. The spectra have a median S/N of 93 measured at a wavelength of $\sim$ 550 nm. Nearly 96\% of the total amount of spectra have been collected within the GAPS programme, but TOI-2076 was also followed-up during Spanish CAT time allocations (CAT19A$\_$162, PI: Nowak; ITP19$\_$1 PI Pall\'e). The merged sample consists of 294 spectra. We excluded from the dataset the spectrum taken at epoch BJD 2459286.697138 with low S/N ($<25$).

We adopted three different algorithms to extract the RVs from the HARPS-N spectra, with the purpose of testing their relative sensitivity to stellar activity contamination. In our study, we analyse all these alternative HARPS-N RV datasets with the same set of test models. We used the standard Data Reduction Software (DRS) pipeline (version 3.7.0) through the \textsc{YABI} workflow interface \citep{YABI}, which is maintained by the Italian center for Astronomical Archive (IA2)\footnote{\url{ https://ia2.inaf.it}}. The RVs and activity diagnostics full width at half maximum (FWHM) and bisector inverse slope (BIS) have been derived from the DRS cross-correlation function (CCF), which has been calculated by adopting a reference mask for a star with spectral type K5, and a half-window width of 30 \kms. 

We extracted a second version of the RV dataset using the code \texttt{SERVAL} (\texttt{spectrum radial velocity analyser}, version dated 26 January 2022; \citealt{2018A&A...609A..12Z}), which adopts a procedure based on template-matching to derive relative RVs\footnote{The code is publicly available at \url{https://github.com/mzechmeister/serval}. We used \texttt{-snmin 10.0 -niter 2 -safemode} as keywords in calling the software procedure, and the keyword \texttt{-oset} to select a specific subset of echelle spectral orders for the RV computation.}. 
Using \texttt{SERVAL}, we calculated the RVs using both all the echelle orders, but also selecting specific echelle orders (wavelength ranges) for the computation of a ``chromatic'' RV datasets, as detailed in Table \ref{tab:rvsummary}. Our goal is to test whether we can reduce the RV scatter due to stellar activity by excluding ``bluer'' orders of the full wavelength range covered by HARPS-N ($\sim$3800\,\AA\, to 6900\,\AA) and, ultimately, to verify if a reduced activity contamination can help to retrieving the planetary-induced Doppler signals. A wavelength dependence in the activity-induced RV scatter could be expected for young and active stars whose photosphere is spot-dominated, with the amplitude decreasing in the redder parts of the spectrum (e.g. \citealt{Prato_2008,Mahmud_2011,carleo2020A&A...638A...5C}). 
For our empirical test, we selected four different subsets of echelle orders, and we focused on the dataset corresponding to the wavelength range from $\sim$5393\,\AA\, to $\sim$6916\,\AA\, to perform an analysis with all the test models adopted in this study (Sect. \ref{sec:rvlcmodel} and Appendix \ref{app:rvmodeldetails}). The other ``chromatic'' datasets have been analysed only with a subset of the test models, to compare the results obtained especially for planet TOI-2076\,d. From Table \ref{tab:rvsummary} we note that the RV rms decreases when ``bluer'' orders are progressively excluded from the RV computation, at the cost of increasing the RV uncertainties $\sigma_{\rm RV}$.

Recently, \cite{Artigau_2022} proposed a line-by-line (LBL) algorithm for high-precision RV measurements. We applied the algorithm \texttt{LBL}\footnote{We used version 0.63 of the publicly available \texttt{Python} code (\url{https://github.com/njcuk9999/lbl}), following a standard recipe to run the algorithm, as described in the online repository, using all the echelle spectral orders.} to our HARPS-N spectra as a further test to evaluate the performances of a different RV extraction method for a star with a high level of activity-induced RV scatter.   

HARPS-N data represents the bulk of the spectroscopic dataset analysed in this work. We also used data collected by CARMENES and NEID spectrographs, that are described in Appendix \ref{app:data}. Fig. \ref{fig:rvtimeseries} shows the time series of the combined RV datasets, and their properties are summarised in Table \ref{tab:rvsummary}. The RV datasets are available online at CDS.

\begin{table*}[h!]
    \centering
    \small
    \caption{Summary of the different RV datasets analysed in this work.}
\begin{threeparttable}    
    \begin{tabular}{ccccc}
         \noalign{\smallskip}
         \hline
          \noalign{\smallskip}
         Instrument & Nr. of data & Time span & RV rms & median $\sigma_{\rm RV}$ \\
         & & [BJD-2450000] & $[\ms]$ & $[\ms]$ \\
         \noalign{\smallskip}
         \hline
         \noalign{\smallskip}
         HARPS-N, all orders (SERVAL) & 293 & 9068.41--10187.36 [1119 d] & 36.0 & 1.5  \\
          \noalign{\smallskip}
          HARPS-N, wavelength range nr. 1 (SERVAL; from $\sim$4014\,\AA\, to $\sim$5393\,\AA)\tnote{a} & 293 & 9068.41--10187.36 [1119 d] & 37.4 & 1.8  \\
          \noalign{\smallskip}
          HARPS-N, wavelength range nr. 2 (SERVAL; from $\sim$4800\,\AA\, to $\sim$6916\,\AA)\tnote{b} & 293 & 9068.41--10187.36 [1119 d] & 34.6 & 1.9  \\
          \noalign{\smallskip}
          HARPS-N, wavelength range nr. 3 (SERVAL; from $\sim$5393\,\AA\, to $\sim$6916\,\AA)\tnote{c} & 293 & 9068.41--10187.36 [1119 d] & 32.9 & 2.8  \\
          \noalign{\smallskip}
           HARPS-N, wavelength range nr. 4 (SERVAL; from $\sim$5695\,\AA\, to $\sim$6916\,\AA)\tnote{d} & 293 & 9068.41--10187.36 [1119 d] & 32.5 & 3.7  \\
          \noalign{\smallskip}
         HARPS-N, all orders (DRS) & 293 & 9068.41--10187.36 [1119 d] & 34.7 & 1.6 \\
          \noalign{\smallskip}
           HARPS-N, all orders (LBL) & 293 & 9068.41--10187.36 [1119 d] & 36.6 & 1.5 \\
          \noalign{\smallskip}
          NEID & 15 & 9348.80--9392.83 [44 d] & 27.4 & 1.7 \\
          \noalign{\smallskip}
         CARMENES-VIS & 22 & 9310.63--9495.29 [185 d] & 42.2 & 5.3 \\
         \noalign{\smallskip}
         \hline
    \end{tabular}
    \begin{tablenotes}
     \item[a] This wavelength range corresponds to pipeline-wise echelle orders from 6 to 43.  
     \item[b] This wavelength range corresponds to pipeline-wise echelle orders from 30 to 69.
     \item[c] This wavelength range corresponds to pipeline-wise echelle orders from 44 to 69.
     \item[d] This wavelength range corresponds to pipeline-wise echelle orders from 50 to 69.
                    \end{tablenotes}
\end{threeparttable}    
    \label{tab:rvsummary}
\end{table*}

\begin{figure}
    \centering
    \includegraphics[width=0.5\textwidth,clip,trim=0.7cm 0.5cm 0.5cm 1cm]{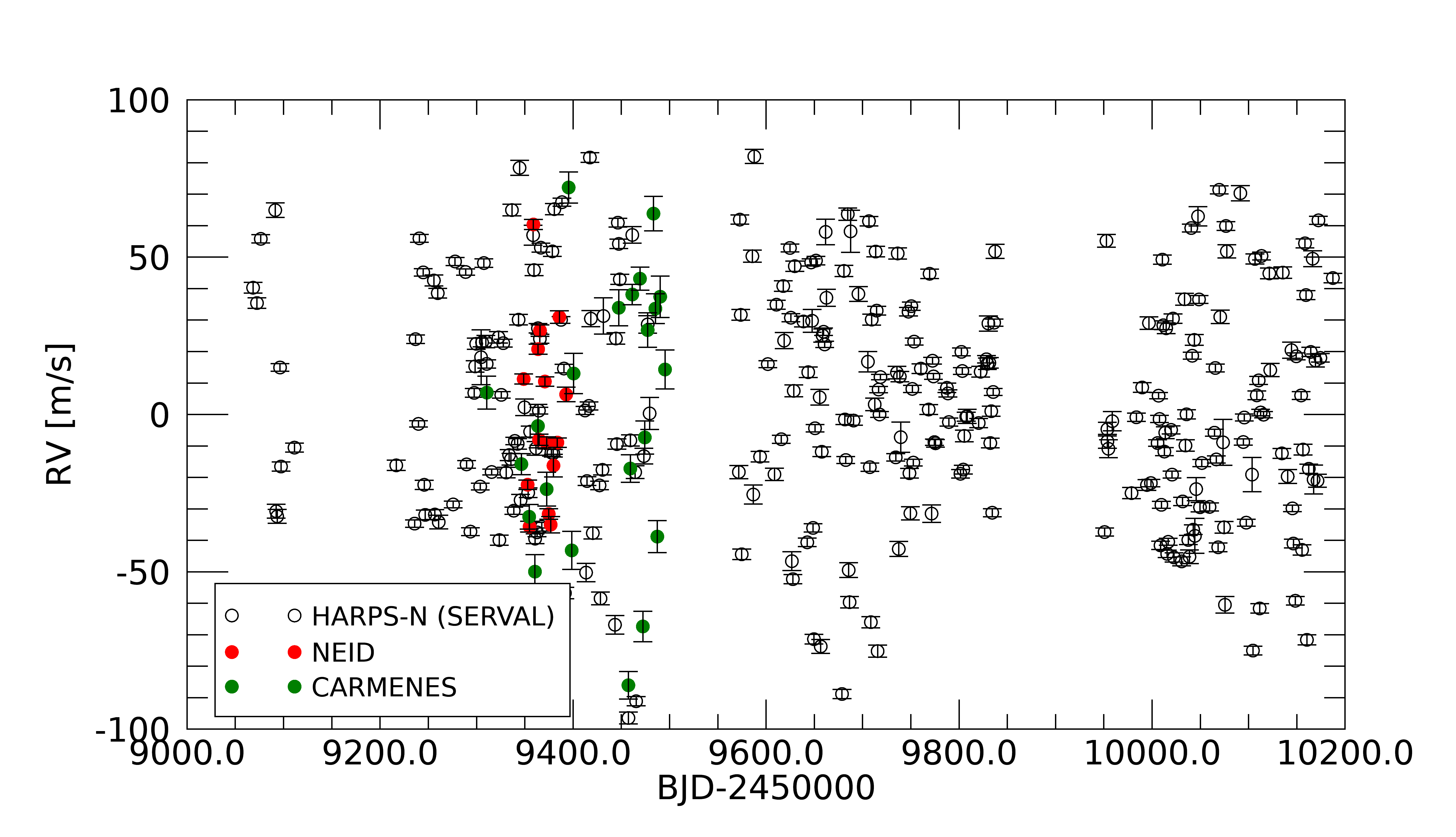}
    \caption{RV time series analysed in this work. For HARPS-N, data extracted with SERVAL (all echelle orders) are shown.}
    \label{fig:rvtimeseries}
\end{figure}


\section{Stellar fundamental parameters} \label{sec:stellarparam}

\subsection{Photometric \teff\ and reddening}
\label{sec:phot_teff}

From the \texttt{colte} package \citep{colte} and considering the \textit{Gaia} DR3 and 2MASS photometry, we estimated a weighted mean temperature of $5180\pm45$\,K from calibration relations based on different colour indices, with \teff\ ranging from $5122\pm56$\,K in the $G-B_{p}$ colour index up to $5227\pm60$\,K for $G-R_{p}$. We use this value as an initial guess for the spectroscopic determination of \teff, as described in Sect. \ref{sec:spec}.  

Reddening toward TOI-2076 is negligible, as resulting from the maps by \citet{pic}
and the expectation from the short distance from the Sun (42.0 pc).
From the spectral energy distribution, there are no indications of the presence
of significant IR excess. 

\subsection{Spectroscopic analysis}
\label{sec:spec}



TOI-2076 is a young ($\sim 300$\,Myr old) star (see Sect. \ref{sec:kinematics}), for which the intense magnetic activity can modify the structure of the stellar photosphere, thus altering the shape of the spectral lines formed in the upper layers. This effect could be particularly evident for iron (Fe) lines, which are typically used to derive the stellar parameters with the standard method based on line equivalent widths (EWs) (see, e.g., \citealt{2019galarza, 2020spina}). As a consequence, when micro-turbulence velocity $\xi$ is derived by imposing the same iron abundance for weak and strong Fe lines, this could lead to an overestimation of $\xi$, and therefore to an underestimation of the iron abundances [Fe/H]. To avoid these issues, we considered the procedure developed by \cite{2020baratella_ges} and based on the use of titanium (Ti) lines to derive the surface gravity ($\log g$) and $\xi$. Titanium lines typically form deeper in the photosphere compared to Fe lines and are less influenced by the effects of the magnetic fields, as seen on Fe transitions which show EWs that are 5--10 m\AA\, larger in young, active solar analogue stars (see Fig. 9 in \cite{2020baratella_ges}), allowing for a measurement of $\xi$ not correlated with stellar activity. We used a combination of Fe and Ti lines to derive the effective temperature ($T_{\mathrm{eff}}$), adopting the Ti and Fe line list published by \cite{2020baratella_gaps}, and measuring the line EWs through the software ARESv2 \citep{2015sousa}, discarding those lines with errors larger than 10\% and with EW>120 m\AA.

As an initial guess for the spectroscopic analysis, we took the photometric  $T_{\mathrm{eff, phot}}$ derived above (i.e. $5180\pm45$\,K), $\log g$ from \textit{Gaia} parallax ($4.59\pm0.08$\,dex), and microturbulence velocity $\xi=0.80\pm0.04$\,\kms from the calibration of \cite{2016ferreira}.



We thus created the 1D LTE model atmosphere linearly interpolated from the ATLAS9 grid of \cite{2003castelli} with new opacities (ODFNEW) and used the driver \textit{abfind} of MOOG \cite{1973sneden} for deriving the following final spectroscopic parameters: $T_{\mathrm{eff,spec}}=5200\pm100$\,K, $\log g_{\mathrm{spec}}=4.52\pm0.05$\,dex, $\xi_{\mathrm{spec}}=0.90\pm0.10$\,\kms, and [Fe/H]=$-0.03\pm0.06$ (see Table\,\ref{t:star_param}). 
Our spectroscopic parameters agree well with the input estimates, and with those published by \cite{osborn2022}. $T_{\mathrm{eff,spec}}$ is in agreement with $T_{\mathrm{eff,photo}}$, although less precise. Nonetheless, we prefer to adopt the more conservative $T_{\mathrm{eff,spec}}$ as our reference value, in that it is derived, together with the other atmospheric parameters, using a method optimised for young and active stars, that robustly account for systematics. Overall, the star has solar elemental abundances (see \citealt{filomeno_subm}).

\subsection{Lithium}
\label{sec:lithium}


We measured the equivalent width of the Li line at 6707.8\,\AA\, using the IRAF task \texttt{splot}. Its value resulted to be $89.8\pm1.0$ m\AA, where the error was estimated from the standard deviation of three EW measurements. Then, we also derived the NLTE Li abundance through the prescriptions by \cite{NLTE_li} and by adopting the spectroscopic parameters from Sect.\,\ref{sec:spec}. Final lithium abundance $\log \mathrm{A(Li)^{NLTE}}$ is $2.11\pm0.10$\,dex, where uncertainty was estimated considering both the errors on EW and on spectroscopic parameters\,(see Table\,\ref{t:star_param}). We find very similar values also applying the spectral synthesis method (see \citealt{Biazzoetal2022}). Our Li abundance is intermediate between the Pleiades open cluster and the Ursa Major association, fully compatible with the group membership and age derived by \citet{nardiello2022} (see also Sect. \ref{sec:kinematics}).



\subsection{Projected rotational velocity}
\label{sec:vsini}


Using the spectroscopic stellar parameters derived above, the driver {\it synth} of MOOG (\citealt{1973sneden}) and the \cite{2003castelli} grid of model atmospheres, we also measured the projected rotational velocity ($v \sin i$) through the synthesis of three spectral regions around 5400, 6200, 6700\,\AA. Assuming a macroturbulence velocity $v_{\rm macro}$=2.05\,km\,s$^{-1}$ from the relations by \cite{Breweretal2016}, we found a final value of \vsini=5.2$\pm$0.4\,\kms (see Table\,\ref{t:star_param}). 

\cite{Raineretal2023} determined a linear relation to derive the projected rotational velocity of a star from the FWHM of the CCF (see Eq. 7 therein calculated for a K5 mask). Using this calibration, we obtain \vsini = $5.2\pm2.4$ \kms, assuming the weighted mean of all the FWHM measurements and the median of the associated errors as a reference value (FWHM=$8.8412\pm0.0032$ \kms), and taking into account the errors of the coefficients in the calibration formula. This value is in agreement with the measurement from the spectral synthesis, although less precise. 

\subsection{Stellar rotation period}
\label{sec:rotation}

\cite{hedges2021} derived a stellar rotation period $P_{\rm rot,\,\star}= 7.27\pm0.23$ days from the analysis of TESS (Sectors 16 and 23) and KELT light curves. We use our extracted TESS light curve, including Sector 50, to redetermine $P_{\rm rot,\,\star}$. The light curve was analysed with the \texttt{CLEAN} algorithm \citep{Roberts87}, and we derived $P_{\rm rot,\,\star}$ = 7.3$\pm$1.1\,d with a very high confidence level in all the individual sectors. An improved period determination comes from the analysis of the combined sectors, that is $P_{\rm rot,\,\star}$ = 7.31$\pm$0.03 d, and we assume this as our reference value. In Fig. \ref{fig:lcperiod} we show the \texttt{CLEAN} periodograms for each Sector and for their combination. We note a significant power peak, although smaller than the primary, at about half the rotation period, which arises from the presence of a secondary minimum in the light curve, whose amplitude evolves with time, reaching the maximum depth in Sector 23.

We also analysed archival light curves from ASAS-SN and SuperWASP as a cross-check. The analysis of the periodogram of the ASAS-SN light curve resulted in a peak at $P=7.53$~d with low power ($Pw\sim 0.04$). From the analysis of the periodogram of the SuperWASP light curve, we measured a single significant peak at a period $P=7.33$~d ($Pw\sim 0.15$). 



\subsection{Coronal and chromospheric activity}
\label{sec:activity}

The HARPS-N data allowed us to determine a mean value of the chromospheric index $\log\,R^{\prime}_\mathrm{HK} = -4.354 \pm 0.025$ (Sect. \ref{sec:freqcontentanalysis}). We employed this value and its uncertainty to estimate the X-ray luminosity of TOI-2076 with the scaling law by \cite{Mama+Hille2008}, obtaining $\log L_{\rm x} = 28.86^{+0.09}_{-0.08}$\,erg\,s$^{-1}$. Using the calibration with the stellar age by the same authors, we found that the predicted \logrhk is in the range $-4.38^{+0.04}_{-0.03}$ for an age of $300 \pm 80$\,Myr, in good agreement with the measurements.

The only direct measurement of the X-ray luminosity of TOI-2076 currently available is based on a faint detection of this source in the ROSAT All-Sky Survey, leading to $\log L_{\rm x} = 28.67 \pm 0.16$\,erg\,s$^{-1}$ \citep{hedges2021}. This value is lower but in agreement within the error bars than that derived from the measured chromospheric Ca\,II H\&K index.

We obtained an alternative estimate of the X-ray luminosity starting from the mean rotation period $P = 7.31 \pm 0.03$ d (Table \ref{t:star_param}) and adopting the activity--rotation relationships by \cite{Pizzo+2003}, that yield $\log L_{\rm x} = 28.75 \pm 0.01$\,erg\,s$^{-1}$, just $\sim 20$\% lower and within $2 \sigma $ with respect to the value derived from the chromospheric index and the stellar age. This is the reference value that we adopted for investigating the photo-evaporation histories of the planets in this system (Sect.\ \ref{sec:atmophotoeva}). 

However, we also verified that this X-ray luminosity corresponds to the 2\% percentile of the distribution expected for stars with the same mass and age of TOI-2076, while the rotation period corresponds to the 45\% percentile  of the related distribution \citep{Johnstone+2021}. In conclusion, the coronal emission level appears to be in the low-end tail of the expected distribution for stars similar to TOI-2076.

\subsection{Membership to group and stellar age}
\label{sec:kinematics}


TOI-2076 has very similar kinematic parameters and age to the young star TOI-1807 \citep{hedges2021}. Their membership to the same comoving group was confirmed by \citet{nardiello2022}, who found that TOI-2076 has a rotation period matching well the rotational sequence of the group \citep[Fig. 7 in ][]{nardiello2022}. In this study we adopt for TOI-2076 the robust age that \citet{nardiello2022} determined for the comoving group, namely 300$\pm$80 Myr.

\subsection{Binarity}
\label{sec:binarity}

Beside the comoving objects discussed in \citet{nardiello2022},
which are at physical separations too wide for being gravitationally bound to TOI-2076, there are no indications of 
close massive companions from \textit{Gaia} astrometry \citep{gaia2016} using data from DR3 (RUWE = 0.98; \citealt{gaia2021,gaia2023}), direct imaging \citep[Gaia and AO observations by][]{hedges2021}, our own 4-yr RV time series, and long-term proper motion differences (less than two $\sigma$ for Tycho2 and \textit{Gaia} DR3).

\subsection{Mass, radius, and luminosity from SED analysis}
\label{sec:massradius}

\begin{figure}
    \centering
    \includegraphics[width=0.5\textwidth]{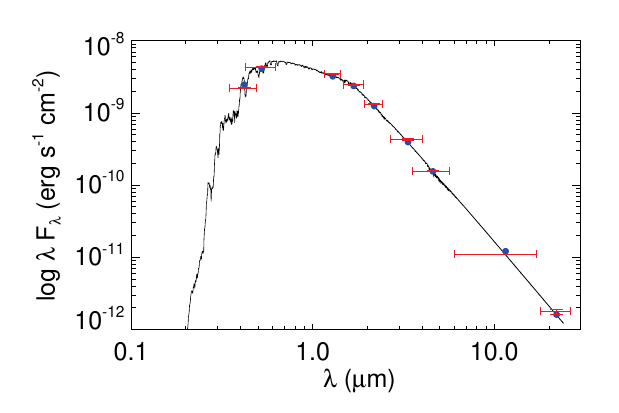}
    \caption{Spectral energy distribution of BD+40\,2790 (TOI-2076) with the best-fit model overplotted (solid line). Red and blue points correspond to the observed and predicted values, respectively.}
    \label{fig:sed}
\end{figure}

We determined the stellar radius and mass with the EXOFASTv2 tool \citep{Eastmanetal2019}, by fitting the stellar Spectral Energy Distribution (SED). The stellar luminosity calculated from the SED has been provided as input to the MIST stellar evolutionary tracks \citep{Dotter2016}. For fitting the SED we considered the Tycho $B$ and $V$ magnitudes \citep{2000}, the 2MASS near-IR $J$, $H$ and $K$ magnitudes \citep{cutri2003}, and the WISE mid-IR $W1$, $W2$, $W3$ and $W4$ magnitudes \citep{cutri2013}. We imposed Gaussian priors on (i) the stellar effective temperature $T_{\rm eff}$ and metallicity ${\rm [Fe/H]}$ from our analysis of the HARPS-N spectra, (ii) the parallax $23.8052\pm0.0125$~mas from the \textit{Gaia} DR3, and (iii) the stellar age $300\pm80$~Myr from \cite{nardiello2022}. We used uninformative priors for all the other parameters. The best fit of the SED is shown in Fig.~\ref{fig:sed}. We found $M_\star=0.849^{+0.027}_{-0.026}~\rm M_\odot$ and $R_\star=0.758\pm0.014~\rm R_\odot$. The derived mass is in agreement with the prediction of the PARSEC (PAdova TRieste Stellar Evolutionary Code) models \citep{2012MNRAS.427..127B}.

Coupling the measured stellar radius and rotation period
yields an equatorial velocity 5.24 \kms. This is very close to
the observed \vsini, further supporting the edge-on inclination of the star obtained through the Rossiter-Mc Laughlin effect by \cite{Frazier_2023}.

\begin{table}[!htb]
   \caption[]{Stellar parameters of 
BD+40 2790 (TOI-2076)}
     \label{t:star_param}
     \tiny
     \centering
       \begin{tabular}{lcc}
         \hline
         \noalign{\smallskip}
         Parameter   &  Value & Ref  \\
         \noalign{\smallskip}
         \hline
         \noalign{\smallskip}
$\alpha$ (J2000)          &  14:29:34.24 &  \citep{gaia2016,gaia2021}  \\
$\delta$ (J2000)          &  +39:47:25.54   &  \citep{gaia2016,gaia2021}  \\
$\mu_{\alpha}$ (mas/yr)  &    -118.071$\pm$0.010  &  \citep{gaia2016,gaia2021}  \\
$\mu_{\delta}$ (mas/yr)  &     -6.860$\pm$0.013  &  \citep{gaia2016,gaia2021} \\
$\pi$  (mas)             &    $23.8052\pm0.0125$ &  \citep{gaia2016,gaia2021,gaia2023}  \\
\noalign{\medskip}
B$_T$ (mag)                &    $10.258\pm0.025$  & \citep{2000}  \\
V$_T$ (mag)                  &    $9.238\pm0.015$     & \citep{2000}  \\
G (mag)                  &    $8.918\pm0.003$  &  \citep{gaia2016,gaia2021,gaia2023}  \\
\textit{Gaia} BP              &   9.349$\pm$0.003  & \citep{gaia2016,gaia2021,gaia2023} \\
\textit{Gaia} RP              &   8.319$\pm$0.004  & \citep{gaia2016,gaia2021,gaia2023}  \\
J$_{\rm 2MASS}$ (mag)    &   $7.613\pm0.019$  & \citep{cutri2003}   \\
H$_{\rm 2MASS}$ (mag)    &   $7.188\pm0.027$  & \citep{cutri2003}  \\
K$_{\rm 2MASS}$ (mag)    &   $7.115\pm0.017$  & \citep{cutri2003}  \\
WISE1 (mag)              &   $7.011\pm0.050$  & \citep{cutri2013} \\
WISE2 (mag)              &   $7.126\pm0.019$  & \citep{cutri2013} \\
WISE3 (mag)              &   $7.092\pm0.016$  & \citep{cutri2013} \\
WISE4 (mag)              &   $7.003\pm0.089$  & \citep{cutri2013} \\
\noalign{\medskip}
T$_{\rm eff}$ (K)        &  5200$\pm$100        & This paper \\  
& & (spectroscopic; adopted) \\
T$_{\rm eff}$ (K)        &  5180$\pm$45      & This paper (photometric) \\  
$\log g$                 &  4.52$\pm$0.05    & This paper  \\ 
${\rm [Fe/H]}$ (dex)     &  -0.03$\pm$0.06      & This paper  \\ 
E(B-V)                   &  0.003$\pm$0.015     & PIC \citep{pic} \\ 
\noalign{\medskip}
$S_{\rm MW}$             &   $0.548\pm0.027$    & This paper (HARPS-N spectra) \\
$\log R^{'}_{\rm HK}$    &  -$4.354\pm0.025$ &  This paper (HARPS-N spectra) \\  
$v\sin{i_{\star}}$ (km/s)      &  5.2$\pm$0.4  & This paper  \\  
$L_{\rm bol}$ (L$_{\odot}$) &    $0.383\pm0.013$    & This paper  \\
${\rm P_{\rm rot,\,\star}}$ (d)  &     7.31$\pm$0.03  &   This paper  \\ 
$L_{\rm X}$  [erg/cm$^2$]   &  $5.6^{+2.5}_{-1.0}\times10^{28}$   & This paper \\ %
$\log L_{\rm X}/L_{\rm bol}$     &  -$4.42^{+0.16}_{-0.09}$  & This paper  \\
A(Li)                    &     2.11$\pm$0.10  &  This paper   \\
\noalign{\medskip}
Mass (M$_{\odot}$)       &    $0.849^{+0.027}_{-0.026}$    & This paper  \\
Radius (R$_{\odot}$)     &    $0.758\pm0.014$    & This paper  \\
Density ($\rm g\,cm^{-3}$)    &    $2.74^{+0.15}_{-0.14}$ & This paper \\
Age  (Myr)               &    $300\pm80$ & \cite{nardiello2022}  \\

         \noalign{\smallskip}
         \hline
      \end{tabular}
\end{table}


\section{Frequency content analysis of RVs and activity diagnostics } \label{sec:freqcontentanalysis}

\begin{figure*}
    \centering
    \includegraphics[width=1\linewidth]{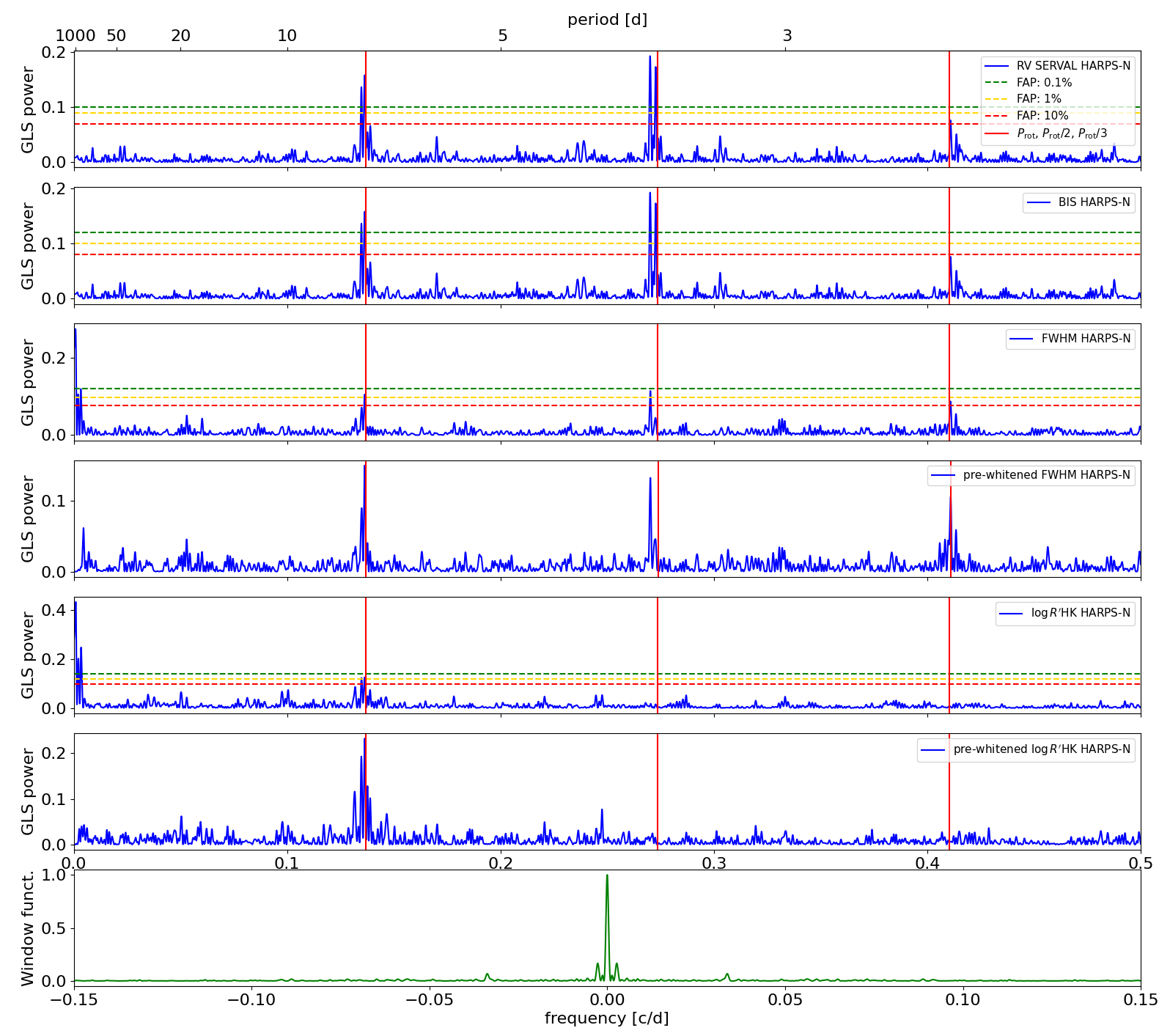}
    \caption{GLS periodograms of one of the RV dataset (SERVAL, all orders) and stellar activity diagnostics extracted from HARPS-N spectra.}
    \label{fig:gls_diagnostics_harpn}
\end{figure*}

In our study, we analysed the time series of three activity diagnostics calculated from HARPS-N spectra, namely BIS, FWHM, and \logrhk. The FWHM and \logrhk time series show by eye a possible long-term modulation/cycle, which is confirmed by calculating the Generalised Lomb-Scargle (GLS; \citealt{2009A&A...496..577Z}) periodograms  shown in Fig. \ref{fig:gls_diagnostics_harpn}. Both show significant peaks, with a false alarm probability (FAP) lower than 0.1$\%$, with the highest power around 1000 days. The statistical FAP has been evaluated through 10\,000 bootstrap (with replacement) simulations. The periodograms of the pre-whitened time series show peaks at the stellar rotation period and its first and second harmonic (for the FWHM), and at the rotation period (for the \logrhk). The periodogram of the BIS index does not show peaks at low frequencies, and it is dominated by the first harmonic of the stellar rotation period.   

\begin{figure}
    \centering
    \includegraphics[width=0.5\textwidth]{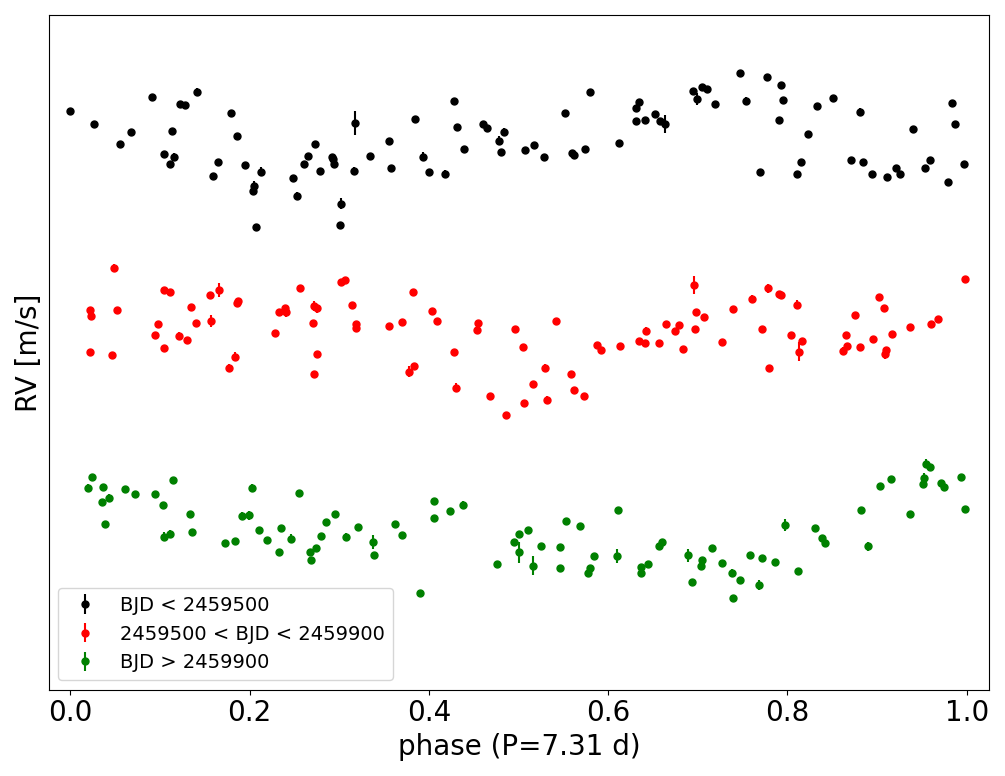}
    \caption{RVs (SERVAL extraction, wavelength range nr. 3) phase-folded to the stellar rotation period (same epoch used as phase=0 for the three series). Data are divided into three sub-samples (with offsets applied for better readability) to show how the variability due to stellar activity changes on a seasonal basis.}
    \label{fig:rvphasefold}
\end{figure}

Figure \ref{fig:gls_rv_harpn} shows the GLS periodograms of HARPS-N RVs extracted with different algorithms (Table \ref{tab:rvsummary}). In all the cases, dominant and statistically significant peaks are found at ${\rm P_{\rm rot,\,\star}}$ and its first harmonic, clearly denoting that the observed RV scatter $>30$ \ms is mostly due to stellar activity, that must be filtered out in the attempt to detect the Doppler signals related to the planets in the system. In Fig. \ref{fig:rvphasefold} we show the RVs phase folded to ${\rm P_{\rm rot,\,\star}}$ for three sub-samples, to highlight how the variability induced by the stellar activity changes on a seasonal basis. There are no significant peaks at low frequencies, contrary to what is observed for the FWHM and \logrhk.


\section{Photometric and spectroscopic data modeling} \label{sec:datamodelling}

\subsection{Transit light curve modeling} \label{sec:transitmodel}
In this study, we do not provide additional photometric transits to those that have already been presented and analysed by \cite{osborn2022} and \cite{Zhang_2023}. Nonetheless, we performed a fit of all the available TESS, CHEOPS, and LCO/MuSCAT3 transit light curves using our own extracted TESS photometric dataset (Sect. \ref{sec:tessdata}), and the CHEOPS and LCO/MuSCAT3 detrended light curves published by \cite{osborn2022}.

We modelled the transits using the publicly available code \texttt{batman} \citep{Kreidberg2015} assuming circular orbits. We explored the full parameter space using the Monte Carlo (MC) nested sampler and Bayesian inference tool \texttt{MULTINEST V3.10} (e.g. \citealt{Feroz2019}), through the \texttt{pyMULTINEST} wrapper \citep{Buchner2014}. In place of the ratio between the planetary semi-major axis and the stellar radius $a_{\rm p}/R_{\star}$ (which is an input parameter required by \texttt{batman}), we used the stellar density $\rho_{*}$ as a free parameter of the model (e.g. \citealt{2007ApJ...664.1190S}). For $\rho_{*}$ we adopted a Gaussian prior based on the mass and radius derived in Sect. \ref{sec:stellarparam}, from which we derived the $a_{\rm p}/R_{\star}$ ratios at each step of the MC sampling. We adopted a quadratic law for the limb darkening (LD), and fitted the coefficients $u_{\rm 1}$ and $u_{\rm 2}$ using the parametrization and uniform priors for the coefficients q$_1$ and q$_2$ given by \cite{kipping2010} (see Eq. 15 and 16 therein). We used a different pair of LD coefficients for each telescope. We used uniform priors for the inclination angle of the orbital planes $\mathcal{U}$(80,90) degrees), and for the relative radius ratios $R_p/R_\star$ ($\mathcal{U}$(0,1)). We included uncorrelated jitter terms and offsets in the model (uniform priors) for each instrument. Following the methodology outlined in \cite{2022MNRAS.516.4432M}, we used \textit{Gaia} DR3 data to measure the dilution factor, and identify nearby contaminating stars that might be blended with the target. The dilution factor is defined as the total flux from contaminant stars that fall into the photometric aperture divided by the flux of the target star. We found that for TOI-2076 an almost negligible dilution factor ($\sim$0.0005), which is so low that it is not necessary to correct the planetary radii measurements derived from the fit of the light curves.

\begin{figure}
    \centering
    \includegraphics[width=0.415\textwidth]{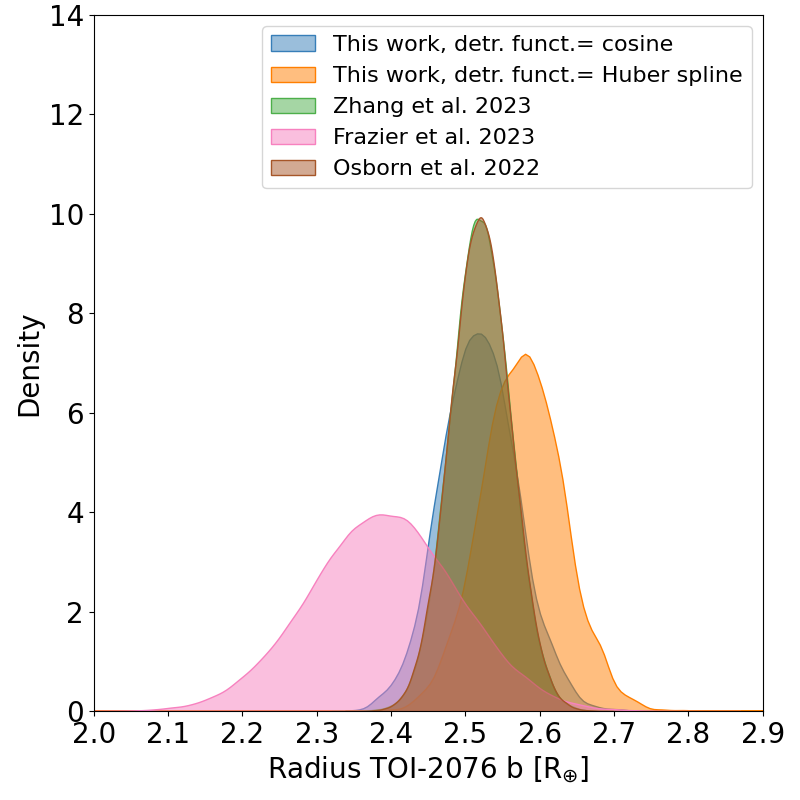}\\
    \includegraphics[width=0.415\textwidth]{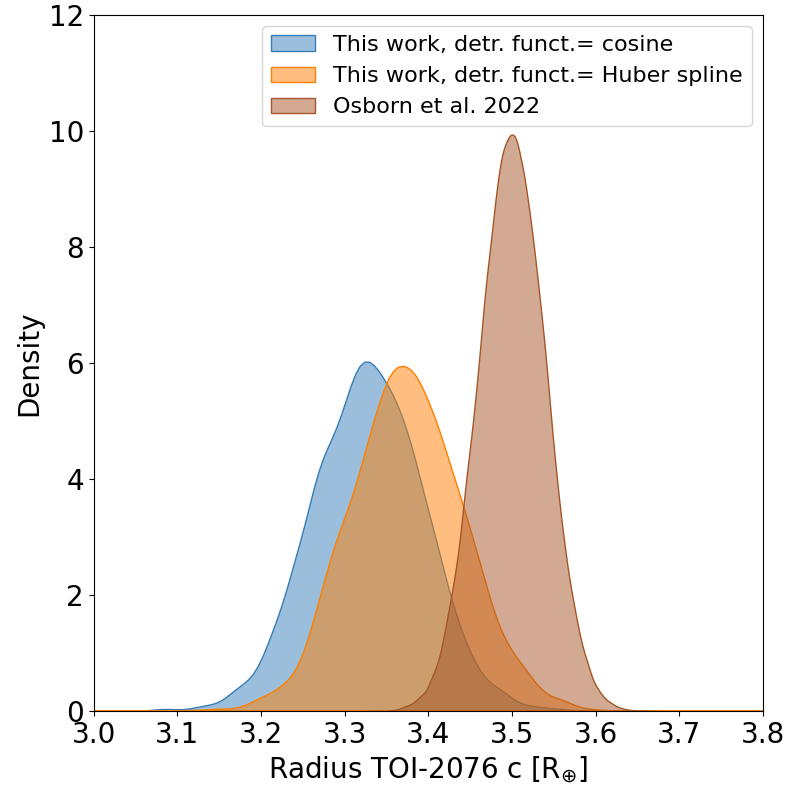}\\
    \includegraphics[width=0.40\textwidth]{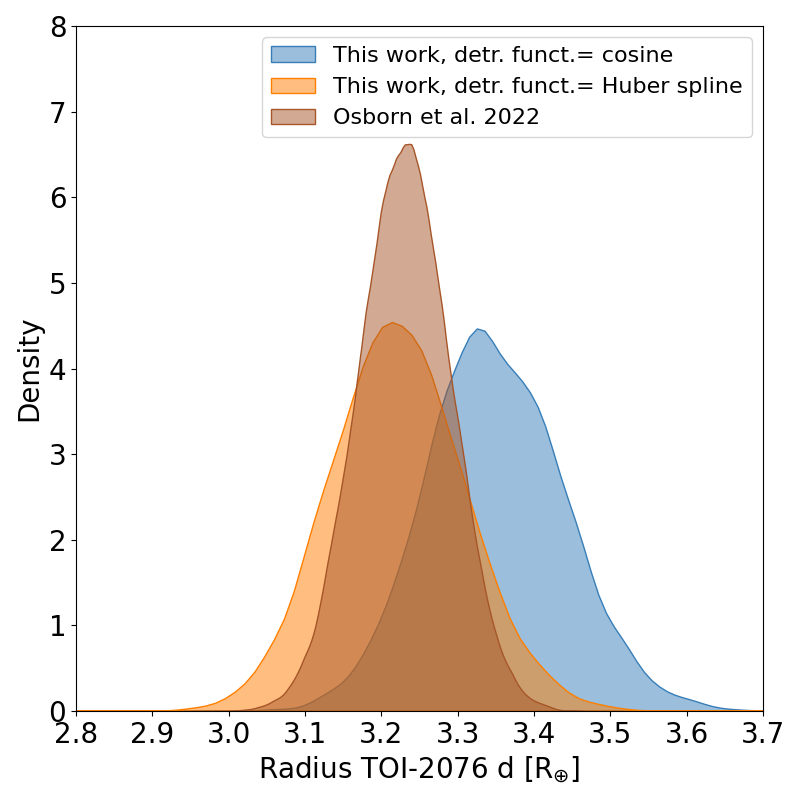}
    \caption{Posterior distributions of the planetary radii derived from the analysis of the available light curves, which include our version of the TESS light curve, which was detrended using two different algorithms, as described in Sect. \ref{sec:tessdata}). We also show the posterior distributions corresponding to the radius measurements available in the literature. The posteriors for planet $b$ corresponding to the measurements of \cite{osborn2022} and \cite{Zhang_2023} are equal and, therefore, are not discernible.}
    \label{fig:raddist}
\end{figure}

Our derived posterior distributions for the planetary radii are shown in Fig. \ref{fig:raddist}, together with those corresponding to the measurements available in the literature. According to our analysis, the planetary radii are $R_b$=2.51$\pm$0.05 \rearth, $R_c$=3.33$\pm$0.07 \rearth, and $R_d$=3.35$^{+0.09}_{-0.08}$ \rearth, for a detrending that employs a sum of sines and cosines, and $R_b$=2.58$\pm$0.05 \rearth, $R_c$=3.37$\pm$0.07 \rearth, and $R_d$=3.22$^{+0.09}_{-0.08}$ \rearth, for a detrending that uses a spline function with a robust Huber estimator. There is an agreement within 1$\sigma$ between our different measurements. In this work, we take the weighted average of the values as our reference planet radii: $R_b$=2.54$\pm$0.04 \rearth, $R_c$=3.35$\pm$0.05 \rearth, and $R_d$=3.29$\pm$0.06 \rearth. We verified that the radius measurements are not influenced by the presence of TTVs for planets $b$ and $c$ (of the order of a few minutes), as reported by \cite{osborn2022}. 
We confirm that all three planets have sub-Neptune sizes, and that the innermost planet $b$ has the smallest radius. We note that, for planet $c$, we find a smaller radius with respect to that measured by \cite{osborn2022}. The values differ at a level of 2.1 and 1.6 $\sigma$, depending on the data detrending method. Nonetheless, our analysis includes one more TESS transit with respect to the sample analysed by \cite{osborn2022}. 

\subsection{RV modeling and planetary mass measurements} \label{sec:rvlcmodel}

We analysed the RVs of TOI-2076 with the goal of measuring the masses of the three planets in the system. We tested several models to try to mitigate the dominant stellar activity signals, all based on Gaussian process (GP) regression (see Appendix \ref{app:rvmodeldetails} for details). As we discussed in Sect. \ref{sec:freqcontentanalysis}, the RV time series is dominated by periodic signals unambiguously ascribable to stellar activity, thus we adopt GP-based models that include the stellar rotation period ${\rm P_{\rm rot,\,\star}}$ as a free parameter. We applied each test model to a time series of RVs extracted from HARPS-N spectra using different algorithms, as described in Sect. \ref{sec:harpsrv}. Some of the analysed datasets include all the RVs considered in this study (HARPS-N, CARMENES and NEID data points), while others are limited to the much larger sample of HARPS-N RVs time series.

Taking into account the large number of test models, RV extraction methods, and free parameters, and that the expected planetary Doppler semi-amplitudes are at least one order of magnitude lower than the dominant activity scatter, we did not perform a joint RV+transit fit, because the much longer computational time does not give better results. Instead, we used the best-fit transit ephemeris derived in Sect. \ref{sec:transitmodel} as Gaussian priors for the RV-only modeling. While this approach made our analysis bearable by greatly reducing the computation time, anyway limiting the modeling to RVs only has no significant impact on the results for the dynamical masses in this case.  

For each test model, we used \texttt{MULTINEST} to explore the full parameter space and compute the Bayesian evidence $\ln\mathcal{Z}$. The MC set-up included 500 live points, a sampling efficiency of 0.5, and a Bayesian evidence tolerance of 0.3. 
The GP regression analysis was performed using the publicly available \textsc{Python} module \texttt{george} v0.2.1 \citep{Ambikasaran2015}, integrated within the \texttt{MULTINEST} framework. To perform the multidimensional GP regression analysis we used the \textsc{Python} module \texttt{pyaneti} \citep{pyaneti,pyaneti2} incorporated into our \texttt{MULTINEST} framework.

For the majority of the tests, we modelled the Doppler signals due to the three planets both with Keplerians and fixing the orbit eccentricities to zero, ignoring terms related to planet-planet interactions. For the orbital ephemeris, we adopted Gaussian priors based on our best-fit solution for the transit light curve modeling. In all the cases, through model comparison using $\ln\mathcal{Z}$, we found that the models with circular orbits are statistically favoured. Moreover, still including GP regression, we also performed an RV modeling using the N-body orbital integrator TRAnsits and Dynamics of Exoplanetary Systems (\texttt{TRADES} v2.20.0\footnote{\url{https://github.com/lucaborsato/trades}}; \citealt{borsato2014A&A...571A..38B}). We limited our analysis to the largest dataset represented by the HARPS-N time series to reduce the computational time, with a negligible loss of information. The free parameters related to each planet that we used as input to \texttt{TRADES} are the planet mass, the orbital period, the mean anomaly, the eccentricity, and the argument of periastron. The longitude of the ascending node has been fixed to 180 degrees, and the stellar mass has been fixed to the best-fit value derived in our work (0.849 M$_{\odot}$). In our analysis, \texttt{TRADES} has been incorporated into the \texttt{MULTINEST}+\texttt{george} framework.   

We report in Table \ref{tab:rvpriors} the list of priors used in our analyses. Concerning the results, the GP quasi-periodic regression is able to recover the stellar rotation period with precision at a level of a hundredth of a day for all the models, using a uniform prior $\mathcal{U}$(0,10) days. Examples of the stellar activity term in the RVs, as fitted by our GP regression, are shown in Fig. \ref{fig:gpactivity1} and \ref{fig:gpactivity2}. The median values of the posteriors of the uncorrelated jitter for HARPS-N ($\sigma_{\rm jit,\,HARPS-N}$) are generally close to 10 \ms for all the test models, with the exception of the models based on multidimensional GP regression (see Appendix \ref{app:rvmodeldetails}), for which $\sigma_{\rm jit,\,HARPS-N}$ is typically halved.   

\begin{figure}[ht!]
    \centering
{\includegraphics[width=0.4\textwidth]{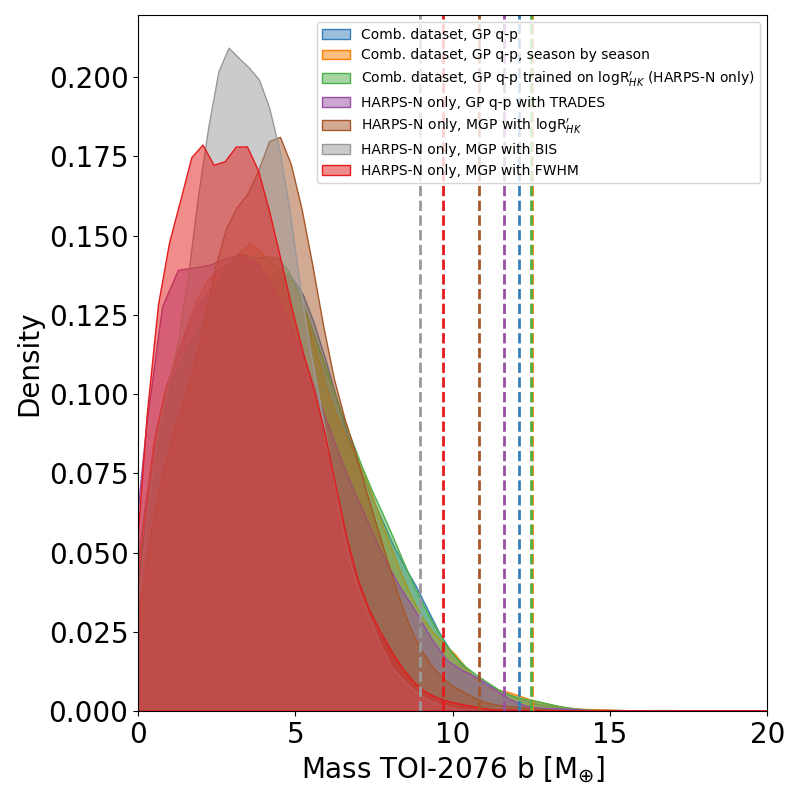}}\\
{\includegraphics[width=0.4\textwidth]{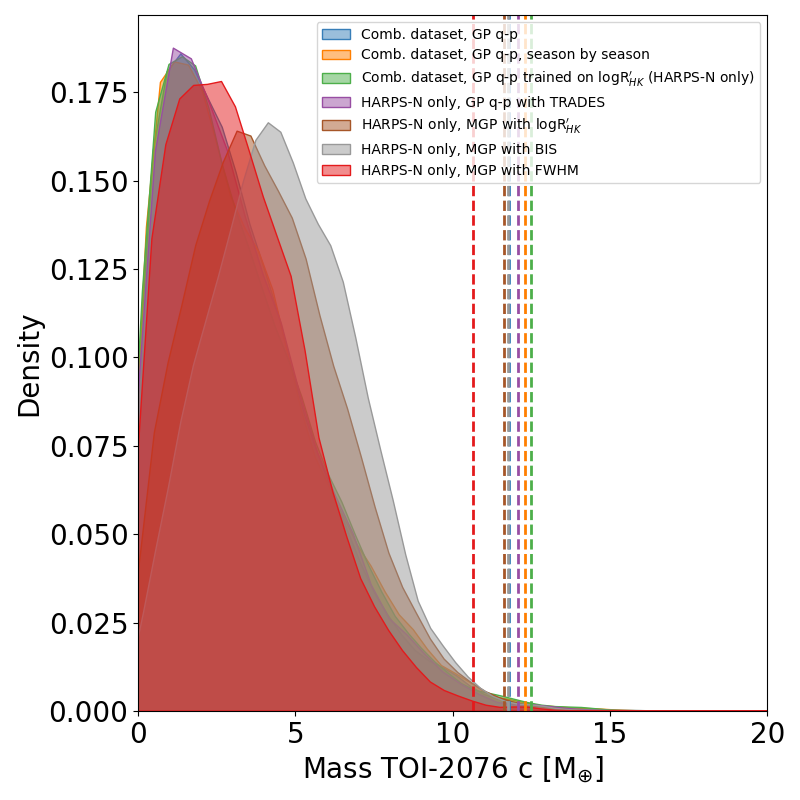}}\\
{\includegraphics[width=0.4\textwidth]{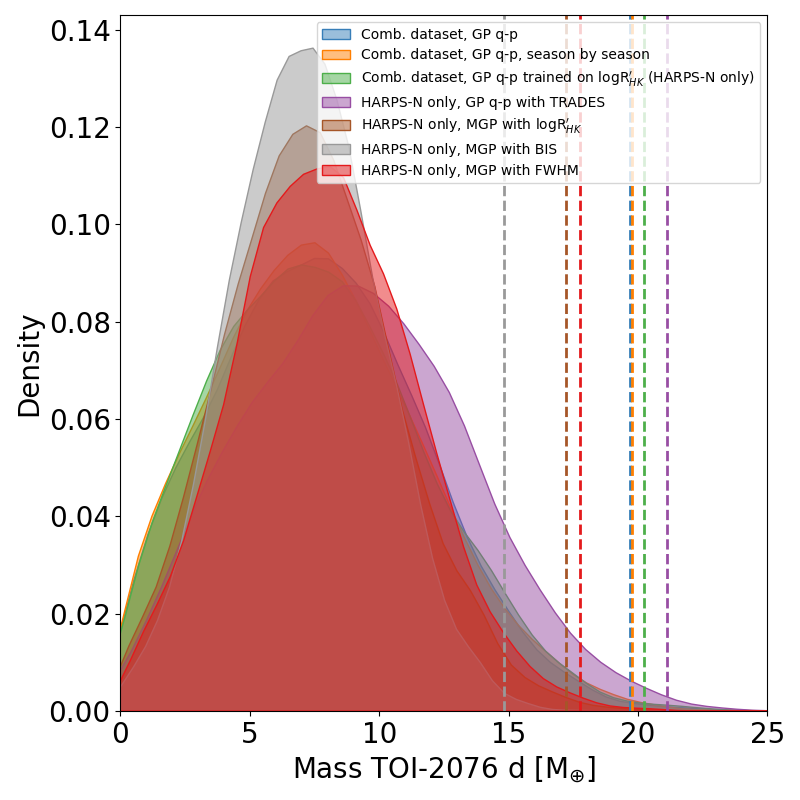}}
    \caption{From top to bottom: posterior mass distributions for the planet BD+40 2790/TOI-2076~b, c, and d. Here we show the results corresponding to the HARPS-N RV dataset extracted with SERVAL, wavelength range nr. 3. The dashed vertical lines indicate the 3$\sigma$ upper limits for each distribution with the same colour.}
    \label{fig:mass_planet}
\end{figure}

We show in Fig. \ref{fig:mass_planet} the posterior distributions of the planetary masses that we obtained for each test model, after selecting one of the different RV extraction methods applied to HARPS-N spectra as an example. Indeed, with a very few noteworthy exceptions, discussed hereafter, the results appear not to be affected by the method used to calculate the HARPS-N RVs. The mass posteriors corresponding to other RV extraction methods are shown in the Appendix (Fig. \ref{fig:mass_planet_b_2}--\ref{fig:mass_planet_d_2}). Concerning planet BD+40 2790~b, all the test models do not result in a statistically significant measurement of the mass (e.g. with a significance level of at least $3\sigma$), independently from the considered HARPS-N RV dataset, although all the posteriors do not peak to zero and generally well match. Our analysis allowed us to derive 3$\sigma$ mass upper limits (model-averaged) in the range $\sim$11--12 \mearth, revealing that BD+40 2790~b is very convincingly less massive than Neptune. A similar conclusion holds for BD+40 2790~c, for which we derive 3$\sigma$ mass upper limits (model-averaged) in the range $\sim$12--13.5 \mearth. We summarise in Table \ref{tab:massupplim} the values of the model-averaged mass upper limits and the corresponding dispersion. We note that the mass upper limits derived for planet $c$ show the lower dispersion within the different models. We also note that we derive a mass for planet $c$ with a significance level greater than 2$\sigma$ using a multidimensional GP regression trained on the BIS activity indicator (grey coloured posteriors), i.e. $m_c$=5.8$^{+2.6}_{-2.5}$, 5.7$^{+2.5}_{-2.4}$, 5.7$^{+2.9}_{-2.6}$ \mearth, for the DRS, SERVAL ``all echelle orders'', and LBL HARPS-N RV dataset respectively. 

For BD+40~2790~d we highlight two main results. Using HARPS-N RVs calculated with the DRS, SERVAL ``all echelle orders'', and LBL pipelines (panels (a), (b), and (d) in Fig. \ref{fig:mass_planet_d}), the peaks of the posterior distributions corresponding to the model which uses the N-body orbital integrator \texttt{TRADES} (purple coloured posteriors) are all shifted toward masses higher than the other models, with best-fit median values that have a statistical significance in the range 1.8--2.1$\sigma$ (respectively, $m_d$=7.9$\pm$4.4, 9.3$^{+4.3}_{-4.4}$, and 8.6$^{+4.7}_{-4.9}$ \mearth). For the case of HARPS-N RVs calculated with SERVAL in the wavelength range nr. 3 (Table \ref{tab:rvsummary}, the posterior distribution corresponding to the \texttt{TRADES}-based model does not change significantly ($m_d$=7.9$\pm$4.4 \mearth), and the posteriors of all the other test models show a very good agreement. These outcomes point out that (i) \texttt{TRADES} finds evidence for a value of $m_d$ in the range $\sim$8--9 \mearth\, independently from which HARPS-N RV ``recipe'' is used in the analysis, and (ii) the accordance among different models when using one specific HARPS-N RV dataset possibly reveals the true mass of BD+40 2790~d. To be conservative, we assume a 3$\sigma$ upper limit (model-averaged) for the mass of BD+40 2790~d in the range $\sim$14--19 \mearth, with a scatter of $\sim$2--3.5 \mearth\, depending on the RV extraction recipe for HARPS-N spectra (Table \ref{tab:massupplim}). However, some models provide hints of the true mass of the planet. Focusing on panel (c) of Fig. \ref{fig:mass_planet_d}, we note that the models based on a multidimensional GP regression provide the most statistically significant mass measurements ($m_d$=7.3$^{+3.4}_{-3.2}$, 7.1$\pm$2.8, and 7.8$^{+3.5}_{-3.3}$ \mearth, respectively for the models that include \logrhk, BIS and FWHM, corresponding to a significance level between 2.3--2.5$\sigma$). In Table \ref{tab:resultmgpmodel} we report those of the MGP model that includes the BIS activity diagnostic, as an example of the results obtained for a specific test model. 

The observed ``chromatic'' dependence of the posterior distributions for the mass of planet $d$ is interesting, and it is confirmed by comparing the results with those of Fig. \ref{fig:additional_mass_planet_d}, which shows the mass posteriors corresponding to HARPS-N RVs extracted with SERVAL in the wavelength ranges 1, 2, and 4, following the nomenclature of Table \ref{tab:rvsummary}. It would be interesting to confirm this ``chromatic'' dependence with a spectroscopic follow-up in the near-IR wavelength range. 

\begin{figure}
    \centering
    \includegraphics[width=0.47\textwidth,clip,trim=1cm 1.5cm 0.5cm 1.5cm]{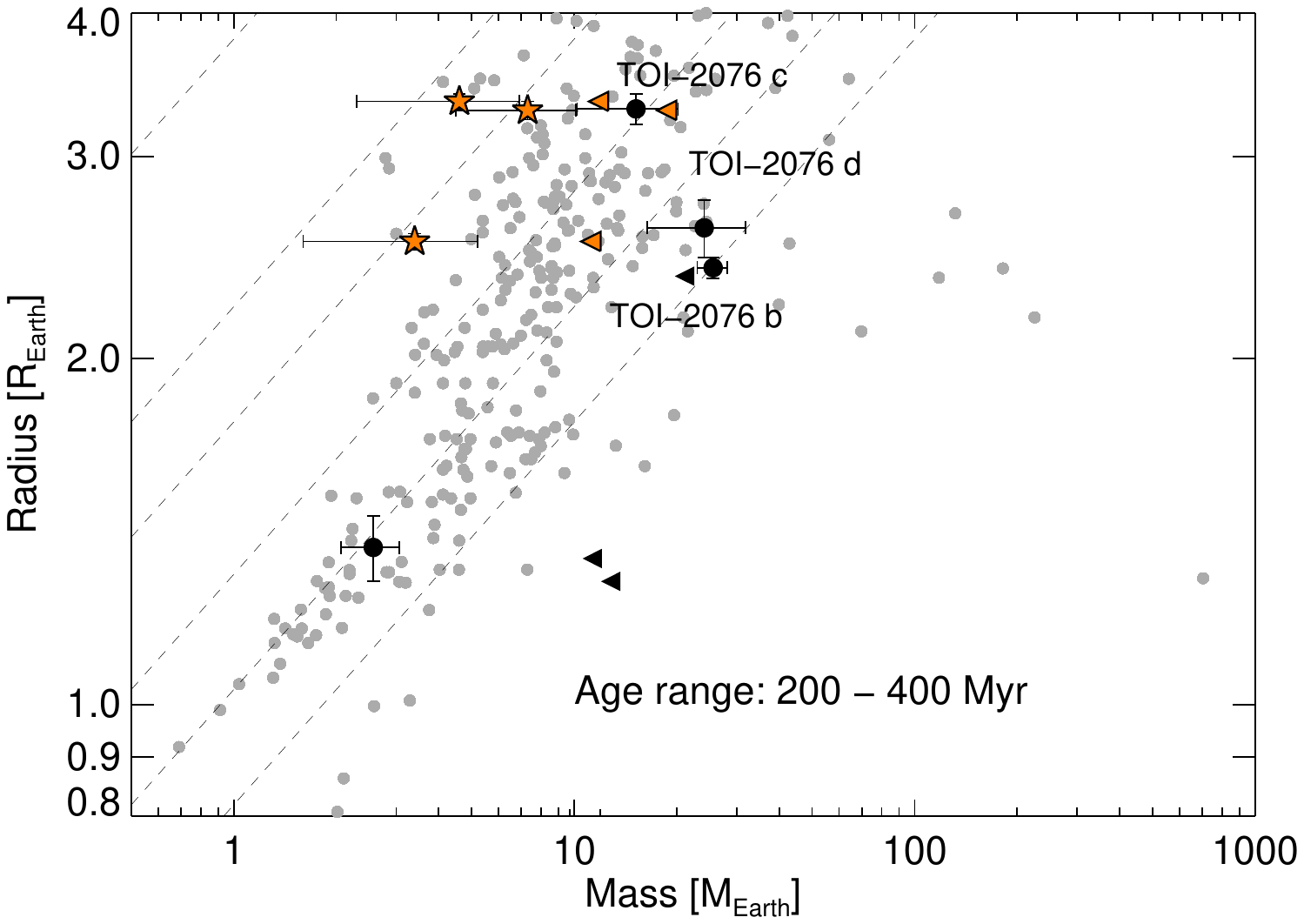}\\
    \caption{Mass-radius diagrams showing planets with radius R$<$4 \rearth, and age in the range 200--400 Myr (black symbols), to which TOI-2076 belongs. Black dots indicate planets with a measured mass, while black triangles represent planets for which only a mass upper limit is available. Grey dots represent a sample of older planets with masses and radii known at least at 30 and 10\%, respectively. To produce this plot, we considered the known planets collected by the Exo-MerCat tool (\citealt{2020A&C....3100370A}, Alei et al. in prep.; user interface available at \url{https://gitlab.com/eleonoraalei/exo-mercat-gui}) by merging the information from the main exoplanets online catalogs (e.g. NASA Exoplanets Archive and Exoplanets Encyclopaedia). The locations of the three planets of the TOI-2076 system are indicated with orange triangles within the appropriate age bin by considering the mass upper limit in Table \ref{tab:massupplim}, while the orange star symbols represent the positions of TOI-2076 b, c, d if we consider the mass obtained with the modeling of the SERVAL RVs (wavelength range nr. 3) with Multi GP using the BIS. Diagonal dashed lines indicate the location of planets with an equal density. }
    \label{fig:massradiusdiag}
\end{figure}

We note that the models based on multidimensional GP regression result with a negative RV acceleration $\dot{\gamma}$, in some cases with a significance close to 3$\sigma$. For instance, we report in Table \ref{tab:resultmgpmodel} $\dot{\gamma}\sim -2.63^{+1.06}_{-0.91} \ms yr^{-1}$. Given the complexity of the RV modeling presented so far, we do not investigate further the reliability and the nature of this result. Data collected over a longer time baseline could help to test whether a negative $\dot{\gamma}$ is the imprint of a long-term activity trend (which we detected in some of the activity indexes, as discussed in Sect. \ref{sec:freqcontentanalysis}), or it is due to an outermost companion (\citealt{hedges2021} excluded nearby companions to contrast limits of 5–8 mag using high-resolution imaging), or it is an artefact of our test models.
We show in Fig. \ref{fig:massradiusdiag} a version of the mass-radius diagram for exoplanets that includes young systems within the age bin 200--400 Myr, within which TOI-2076 lies. The diagram allows for a comparison of the TOI-2076 system with a sub-sample of older (mature) exoplanets for which precise measurements of mass and radius are currently available.


\section{Planetary atmospheric photo-evaporation} \label{sec:atmophotoeva}

We investigated a possible evolution of the planets $b$, $c$ and $d$ on the mass-radius parameter space by evaluating the mass loss rate of the planetary atmospheres using the 
ATES photoionization hydrodynamic code by \cite{Caldiroli+2021,Caldiroli+2022}, who provide an analytic expression for the planetary mass loss rate. 
This expression depends on the planetary 
mean density, gravitational potential energy,
and stellar high-energy flux, and we take into account the temporal evolution of all these quantities, and in particular of the X-rays and EUV irradiation.

Our code is coupled with the planetary core-envelope model introduced by \cite{LopFor14}. 
This model, developed for H-He dominated atmospheres, provides the envelope radius $R_{\rm env}$ as a function of planetary mass $M_{\rm p}$, atmospheric mass fraction $f_{\rm atm}$, the stellar bolometric flux incident on the planet, and the age of the system. 
The model allows us to follow the temporal evolution of the planet's radius, accounting for the cooling and contraction of the envelope and, indirectly, for the mass loss. 
Our photo-evaporation code takes into account also the variation of the stellar bolometric flux and effective temperature, computed by means of the MESA Stellar Tracks (MIST; \citealt{choi+2016}), which are important in determining the planetary radius.
This approach was initially introduced by \cite{Locci19}, where population studies were also conducted, and it was subsequently updated for the analysis of individual systems (e.g.\ \citealt{2022ApJ...925..172M}).

For modeling the XUV irradiation at different ages we adopted two different descriptions (Fig.\ \ref{fig:lxuv}): first, we employed an analytical description as a broken power law of the stellar X-ray emission (5--100\,\AA) vs.\ age \citep{Penz08a}, and we computed the stellar EUV luminosity (100--920\,\AA) using the scaling law proposed by \cite{SF22} (SF22 in the following), which is an updated version of the better-known relation by \cite{SF11}; next, we performed simulations in which the evolution of the stellar X-ray and EUV stellar fluxes is described following the semi-empirical approach by \cite{Johnstone+2021} (Jo21 in the following).

Since our analysis of the RVs mainly resulted in the determination of mass upper limits,
to simulate the planetary evolution due to photo-evaporation 
we explored different sets of possible current masses for each planet, below the values reported in Table \ref{tab:massupplim} and down to 4 \mearth, accounting for possible values of the real masses that are not ruled out from our dynamical analysis.

Given the planetary mass and radius, the code first calculates the mass and radius of the core, $M_{\rm core}$ and $R_{\rm core}$, the radius of the gas envelope, $R_{\rm env}$, and the atmospheric fraction, $f_{\rm atm}$. For this aim, 
we followed the approach in \cite{Fernandez24} and solved a system of 4 equations in 4 unknowns. The unknowns are the four quantities listed above, while the equations include the relation $R_{\rm P} = R_{\rm core} + R_{\rm env}$, an equation linking the planet's mass to the core mass and atmospheric fraction, $f_{\rm atm} = M_{\rm env}/(M_{\rm core} + M_{\rm env})$, the equation for calculating the envelope radius proposed by \cite{LopFor14}, and finally an equation proposed by \cite{Fortney2007} that relates the core radius to the core mass. 
The last equation also allows us to choose between different core compositions, such as a rock/iron or ice/rocky core, with different relative fractions. 
Once the initial values of the 4 unknowns are calculated, the simulation begins by evolving the planetary atmosphere over time, as described earlier, keeping fixed the core mass and size, and the star-planet distance as well.

In this study, we investigated the evolutionary history both in the future and in the past, as done, e.g., by \cite{2022ApJ...925..172M}, \cite{damasso_2023A&A...672A.126D} and \cite{2024A&A...682A.129M}.
The simulations start at 10 Myr, when we assume that the protoplanetary disk is fully dissipated and the planets reached their final orbits. They end at 5 Gyr when we assume that the level of stellar activity has decreased to a point where it no longer significantly influences planetary evolution.

\begin{figure*} 
\begin{tabular}{ccc} 
\includegraphics[width=7.cm]{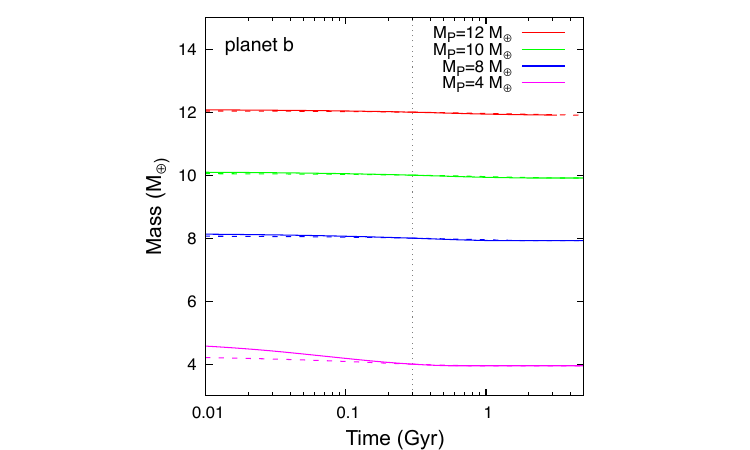} & \hspace{-2cm}\includegraphics[width=7.cm]{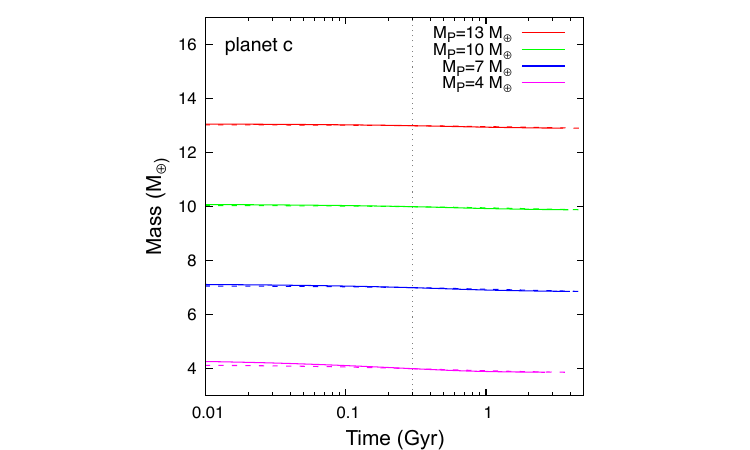} & \hspace{-2cm}\includegraphics[width=7.cm]{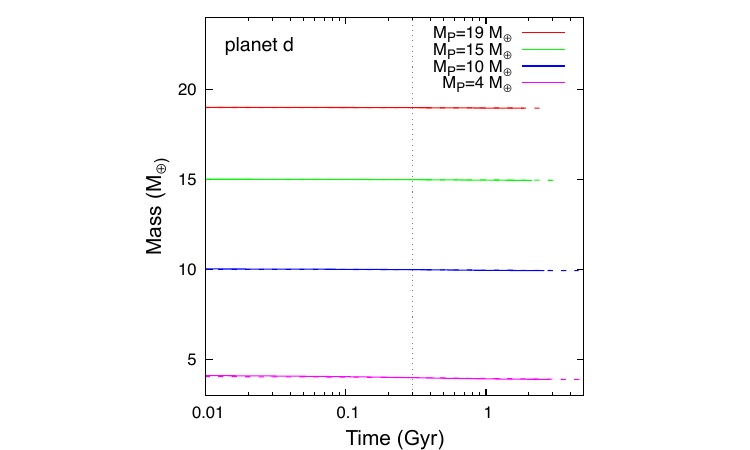} \\
\includegraphics[width=7.cm]{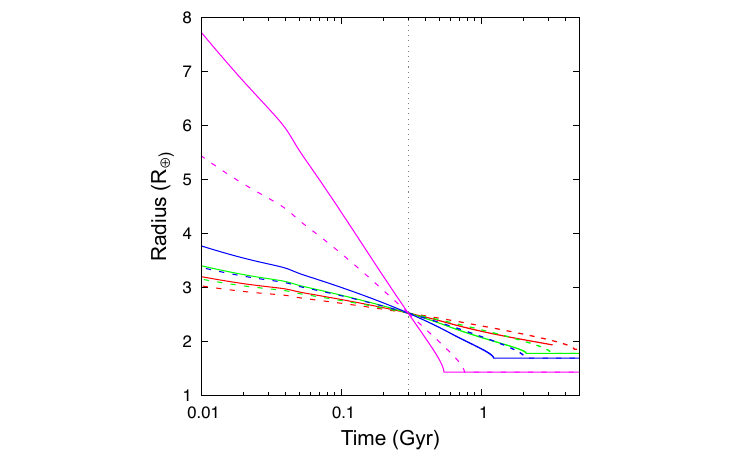} & \hspace{-2cm}  \includegraphics[width=7.cm]{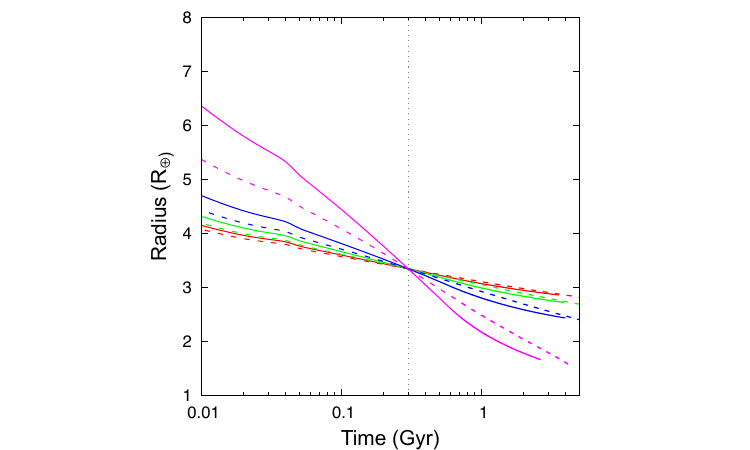}  &  \hspace{-2cm}\includegraphics[width=7.cm]{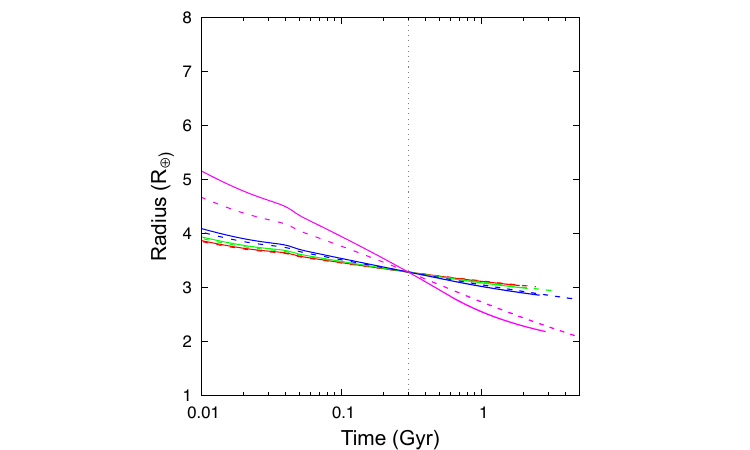}\\
\includegraphics[width=7.cm]{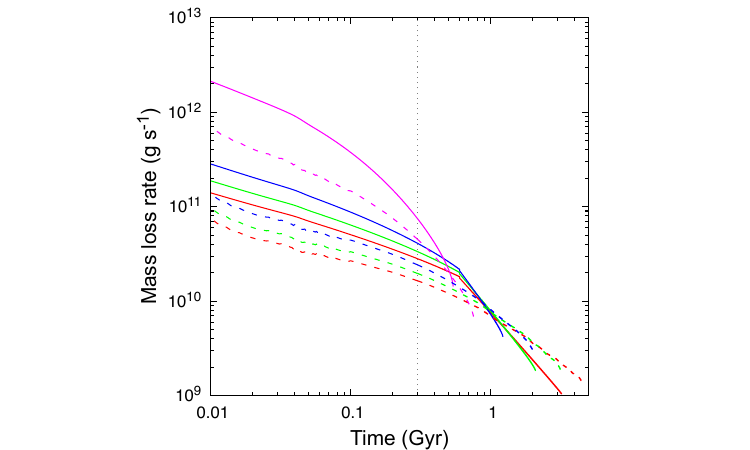} & \hspace{-2cm}\includegraphics[width=7.cm]{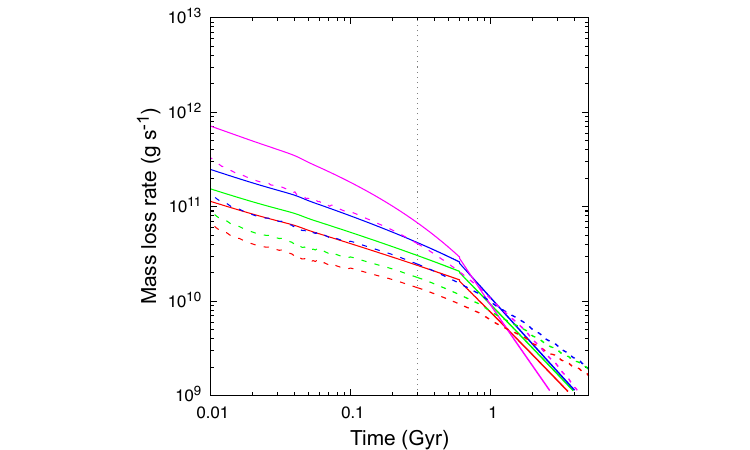}  & \hspace{-2cm}\includegraphics[width=7.cm]{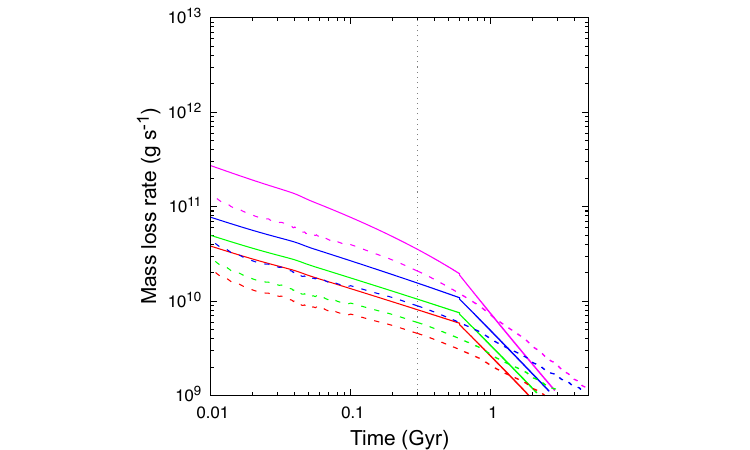}\\
\end{tabular}
\caption{Temporal evolution of mass, radius, and mass loss rate for three different values of planetary mass for each planet in the TOI-2076 system. The left panels show the evolution of planetary parameters for planet $b$, the middle panels for planet $c$, and the right panel for planet $d$. Solid lines represent models in which we describe the stellar X-ray flux evolution using the law proposed by \cite{Penz08a}, and EUV radiation using the law developed by \cite{SF22}. Dashed lines represent models where the XUV evolution is described using the \cite{Johnstone+2021} description. The vertical dotted grey line is located at the current stellar age.}
\label{fig:evap}
\end{figure*}

\begin{figure}
\centering
\includegraphics[width=10cm, trim = 1cm 13cm 2cm 3cm]
{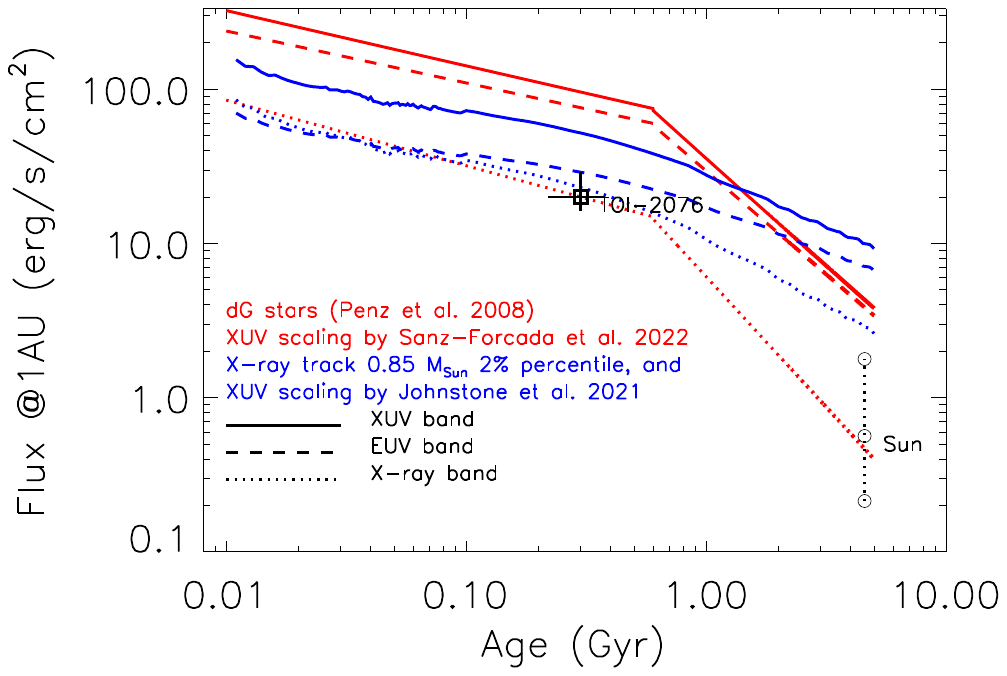} 
\caption{Temporal evolution of the X-ray (5--100\,\AA), EUV (100--920\,\AA), and total XUV flux at 1\,AU. The square symbol indicates the X-ray flux of TOI-2076 at the current stellar age. Different evolutionary models and X-ray to EUV scaling laws are compared (see legend).}
\label{fig:lxuv}
\end{figure}

The results are shown in Figure \ref{fig:evap}, where we present the time evolution of mass, radius, and mass loss rate. The planetary parameters at the current age, at 10\,Myr, and at 1 Gyr are also reported in Table \ref{table:p08sf22results} for the two assumed X-EUV scaling laws. 
In general, we found that all planets behave in the low-gravity regime of the atmospheric hydrodynamic outflow, which occurs when the volume-averaged mean excess energy due to photo-heating exceeds the gravitational binding energy \citep{Caldiroli+2022}. 
In this regime, the photo-evaporation efficiency, $\eta$, is relatively high, but it never reaches the theoretical energy-limited maximum \citep{Erkaev+2007} due to advective and radiative cooling. 
Moreover, we verified that the \cite{Penz08a} evolution of the X-ray luminosity, coupled with the SF22 X-EUV scaling law, produces mass loss rates a factor $\sim 2$ higher than Jo21 at early ages, while the Jo21 approach sustains higher mass loss rates at ages $\gtrsim 1$\,Gyr.
In the following, we describe in more detail the case for each planet.

\textit{Planet b}.
We explored the photo-evaporation history for four possible values of the planetary mass, namely 4, 8, 10, and 12\,\mearth.
We determined core radii in the range
$R_{\rm core} =$ 1.4--1.9\,\rearth, and 
atmospheric mass fractions $f_{\rm env} \sim 0.8$--$1.2\% \mpl$ at present age.

The X-ray flux received by the planet is $F_{\rm x} \sim 2.6 \times 10^3$\,erg/s/cm$^2$. 
Assuming the SF22 scaling law, the EUV flux is $F_{\rm euv} \sim 9.8 \times 10^3$\,erg/s/cm$^2$, and the current photo-evaporation rate results $\Dot{M} \sim 2.8 \times 10^{10}$\,g/s for the highest mass, and $\sim 7.5 \times 10^{10}$\,g/s for the lowest mass considered, corresponding to 0.15--0.22\,\mearth/Gyr.
Instead, the Jo21 X-EUV relation yields $F_{\rm euv} \sim 3.4 \times 10^3$\,erg/s/cm$^2$, and $\Dot{M}$ in the range 1.4--$4.5 \times 10^{10}$\,g/s, which is a factor $\sim 2$ lower than the previous result.

At 10\,Myr the photo-evaporation rate was a factor 5--7 higher, but the total mass lost up to now is $\sim 0.1$\,\mearth\, in all cases, and the planet maintained nearly the same core-envelope structure throughout its lifetime. In fact, we found that planet b probably started with a mass $\sim 10$\,\% higher than the assumed value and a radius between 3 and 8\,\rearth, depending on the assumed mass and evolutionary history of the XUV irradiation.

We found that the innermost planet $b$ will lose completely its gaseous envelope within the age range 0.5--3\,Gyr, unless its mass is very close to the upper limit of 12\,M$_\oplus$.

Observations of absorption in the He\,I 108.3\,nm line were presented by \cite{Zhang_2023} and \cite{gaidos2023MNRAS.518.3777G}
during a single transit of the innermost planet TOI-2076\,b, in agreement with the ongoing photo-evaporation predicted by our model. \cite{Zhang_2023} presented estimates of the mass loss rate, assuming a mass of 9\,M$_\oplus$ and an XUV flux of $9.5 \times 10^3$\,erg/cm$^2$/s. 
They estimated a time scale of 0.6--0.7\,Gyr for losing 1\% of the planetary mass, which we translate into $\Dot{M} = 2.4$--$2.8 \times 10^{10}$\,g/s, or 0.13--0.15\,\mearth/Gyr. 
In spite of several differences in the computation\footnote{\cite{Zhang_2023} assumed an X-ray luminosity $\sim 20$\% lower than our estimate, but a higher flux at the planet due to the wrong adoption of a semi-major axis of 0.0631\,au for planet $b$.}, this estimate falls in between the two estimates that we derived assuming the same planetary mass, and the two alternative X-EUV scaling laws. Indeed, we obtained XUV fluxes at the planet from $6.0 \times 10^3$\,erg/cm$^2$/s \citep{Johnstone+2021} to $1.2 \times 10^4$\,erg/cm$^2$/s \citep{SF22}, that imply mass loss rates from $2.0 \times 10^{10}$ to $3.7 \times 10^{10}$\,g/s, or 0.1--0.2\,\mearth/Gyr. 
We conclude that a factor 2 uncertainty in these estimates is a good measure of the systematics in this kind of computation. 
On the other hand, \cite{gaidos2023MNRAS.518.3777G} used the 1-D hydrodynamic escape model of \cite{kuby+2018a,kuby+2018b}, and derived a mass loss rate of 0.77\,\mearth/Gyr, assuming a planetary mass of 6.8\,\mearth\footnote{As in \cite{Zhang_2023}, we note that they adopted an incorrect orbital semi-major axis in their computation.}. 

\textit{Planet $c$}.
We explored four different values for the mass also in this case, but with a slightly larger range: 4, 7, 10, and 13\,\mearth.

Although planet $c$ receives a dose of XUV radiation $\sim 2.6$ times lower than planet $b$, the mass loss rates of the two planets are very similar at any age. This is explained by the lower planetary density of planet $c$, which almost compensates the difference in XUV flux.
In fact, the actual mass could be similar to that of planet $b$, but its radius is about 30\% larger, which implies a shallower gravitational potential well.

Our model predicts an atmospheric mass fraction $f_{\rm atm} \sim$ 3.3--3.5$\%$ at current age, and $\sim 4.2$--9.5\% at 10\,Myr, assuming SF22, or 3.6--6.4\% following Jo21.
The larger value of $f_{\rm atm}$ with respect to planet $b$ implies that planet $c$ will not lose its atmosphere completely within 5 Gyr.

\textit{Planet $d$}.
This planet receives an XUV irradiation about a factor of 5 lower than planet $b$. 
We explored a mass range between 4\,\mearth\, and our derived upper limit of 19\,\mearth.
This outermost planet generally shows small variations of mass and radius during the time span of the simulations. 
For a mass $\sim 19$\,\mearth\, this planet remains almost stable against hydrodynamic evaporation, losing only negligible fractions of its envelope due to hydrostatic Jeans escape. In this case, the planet's radius evolves only due to the gravitational contraction of the envelope. However, for a mass of $\sim$4 \mearth\, our model predicts a contraction of $\sim40\%$ of the planet's radius up to an age of 5 Gyr.


\newpage 

\begin{table*}
\centering
\small
\caption{Summary of the results derived from the mass posterior distributions shown in Fig. \ref{fig:mass_planet} and \ref{fig:mass_planet_b_2}--\ref{fig:mass_planet_d_2}, listed according to the HARPS-N RV extraction recipes described in Table \ref{tab:rvsummary}.}
\label{tab:massupplim}
    \begin{tabular}{lccc}
    \hline
    \textbf{HARPS-N RV dataset} & \textbf{Model-averaged}  & \textbf{Mean mass} & \textbf{Mass upper limit} \\ 
    & \textbf{best-fit mass\tablefootmark{a}} & \textbf{upper limit [\mearth]\tablefootmark{b}} & \textbf{rms [\mearth]\tablefootmark{b}} \\
    \hline
    \noalign{\smallskip}
    \textit{TOI-2076\,b}\\
    \noalign{\smallskip}
    SERVAL, all orders & $3.9_{-2.3}^{+2.6}$ & $m_b<11.4$ & 1.5 \\
    \noalign{\smallskip}
    SERVAL, wavelength range nr. 3 & $3.8_{-2.3}^{+2.6}$ & $<11.2$ & 1.3\\
    \noalign{\smallskip}
    DRS, all orders & $3.6_{-2.1}^{+2.6}$ & $<10.8$ & 1.2 \\
    \noalign{\smallskip}
    LBL, all orders & $3.7_{-2.2}^{+2.7}$ & $<11.8$ & 1.3\\
    \noalign{\smallskip}
    \textit{TOI-2076\,c}\\
    \noalign{\smallskip}
    SERVAL, all orders & $4.0_{-2.3}^{+2.9}$ & $m_c<12.7$ & 1.1 \\
    \noalign{\smallskip}
    SERVAL, wavelength range nr. 3 & $3.2_{-2.0}^{+2.8}$ & $<11.8$ & 0.6 \\
    \noalign{\smallskip}
    DRS, all orders & $4.0_{-2.3}^{+2.9}$ & $<12.8$ & 0.8 \\
    \noalign{\smallskip}
    LBL, all orders & $3.9_{-2.4}^{+3.1}$ & $<13.4$ & 0.6 \\
    \noalign{\smallskip}
    \textit{TOI-2076\,d}\\
    \noalign{\smallskip}
    SERVAL, all orders & $4.8_{-2.8}^{+3.5}$ & $m_d<15.2$ & 3.5 \\
    \noalign{\smallskip}
    SERVAL, wavelength range nr. 3 & $7.7_{-3.7}^{+3.8}$ & $<18.7$ & 2.0 \\
    \noalign{\smallskip}
    DRS, all orders & $3.9_{-2.5}^{+3.2}$ & $<13.7$ & 3.4 \\
    \noalign{\smallskip}
    LBL, all orders & $4.3_{-2.8}^{+3.6}$ & $<15.2$ & 3.2 \\
    \noalign{\smallskip}
    \hline
    \end{tabular}
    \tablefoot{\small
				\tablefoottext{a}{Model-averaged values of the $50^{\rm th}$ percentiles and corresponding error bars.}
    \tablefoottext{b}{Model-averaged values and rms of the planetary mass $3\sigma$ upper limits.}
			}
\end{table*}

	\begin{table}[htbp]
		\caption{Best-fit values of the free parameters of the model that includes a multidimensional GP regression, using the activity diagnostic BIS and the HARPS-N RVs calculated with \texttt{SERVAL} for the wavelength range nr. 3 (Table \ref{tab:rvsummary}).}
		\label{tab:resultmgpmodel}
		\begin{center}
			\begin{tabular}{ll}
				\hline
				\textbf{Parameter}   & \textbf{Best-fit value}\tablefootmark{a}\\
				\hline
				\noalign{\smallskip} 
				\textit{Multidimensional GP parameters:} & \\
				\noalign{\smallskip}
                    $A_{\rm RV}$\tablefootmark{b} [m/s] & $3.3^{+1.8}_{-1.4}$ \\
                    \noalign{\smallskip}
                    $B_{\rm RV}$ [m/s] & $31.4^{+3.3}_{-2.8}$  \\
                    \noalign{\smallskip}
                    $A_{\rm BIS}$ [m/s] & $2.5^{+1.6}_{-1.5}$ \\
				\noalign{\smallskip}
                    $B_{\rm BIS}$ [m/s] & $-33.5^{+1.6}_{-1.5}$ \\
				\noalign{\smallskip}
				$\lambda$ [d] & $28.1^{\rm +1.7}_{\rm -1.5}$ \\
				\noalign{\smallskip}
				$w$ & $0.46^{+0.03}_{-0.02}$ \\ 
				\noalign{\smallskip}
				$\theta$ [d] & $7.33\pm0.01$ \\      
				\noalign{\smallskip}
				\textit{Planet-related parameters:} &  \\
				\noalign{\smallskip}  
				$K_b$ [\ms] & $1.1\pm0.6$\\
				\noalign{\smallskip} 
				orbital period, $P_b$ [d] & $10.35523\pm0.00001$ \\
				\noalign{\smallskip}
				T$_{conj,\,b}$ [BJD-2450000] & $8950.8289\pm0.0005$ \\
				\noalign{\smallskip}
                    semi-major axis\tablefootmark{c}, $a_b$ [au] & $0.0880\pm0.0009$ \\
				\noalign{\smallskip}
                    mass, $m_b$ [\mearth] & $3.4\pm1.8$ ($<$9.0, 3$\sigma$)\\
                    \noalign{\smallskip}    
                    bulk density, $\rho_b$ [\gcm] & $1.2^{+0.7}_{-0.6}$ \\
                    \noalign{\smallskip}    
				$K_c$ [\ms] & $1.2\pm0.6$ \\
				\noalign{\smallskip} 
				orbital period, $P_c$ [d] & $21.01549\pm0.00003$ \\
				\noalign{\smallskip}
				T$_{conj,\,c}$ [BJD-2450000] & $8937.8283\pm0.0007$ \\
				\noalign{\smallskip}
                    semi-major axis\tablefootmark{c}, $a_c$ [au] & $0.1411^{+0.0015}_{-0.0014}$ \\
				\noalign{\smallskip}
                    mass, $m_c$ [\mearth] & $4.6\pm2.3$ ($<$11.8, 3$\sigma$) \\
				\noalign{\smallskip}
                     bulk density, $\rho_c$ [\gcm] & $0.7^{+0.4}_{-0.3}$ \\
                    \noalign{\smallskip}    
				$K_d$ [\ms] & $1.6\pm0.6$ \\
				\noalign{\smallskip} 
				orbital period, $P_d$ [d] & $35.12551\pm0.00007$ \\
				\noalign{\smallskip}
				T$_{conj,\,d}$ [BJD-2450000] & $8938.296\pm0.001$ \\
				\noalign{\smallskip}
                    semi-major axis\tablefootmark{c}, $a_d$ [au] & $0.1988^{+0.0021}_{-0.0020}$ \\
				\noalign{\smallskip}
                    mass, $m_d$ [\mearth] & $7.3\pm2.8$ ($<$14.8, 3$\sigma$) \\
                    \noalign{\smallskip}
                     bulk density, $\rho_d$ [\gcm] & $1.1^{+0.5}_{-0.4}$ \\
                    \noalign{\smallskip}     
				acceleration, $\dot{\gamma}$ [$\ms d^{-1}$] &  $-0.0072^{+0.0029}_{-0.0025}$ \\
				\noalign{\smallskip} 
				\textit{RV and BIS-related parameters} &  \\
				\noalign{\smallskip}          
				$\sigma_{\rm jit\,RV,\: HARPS-N}$ [\ms] & $3.9_{\rm -0.9}^{\rm +0.8}$ \\ 
				\noalign{\smallskip}
				$\gamma_{\rm RV,\, HARPS-N}$ [\ms] & $2.8\pm0.8$ \\
				\noalign{\smallskip}
                    $\sigma_{\rm jit\,BIS,\: HARPS-N}$ [\ms] & $16.2\pm0.8$ \\ 
				\noalign{\smallskip}
				$\gamma_{\rm BIS,\, HARPS-N}$ [\ms] & $20.8\pm1.2$ \\
				\noalign{\smallskip}
				\hline
				\hline
			\end{tabular}
			\tablefoot{\tiny
				\tablefoottext{a}{The uncertainties are given as the $16^{\rm th}$ and $84^{\rm th}$ percentiles of the posterior distributions. For some of the parameters, we provide the $95^{\rm th}$ percentile in parenthesis.}
				\tablefoottext{b}{The posteriors for $A_{\rm RV}$, $B_{\rm RV}$, $A_{\rm BIS}$, and $B_{\rm BIS}$ are bimodal. $A_{\rm RV}$ and $B_{\rm RV}$ are positively correlated, and we have selected values $>0$ to calculate the percentiles. $A_{\rm BIS}$ and $B_{\rm BIS}$ are anti-correlated, and we have selected positive values for $A_{\rm BIS}$ and negative values for $B_{\rm BIS}$ to calculate their percentiles. }
               \tablefoottext{c}{We note that our derived semi-major axis values are larger than those reported by \cite{hedges2021,osborn2022}, and \cite{Zhang_2023,gaidos2023MNRAS.518.3777G} for planet $b$.}
			}
		\end{center}
	\end{table}

\newpage


\section{Summary and conclusions} \label{sec:conclusions}

In this work, we used a large dataset of high-precision radial velocities of the $\sim$300 Myr old star BD+40\,2790 (TOI-2076) with the main goal of measuring the dynamical masses of the three transiting sub-Neptunes discovered around it. The time variations of the RVs are dominated by signals clearly related to the stellar magnetic activity at a level of $\sim30\ms$, as it is expected for such a young star. This fact, coupled with the presence of multiple planetary Doppler signals embedded in the data, makes the mass measurements a challenging task. We undertook a complex analysis endeavour, by testing up to seven models in the attempt to filter the activity signals out with different approaches, and applying them to HARPS-N RV datasets extracted using four different algorithms, for a total of 28 mass determinations per planet. We considered this approach as one potentially very promising, although very time-consuming, because it calls into question two particularly sensitive aspects when dealing with young and active stars with planets which have not been yet investigated for a large sample of targets. However, this same approach can lead to different results that are not straightforward to compare, making any final decision on which masses to adopt quite complicated.
Therefore, in this perspective and for the specific case of TOI-2076, we decided that a good way to present our results is by showing all the 28 mass posteriors (Fig. \ref{fig:mass_planet} and \ref{fig:mass_planet_b_2}--\ref{fig:mass_planet_d_2}), and summarising the measurements as done in Table \ref{tab:massupplim}, where model-averaged values and $3\sigma$ upper limits of the planetary masses are listed for each RV extraction algorithm that we tested in this study. For all three planets, none of the best-fit mass measurements reaches a $3\sigma$ significance level, therefore our more conservative result is represented by the upper limits, from which we can conclude with a high confidence level that the three planets have sub-Neptune masses. On a model-by-model basis, although it is not possible a priori to assume that one model will provide more trustworthy results than others, nonetheless our multi-model approach reveals that in some cases individual posteriors could provide constraints on the masses of the planets better than just 3$\sigma$ upper limits. For instance, none of the posteriors in Figs. \ref{fig:mass_planet} (first and second panel) and \ref{fig:mass_planet_b_2}--\ref{fig:mass_planet_c_2} peaks at null values in all the cases, and this suggests that the current masses of planets $b$ and $c$ could actually be in the range 4--5 \mearth. For the mass of planet $d$, our approach to the data analysis led to the promising value of $\sim$7--8 \mearth, which is a range supported by all the models when we analyse the RVs extracted after excluding some echelle orders at lower wavelengths from the RV computation. For one model in particular, the statistical significance level of the mass goes up to $2.6\sigma$ (Table \ref{tab:resultmgpmodel}). 
This result would suggest that BD+40 2790 may be a promising target for spectroscopic follow-up in the near-IR for at least confirming the mass of planet $d$, although for a K-type star such as TOI-2076 the expected RV internal precision attainable with new-generation and high-precision instruments at 4m-class telescopes such as SPIRou \citep{2020MNRAS.498.5684D} is lower than that in the optical \citep{Reiners_2020}. Possibly, a TTV follow-up could be a more promising route towards accurate and precise mass measurements.
If we assume the results in Table \ref{tab:resultmgpmodel} as a reliable characterisation of the system, the three planets could be classified as low-density sub-Neptunes. A comparison with a population of mature planets with similar masses (within the error bars) suggests in particular that the radii of planets $b$ and $c$ are inflated (star-like symbols in the second panel of Fig. \ref{fig:massradiusdiag}). Their larger radii could be due to bloated and H-He dominated atmospheres, and ultimately to the young age of the system. If so, the current locations of the planets in the mass-radius diagram should change in the future and move to lower radius values typical of the older population, as expected if their atmospheres will experience mass loss and contraction through photo-evaporation.  

In this regard, in the last part of our work we discussed possible evolutionary pathways of the planetary atmospheres by estimating the mass loss rate through photo-evaporation over time, for a grid of possible current masses selected on the base of our derived planetary mass upper limits. Through this theoretical analysis, we derived the predicted values of planetary mass and radius starting from an age of 10 Myr up to a few Gyr. The simulations show that planets $b$ and $c$ are nowadays experiencing atmospheric mass loss driven by photo-evaporation, and this is a general conclusion that does not depend on their actual masses. We predict for planet $c$ an atmospheric mass fraction larger than for planet $b$, and a mass loss rate slightly lower, hence we expect that the HeI line absorption could be detectable for this planet, similarly to what has been claimed for planet $b$, although the origin of the observed absorption remains controversial \citep{Zhang_2023,gaidos2023MNRAS.518.3777G}. We predict that the radius of planet $d$ will be reduced for the effects of to gravitational contraction of the envelope, rather than photo-evaporation mass loss. 

Based on the same simulations, we can predict what the locations of each planet in the mass-radius diagram of Fig. \ref{fig:massradiusdiag} could be in the future, as a consequence of size contraction mainly driven by photo-evaporation. For instance, at 5 Gyr from now the final radius of planet $b$ is expected in the range $\sim$1.5--2 \rearth; for planet $c$, the radius should settle within $\sim$2.5--3 \rearth, or to a value lower than 2 \rearth\, if the mass is $\sim4$ \mearth; for planet $d$, the expected final radius should be between 2--3 \rearth. Hence, the predicted final positions are expected to be within the regions occupied by the population of mature exoplanets with similar masses.

With constituent planetary sizes $R_{b} = 2.54 \pm 0.04$ $R_{\oplus}$, $R_{c} = 3.35 \pm 0.05$ $R_{\oplus}$, and $R_{b} = 3.29 \pm 0.06$ $R_{\oplus}$, as well as period ratios of $P_{c}/P_{b} \approx 2.01$ and $P_{d}/P_{c} \approx 1.67$, TOI-2076 serves as a prototypical example of  the so-called ``peas-in-a-pod'' configuration. It has been observed that sub-Neptunes orbiting the same star display a striking uniformity in their planetary size, mass, and orbital spacing, such that the overall system architecture resembles peas-in-a-pod (\citealt{millholland}; \citealt{wang_2017}; \citealt{weiss_rad}; \citealt{goyal}). It has also been demonstrated that this size uniformity is itself enhanced within systems containing at least one planetary pair close to first-order mean motion resonance (MMR; \citealt{goyal2}), and, with the proximity of the inner two planets to the 2:1 MMR, the TOI-2076 system fits well within that framework. given the especially strong uniformity in planetary size. Additionally, the outer two planets lie close to the 5:3 MMR and harbor nearly identical radii, providing evidence that planetary size uniformity may hold correspondence with higher-order resonances as well. These highly uniform architectures are believed to emerge as a common and natural consequence of the planetary assembly process, with such possible mechanisms including energy optimization of the pairwise planetary mass budget (\citealt{adams1}; \citealt{adams2}), construction of close-in resonant chains by a convergent migration process (\citealt{terquem}; \citealt{morrison}; \citealt{broz}), sequential formation from a narrow planetesimal ring \citep{batygin_ring}, and planetesimal trapping at disk pressure bumps \citep{xu_bump}. The vast majority of multi-planet configurations are observed to harbor non-resonant architectures \citep{fab}, perhaps owing to widespread dynamical instabilities or giant impacts following the disk epoch (\citealt{izidoro}; \citealt{goldberg}; \citealt{lammers}) that fully disrupt these resonant chains. As such, it may be the case that near-resonant systems such as TOI-2076 instead experienced a more quiescent evolutionary history wherein planetary spacing was gently broadened during the disk lifetime (\citealt{terquem2}; \citealt{choksi}), perhaps as a result of stochastic turbulence during migration (\citealt{batygin_adams}; \citealt{goldberg2}), or secular forcing from a perturbing companion planet \citep{choksi2}. Accordingly, the absence of large-scale dynamical disruptions may have allowed TOI-2076 to retain a greater imprint of its primordial high-uniformity resonant state.

\begin{acknowledgements}

This work has been supported by the PRIN-INAF 2019 "Planetary systems at young ages (PLATEA)" and ASI-INAF agreement n.2018-16-HH.0. A. Ma. also acknowledges partial support from the PRIN-INAF 2019 "HOT-ATMOS". We acknowledge financial support from the Agencia Estatal de Investigaci\'on of the Ministerio de Ciencia e Innovaci\'on MCIN/AEI/10.13039/501100011033 and the ERDF “A way of making Europe” through projects PID2021-125627OB-C32 and PID2022-137241NB-C41, and from the Centre of Excellence “Severo Ochoa” award to the Instituto de Astrofisica de Canarias.
S.W. acknowledges support from Heising-Simons Foundation Grant $\#$2023-4050.
G.N. thanks for the research funding from the Ministry of Education and Science programme the "Excellence Initiative - Research University" conducted at the Centre of Excellence in Astrophysics and Astrochemistry of the Nicolaus Copernicus University in Toru\'n, Poland. L.M. acknowledges financial contribution from PRIN MUR 2022 project 2022J4H55R. L.B. and T.Z. acknowledge the support by the CHEOPS ASI-INAF agreement n. 2019-29-HH.0. T.Z. acknowledges NVIDIA Academic Hardware Grant Program for the use of the Titan V GPU card, and the Italian MUR Departments of Excellence grant 2023-2027 “Quantum Frontiers”. This paper contains data taken with the NEID instrument, which was funded by the NASA-NSF Exoplanet Observational Research (NN-EXPLORE) partnership and built by Pennsylvania State University. NEID is installed on the WIYN telescope, which is operated by the National Optical Astronomy Observatory, and the NEID archive is operated by the NASA Exoplanet Science Institute at the California Institute of Technology. NN-EXPLORE is managed by the Jet Propulsion Laboratory, California Institute of Technology under contract with the National Aeronautics and Space Administration. This work has made use of data from the European Space Agency (ESA) mission
{\it Gaia} (\url{https://www.cosmos.esa.int/gaia}), processed by the {\it Gaia}
Data Processing and Analysis Consortium (DPAC,
\url{https://www.cosmos.esa.int/web/gaia/dpac/consortium}). Funding for the DPAC
has been provided by national institutions, in particular the institutions
participating in the {\it Gaia} Multilateral Agreement.
\end{acknowledgements}

%
%

\bibliographystyle{aa}
\bibliography{toi2076}

\begin{thebibliography}{135}
\expandafter\ifx\csname natexlab\endcsname\relax\def\natexlab#1{#1}\fi

\bibitem[{{Adams} {et~al.}(2020{\natexlab{a}}){Adams}, {Batygin}, \&
  {Bloch}}]{adams2}
{Adams}, F.~C., {Batygin}, K., \& {Bloch}, A.~M. 2020{\natexlab{a}}, \mnras,
  494, 2289

\bibitem[{{Adams} {et~al.}(2020{\natexlab{b}}){Adams}, {Batygin}, {Bloch}, \&
  {Laughlin}}]{adams1}
{Adams}, F.~C., {Batygin}, K., {Bloch}, A.~M., \& {Laughlin}, G.
  2020{\natexlab{b}}, \mnras, 493, 5520

\bibitem[{{Alei} {et~al.}(2020){Alei}, {Claudi}, {Bignamini}, \&
  {Molinaro}}]{2020A&C....3100370A}
{Alei}, E., {Claudi}, R., {Bignamini}, A., \& {Molinaro}, M. 2020, Astronomy
  and Computing, 31, 100370

\bibitem[{{Ambikasaran} {et~al.}(2015){Ambikasaran}, {Foreman-Mackey},
  {Greengard}, {Hogg}, \& {O'Neil}}]{Ambikasaran2015}
{Ambikasaran}, S., {Foreman-Mackey}, D., {Greengard}, L., {Hogg}, D.~W., \&
  {O'Neil}, M. 2015, IEEE Transactions on Pattern Analysis and Machine
  Intelligence, 38 [\eprint[arXiv]{1403.6015}]

\bibitem[{{Artigau} {et~al.}(2022){Artigau}, {Cadieux}, {Cook}, {Doyon},
  {Vandal}, {Donati}, {Moutou}, {Delfosse}, {Fouqu{\'e}}, {Martioli}, {Bouchy},
  {Parsons}, {Carmona}, {Dumusque}, {Astudillo-Defru}, {Bonfils}, \&
  {Mignon}}]{Artigau_2022}
{Artigau}, {\'E}., {Cadieux}, C., {Cook}, N.~J., {et~al.} 2022, \aj, 164, 84

\bibitem[{{Baranne} {et~al.}(1996){Baranne}, {Queloz}, {Mayor}, {Adrianzyk},
  {Knispel}, {Kohler}, {Lacroix}, {Meunier}, {Rimbaud}, \&
  {Vin}}]{baranne1996A&AS..119..373B}
{Baranne}, A., {Queloz}, D., {Mayor}, M., {et~al.} 1996, \aaps, 119, 373

\bibitem[{{Baratella} {et~al.}(2020{\natexlab{a}}){Baratella}, {D'Orazi},
  {Biazzo}, {Desidera}, {Gratton}, {Benatti}, {Bignamini}, {Carleo}, {Cecconi},
  {Claudi}, {Cosentino}, {Ghedina}, {Harutyunyan}, {Lanza}, {Malavolta},
  {Maldonado}, {Mallonn}, {Messina}, {Micela}, {Molinari}, {Poretti},
  {Scandariato}, \& {Sozzetti}}]{2020baratella_gaps}
{Baratella}, M., {D'Orazi}, V., {Biazzo}, K., {et~al.} 2020{\natexlab{a}},
  \aap, 640, A123

\bibitem[{{Baratella} {et~al.}(2020{\natexlab{b}}){Baratella}, {D'Orazi},
  {Carraro}, {Desidera}, {Randich}, {Magrini}, {Adibekyan}, {Smiljanic},
  {Spina}, {Tsantaki}, {Tautvai{\v{s}}ien{\.{e}}}, {Sousa}, {Jofr{\'e}},
  {Jim{\'e}nez-Esteban}, {Delgado-Mena}, {Martell}, {Van der Swaelmen},
  {Roccatagliata}, {Gilmore}, {Alfaro}, {Bayo}, {Bensby}, {Bragaglia},
  {Franciosini}, {Gonneau}, {Heiter}, {Hourihane}, {Jeffries}, {Koposov},
  {Morbidelli}, {Prisinzano}, {Sacco}, {Sbordone}, {Worley}, {Zaggia}, \&
  {Lewis}}]{2020baratella_ges}
{Baratella}, M., {D'Orazi}, V., {Carraro}, G., {et~al.} 2020{\natexlab{b}},
  \aap, 634, A34

\bibitem[{{Barrag{\'a}n} {et~al.}(2022{\natexlab{a}}){Barrag{\'a}n}, {Aigrain},
  {Rajpaul}, \& {Zicher}}]{pyaneti2}
{Barrag{\'a}n}, O., {Aigrain}, S., {Rajpaul}, V.~M., \& {Zicher}, N.
  2022{\natexlab{a}}, \mnras, 509, 866

\bibitem[{{Barrag{\'a}n} {et~al.}(2022{\natexlab{b}}){Barrag{\'a}n},
  {Armstrong}, {Gandolfi}, {Carleo}, {Vidotto}, {Villarreal D'Angelo},
  {Oklop{\v{c}}i{\'c}}, {Isaacson}, {Oddo}, {Collins}, {Fridlund}, {Sousa},
  {Persson}, {Hellier}, {Howell}, {Howard}, {Redfield}, {Eisner}, {Georgieva},
  {Dragomir}, {Bayliss}, {Nielsen}, {Klein}, {Aigrain}, {Zhang}, {Teske},
  {Twicken}, {Jenkins}, {Esposito}, {Van Eylen}, {Rodler}, {Adibekyan},
  {Alarcon}, {Anderson}, {Akana Murphy}, {Barrado}, {Barros}, {Benneke},
  {Bouchy}, {Bryant}, {Butler}, {Burt}, {Cabrera}, {Casewell}, {Chaturvedi},
  {Cloutier}, {Cochran}, {Crane}, {Crossfield}, {Crouzet}, {Collins}, {Dai},
  {Deeg}, {Deline}, {Demangeon}, {Dumusque}, {Figueira}, {Furlan}, {Gnilka},
  {Goad}, {Goffo}, {Guti{\'e}rrez-Canales}, {Hadjigeorghiou}, {Hartman},
  {Hatzes}, {Harris}, {Henderson}, {Hirano}, {Hojjatpanah}, {Hoyer},
  {Kab{\'a}th}, {Korth}, {Lillo-Box}, {Luque}, {Marmier}, {Mo{\v{c}}nik},
  {Muresan}, {Murgas}, {Nagel}, {Osborne}, {Osborn}, {Osborn}, {Palle},
  {Raimbault}, {Ricker}, {Rubenzahl}, {Stockdale}, {Santos}, {Scott},
  {Schwarz}, {Shectman}, {Raimbault}, {Seager}, {S{\'e}gransan}, {Serrano},
  {Skarka}, {Smith}, {{\v{S}}ubjak}, {Tan}, {Udry}, {Watson}, {Wheatley},
  {West}, {Winn}, {Wang}, {Wolfgang}, \& {Ziegler}}]{2022MNRAS.514.1606B}
{Barrag{\'a}n}, O., {Armstrong}, D.~J., {Gandolfi}, D., {et~al.}
  2022{\natexlab{b}}, \mnras, 514, 1606

\bibitem[{Barrag\'an {et~al.}(2019)Barrag\'an, Gandolfi, \&
  Antoniciello}]{pyaneti}
Barrag\'an, O., Gandolfi, D., \& Antoniciello, G. 2019, \mnras, 482, 1017

\bibitem[{Barragán {et~al.}(2023)Barragán, Gillen, Aigrain, Meech, Klein,
  Nielsen, Yu, O’Sullivan, Nicholson, \&
  Lillo-Box}]{barragan_10.1093/mnras/stad1139}
Barragán, O., Gillen, E., Aigrain, S., {et~al.} 2023, Monthly Notices of the
  Royal Astronomical Society, 522, 3458

\bibitem[{{Baruteau} {et~al.}(2014){Baruteau}, {Crida}, {Paardekooper},
  {Masset}, {Guilet}, {Bitsch}, {Nelson}, {Kley}, \&
  {Papaloizou}}]{2014prpl.conf..667B}
{Baruteau}, C., {Crida}, A., {Paardekooper}, S.~J., {et~al.} 2014, in
  Protostars and Planets VI, ed. H.~{Beuther}, R.~S. {Klessen}, C.~P.
  {Dullemond}, \& T.~{Henning}, 667--689

\bibitem[{{Batygin} \& {Adams}(2017)}]{batygin_adams}
{Batygin}, K. \& {Adams}, F.~C. 2017, \aj, 153, 120

\bibitem[{{Batygin} \& {Morbidelli}(2023)}]{batygin_ring}
{Batygin}, K. \& {Morbidelli}, A. 2023, Nature Astronomy, 7, 330

\bibitem[{{Benatti} {et~al.}(2023{\natexlab{a}}){Benatti}, {Desidera}, \& {GAPS
  Young Objects Team}}]{2023MmSAI..94b.195B}
{Benatti}, S., {Desidera}, S., \& {GAPS Young Objects Team}.
  2023{\natexlab{a}}, in Memorie della Societa Astronomica Italiana, Vol.~94,
  195

\bibitem[{{Benatti} {et~al.}(2023{\natexlab{b}}){Benatti}, {Desidera}, \&
  Team}]{BenattiSAIt}
{Benatti}, S., {Desidera}, S., \& Team, G.-Y. 2023{\natexlab{b}}, in HACK100,
  ed. P.~{Bonifacio} \& P.~{Molaro}, Vol.~94, 195

\bibitem[{{Biazzo} {et~al.}(2022){Biazzo}, {D'Orazi}, {Desidera}, {Turrini},
  {Benatti}, {Gratton}, {Magrini}, {Sozzetti}, {Baratella}, {Bonomo}, {Borsa},
  {Claudi}, {Covino}, {Damasso}, {Di Mauro}, {Lanza}, {Maggio}, {Malavolta},
  {Maldonado}, {Marzari}, {Micela}, {Poretti}, {Vitello}, {Affer}, {Bignamini},
  {Carleo}, {Cosentino}, {Fiorenzano}, {Giacobbe}, {Harutyunyan}, {Leto},
  {Mancini}, {Molinari}, {Molinaro}, {Nardiello}, {Nascimbeni}, {Pagano},
  {Pedani}, {Piotto}, {Rainer}, \& {Scandariato}}]{Biazzoetal2022}
{Biazzo}, K., {D'Orazi}, V., {Desidera}, S., {et~al.} 2022, \aap, 664, A161

\bibitem[{{Borsato} {et~al.}(2014){Borsato}, {Marzari}, {Nascimbeni}, {Piotto},
  {Granata}, {Bedin}, \& {Malavolta}}]{borsato2014A&A...571A..38B}
{Borsato}, L., {Marzari}, F., {Nascimbeni}, V., {et~al.} 2014, \aap, 571, A38

\bibitem[{{Bressan} {et~al.}(2012){Bressan}, {Marigo}, {Girardi}, {Salasnich},
  {Dal Cero}, {Rubele}, \& {Nanni}}]{2012MNRAS.427..127B}
{Bressan}, A., {Marigo}, P., {Girardi}, L., {et~al.} 2012, \mnras, 427, 127

\bibitem[{{Brewer} {et~al.}(2016){Brewer}, {Fischer}, {Valenti}, \&
  {Piskunov}}]{Breweretal2016}
{Brewer}, J.~M., {Fischer}, D.~A., {Valenti}, J.~A., \& {Piskunov}, N. 2016,
  \apjs, 225, 32

\bibitem[{{Bro{\v{z}}} {et~al.}(2021){Bro{\v{z}}}, {Chrenko}, {Nesvorn{\'y}},
  \& {Dauphas}}]{broz}
{Bro{\v{z}}}, M., {Chrenko}, O., {Nesvorn{\'y}}, D., \& {Dauphas}, N. 2021,
  Nature Astronomy, 5, 898

\bibitem[{{Buchner} {et~al.}(2014){Buchner}, {Georgakakis}, {Nandra}, {Hsu},
  {Rangel}, {Brightman}, {Merloni}, {Salvato}, {Donley}, \&
  {Kocevski}}]{Buchner2014}
{Buchner}, J., {Georgakakis}, A., {Nandra}, K., {et~al.} 2014, \aap, 564, A125

\bibitem[{{Butters} {et~al.}(2010){Butters}, {West}, {Anderson}, {Collier
  Cameron}, {Clarkson}, {Enoch}, {Haswell}, {Hellier}, {Horne}, {Joshi},
  {Kane}, {Lister}, {Maxted}, {Parley}, {Pollacco}, {Smalley}, {Street},
  {Todd}, {Wheatley}, \& {Wilson}}]{2010A&A...520L..10B}
{Butters}, O.~W., {West}, R.~G., {Anderson}, D.~R., {et~al.} 2010, \aap, 520,
  L10

\bibitem[{{Caldiroli} {et~al.}(2021){Caldiroli}, {Haardt}, {Gallo}, {Spinelli},
  {Malsky}, \& {Rauscher}}]{Caldiroli+2021}
{Caldiroli}, A., {Haardt}, F., {Gallo}, E., {et~al.} 2021, \aap, 655, A30

\bibitem[{{Caldiroli} {et~al.}(2022){Caldiroli}, {Haardt}, {Gallo}, {Spinelli},
  {Malsky}, \& {Rauscher}}]{Caldiroli+2022}
{Caldiroli}, A., {Haardt}, F., {Gallo}, E., {et~al.} 2022, \aap, 663, A122

\bibitem[{{Cale} {et~al.}(2021){Cale}, {Reefe}, {Plavchan}, {Tanner}, {Gaidos},
  {Gagn{\'e}}, {Gao}, {Kane}, {B{\'e}jar}, {Lodieu}, {Anglada-Escud{\'e}},
  {Ribas}, {Pall{\'e}}, {Quirrenbach}, {Amado}, {Reiners}, {Caballero}, {Rosa
  Zapatero Osorio}, {Dreizler}, {Howard}, {Fulton}, {Xuesong Wang}, {Collins},
  {El Mufti}, {Wittrock}, {Gilbert}, {Barclay}, {Klein}, {Martioli},
  {Wittenmyer}, {Wright}, {Addison}, {Hirano}, {Tamura}, {Kotani}, {Narita},
  {Vermilion}, {Lee}, {Geneser}, {Teske}, {Quinn}, {Latham}, {Esquerdo},
  {Calkins}, {Berlind}, {Zohrabi}, {Stibbards}, {Kotnana}, {Jenkins},
  {Twicken}, {Henze}, {Kidwell}, {Burke}, {Villase{\~n}or}, \&
  {Boyd}}]{2021AJ....162..295C}
{Cale}, B.~L., {Reefe}, M., {Plavchan}, P., {et~al.} 2021, \aj, 162, 295

\bibitem[{{Canocchi, G.} {et~al.}(2023){Canocchi, G.}, {Malavolta, L.},
  {Pagano, I.}, {Barragán, O.}, {Piotto, G.}, {Aigrain, S.}, {Desidera, S.},
  {Grziwa, S.}, {Cabrera, J.}, \& {Rauer, H.}}]{canocchi23}
{Canocchi, G.}, {Malavolta, L.}, {Pagano, I.}, {et~al.} 2023, A\&A, 672, A144

\bibitem[{{Carleo} {et~al.}(2021){Carleo}, {Desidera}, {Nardiello},
  {Malavolta}, {Lanza}, {Livingston}, {Locci}, {Marzari}, {Messina}, {Turrini},
  {Baratella}, {Borsa}, {D'Orazi}, {Nascimbeni}, {Pinamonti}, {Rainer}, {Alei},
  {Bignamini}, {Gratton}, {Micela}, {Montalto}, {Sozzetti}, {Squicciarini},
  {Affer}, {Benatti}, {Biazzo}, {Bonomo}, {Claudi}, {Cosentino}, {Covino},
  {Damasso}, {Esposito}, {Fiorenzano}, {Frustagli}, {Giacobbe}, {Harutyunyan},
  {Leto}, {Magazz{\`u}}, {Maggio}, {Mainella}, {Maldonado}, {Mallonn},
  {Mancini}, {Molinari}, {Molinaro}, {Pagano}, {Pedani}, {Piotto}, {Poretti},
  {Redfield}, \& {Scandariato}}]{2021A&A...645A..71C}
{Carleo}, I., {Desidera}, S., {Nardiello}, D., {et~al.} 2021, \aap, 645, A71

\bibitem[{{Carleo} {et~al.}(2020){Carleo}, {Malavolta}, {Lanza}, {Damasso},
  {Desidera}, {Borsa}, {Mallonn}, {Pinamonti}, {Gratton}, {Alei}, {Benatti},
  {Mancini}, {Maldonado}, {Biazzo}, {Esposito}, {Frustagli},
  {Gonz{\'a}lez-{\'A}lvarez}, {Micela}, {Scandariato}, {Sozzetti}, {Affer},
  {Bignamini}, {Bonomo}, {Claudi}, {Cosentino}, {Covino}, {Fiorenzano},
  {Giacobbe}, {Harutyunyan}, {Leto}, {Maggio}, {Molinari}, {Nascimbeni},
  {Pagano}, {Pedani}, {Piotto}, {Poretti}, {Rainer}, {Redfield}, {Baffa},
  {Baruffolo}, {Buchschacher}, {Billotti}, {Cecconi}, {Falcini}, {Fantinel},
  {Fini}, {Galli}, {Ghedina}, {Ghinassi}, {Giani}, {Gonzalez}, {Gonzalez},
  {Guerra}, {Hernandez Diaz}, {Hernandez}, {Iuzzolino}, {Lodi}, {Oliva},
  {Origlia}, {Perez Ventura}, {Puglisi}, {Riverol}, {Riverol}, {San Juan},
  {Sanna}, {Scuderi}, {Seemann}, {Sozzi}, \&
  {Tozzi}}]{carleo2020A&A...638A...5C}
{Carleo}, I., {Malavolta}, L., {Lanza}, A.~F., {et~al.} 2020, \aap, 638, A5

\bibitem[{{Casagrande} {et~al.}(2021){Casagrande}, {Lin}, {Rains}, {Liu},
  {Buder}, {Horner}, {Asplund}, {Lewis}, {Martell}, {Nordlander}, {Stello},
  {Ting}, {Wittenmyer}, {Bland-Hawthorn}, {Casey}, {De Silva}, {D'Orazi},
  {Freeman}, {Hayden}, {Kos}, {Lind}, {Schlesinger}, {Sharma}, {Simpson},
  {Zucker}, \& {Zwitter}}]{colte}
{Casagrande}, L., {Lin}, J., {Rains}, A.~D., {et~al.} 2021, \mnras, 507, 2684

\bibitem[{{Castelli} \& {Kurucz}(2003)}]{2003castelli}
{Castelli}, F. \& {Kurucz}, R.~L. 2003, in Modelling of Stellar Atmospheres,
  ed. N.~{Piskunov}, W.~W. {Weiss}, \& D.~F. {Gray}, Vol. 210, A20

\bibitem[{{Chatterjee} {et~al.}(2008){Chatterjee}, {Ford}, {Matsumura}, \&
  {Rasio}}]{2008ApJ...686..580C}
{Chatterjee}, S., {Ford}, E.~B., {Matsumura}, S., \& {Rasio}, F.~A. 2008, \apj,
  686, 580

\bibitem[{{Choi} {et~al.}(2016){Choi}, {Dotter}, {Conroy}, {Cantiello},
  {Paxton}, \& {Johnson}}]{choi+2016}
{Choi}, J., {Dotter}, A., {Conroy}, C., {et~al.} 2016, \apj, 823, 102

\bibitem[{{Choksi} \& {Chiang}(2020)}]{choksi}
{Choksi}, N. \& {Chiang}, E. 2020, \mnras, 495, 4192

\bibitem[{{Choksi} \& {Chiang}(2023)}]{choksi2}
{Choksi}, N. \& {Chiang}, E. 2023, \mnras, 522, 1914

\bibitem[{{Cosentino} {et~al.}(2012){Cosentino}, {Lovis}, {Pepe}, {Collier
  Cameron}, {Latham}, {Molinari}, {Udry}, {Bezawada}, {Black}, {Born},
  {Buchschacher}, {Charbonneau}, {Figueira}, {Fleury}, {Galli}, {Gallie},
  {Gao}, {Ghedina}, {Gonzalez}, {Gonzalez}, {Guerra}, {Henry}, {Horne},
  {Hughes}, {Kelly}, {Lodi}, {Lunney}, {Maire}, {Mayor}, {Micela}, {Ordway},
  {Peacock}, {Phillips}, {Piotto}, {Pollacco}, {Queloz}, {Rice}, {Riverol},
  {Riverol}, {San Juan}, {Sasselov}, {Segransan}, {Sozzetti}, {Sosnowska},
  {Stobie}, {Szentgyorgyi}, {Vick}, \& {Weber}}]{Cosentino2012}
{Cosentino}, R., {Lovis}, C., {Pepe}, F., {et~al.} 2012, Society of
  Photo-Optical Instrumentation Engineers (SPIE) Conference Series, Vol. 8446,
  {Harps-N: the new planet hunter at TNG}, 84461V

\bibitem[{{Cossou} {et~al.}(2014){Cossou}, {Raymond}, {Hersant}, \&
  {Pierens}}]{2014A&A...569A..56C}
{Cossou}, C., {Raymond}, S.~N., {Hersant}, F., \& {Pierens}, A. 2014, \aap,
  569, A56

\bibitem[{{Covino} {et~al.}(2013){Covino}, {Esposito}, {Barbieri}, {Mancini},
  {Nascimbeni}, {Claudi}, {Desidera}, {Gratton}, {Lanza}, {Sozzetti}, {Biazzo},
  {Affer}, {Gandolfi}, {Munari}, {Pagano}, {Bonomo}, {Collier Cameron},
  {H{\'e}brard}, {Maggio}, {Messina}, {Micela}, {Molinari}, {Pepe}, {Piotto},
  {Ribas}, {Santos}, {Southworth}, {Shkolnik}, {Triaud}, {Bedin}, {Benatti},
  {Boccato}, {Bonavita}, {Borsa}, {Borsato}, {Brown}, {Carolo}, {Ciceri},
  {Cosentino}, {Damasso}, {Faedi}, {Mart{\'\i}nez Fiorenzano}, {Latham},
  {Lovis}, {Mordasini}, {Nikolov}, {Poretti}, {Rainer}, {Rebolo L{\'o}pez},
  {Scandariato}, {Silvotti}, {Smareglia}, {Alcal{\'a}}, {Cunial}, {Di
  Fabrizio}, {Di Mauro}, {Giacobbe}, {Granata}, {Harutyunyan}, {Knapic},
  {Lattanzi}, {Leto}, {Lodato}, {Malavolta}, {Marzari}, {Molinaro},
  {Nardiello}, {Pedani}, {Prisinzano}, \& {Turrini}}]{2013A&A...554A..28C}
{Covino}, E., {Esposito}, M., {Barbieri}, M., {et~al.} 2013, \aap, 554, A28

\bibitem[{{Cutri} {et~al.}(2003){Cutri}, {Skrutskie}, {van Dyk}, {Beichman},
  {Carpenter}, {Chester}, {Cambresy}, {Evans}, {Fowler}, {Gizis}, {Howard},
  {Huchra}, {Jarrett}, {Kopan}, {Kirkpatrick}, {Light}, {Marsh}, {McCallon},
  {Schneider}, {Stiening}, {Sykes}, {Weinberg}, {Wheaton}, {Wheelock}, \&
  {Zacarias}}]{cutri2003}
{Cutri}, R.~M., {Skrutskie}, M.~F., {van Dyk}, S., {et~al.} 2003, VizieR Online
  Data Catalog, II/246

\bibitem[{{Cutri} {et~al.}(2021){Cutri}, {Wright}, {Conrow}, {Fowler},
  {Eisenhardt}, {Grillmair}, {Kirkpatrick}, {Masci}, {McCallon}, {Wheelock},
  {Fajardo-Acosta}, {Yan}, {Benford}, {Harbut}, {Jarrett}, {Lake}, {Leisawitz},
  {Ressler}, {Stanford}, {Tsai}, {Liu}, {Helou}, {Mainzer}, {Gettngs},
  {Gonzalez}, {Hoffman}, {Marsh}, {Padgett}, {Skrutskie}, {Beck}, {Papin}, \&
  {Wittman}}]{cutri2013}
{Cutri}, R.~M., {Wright}, E.~L., {Conrow}, T., {et~al.} 2021, VizieR Online
  Data Catalog, II/328

\bibitem[{{Damasso} {et~al.}(2023){Damasso}, {Locci}, {Benatti}, {Maggio},
  {Nardiello}, {Baratella}, {Biazzo}, {Bonomo}, {Desidera}, {D'Orazi},
  {Mallonn}, {Lanza}, {Sozzetti}, {Marzari}, {Borsa}, {Maldonado}, {Mancini},
  {Poretti}, {Scandariato}, {Bignamini}, {Borsato}, {Capuzzo Dolcetta},
  {Cecconi}, {Claudi}, {Cosentino}, {Covino}, {Fiorenzano}, {Harutyunyan},
  {Mann}, {Micela}, {Molinari}, {Molinaro}, {Pagano}, {Pedani}, {Pinamonti},
  {Piotto}, \& {Stoev}}]{damasso_2023A&A...672A.126D}
{Damasso}, M., {Locci}, D., {Benatti}, S., {et~al.} 2023, \aap, 672, A126

\bibitem[{{Donati} {et~al.}(2020){Donati}, {Kouach}, {Moutou}, {Doyon},
  {Delfosse}, {Artigau}, {Baratchart}, {Lacombe}, {Barrick}, {H{\'e}brard},
  {Bouchy}, {Saddlemyer}, {Par{\`e}s}, {Rabou}, {Micheau}, {Dolon}, {Reshetov},
  {Challita}, {Carmona}, {Striebig}, {Thibault}, {Martioli}, {Cook},
  {Fouqu{\'e}}, {Vermeulen}, {Wang}, {Arnold}, {Pepe}, {Boisse}, {Figueira},
  {Bouvier}, {Ray}, {Feugeade}, {Morin}, {Alencar}, {Hobson}, {Castilho},
  {Udry}, {Santos}, {Hernandez}, {Benedict}, {Vall{\'e}e}, {Gallou}, {Dupieux},
  {Larrieu}, {Perruchot}, {Sottile}, {Moreau}, {Usher}, {Baril}, {Wildi},
  {Chazelas}, {Malo}, {Bonfils}, {Loop}, {Kerley}, {Wevers}, {Dunn}, {Pazder},
  {Macdonald}, {Dubois}, {Carri{\'e}}, {Valentin}, {Henault}, {Yan}, \&
  {Steinmetz}}]{2020MNRAS.498.5684D}
{Donati}, J.~F., {Kouach}, D., {Moutou}, C., {et~al.} 2020, \mnras, 498, 5684

\bibitem[{{Dotter}(2016)}]{Dotter2016}
{Dotter}, A. 2016, \apjs, 222, 8

\bibitem[{{Dutra-Ferreira} {et~al.}(2016){Dutra-Ferreira}, {Pasquini},
  {Smiljanic}, {Porto de Mello}, \& {Steffen}}]{2016ferreira}
{Dutra-Ferreira}, L., {Pasquini}, L., {Smiljanic}, R., {Porto de Mello}, G.~F.,
  \& {Steffen}, M. 2016, \aap, 585, A75

\bibitem[{{Eastman} {et~al.}(2019){Eastman}, {Rodriguez}, {Agol}, {Stassun},
  {Beatty}, {Vanderburg}, {Gaudi}, {Collins}, \& {Luger}}]{Eastmanetal2019}
{Eastman}, J.~D., {Rodriguez}, J.~E., {Agol}, E., {et~al.} 2019, arXiv
  e-prints, arXiv:1907.09480

\bibitem[{{Erkaev} {et~al.}(2007){Erkaev}, {Kulikov}, {Lammer}, {Selsis},
  {Langmayr}, {Jaritz}, \& {Biernat}}]{Erkaev+2007}
{Erkaev}, N.~V., {Kulikov}, Y.~N., {Lammer}, H., {et~al.} 2007, \aap, 472, 329

\bibitem[{{Fabrycky} {et~al.}(2014){Fabrycky}, {Lissauer}, {Ragozzine}, {Rowe},
  {Steffen}, {Agol}, {Barclay}, {Batalha}, {Borucki}, {Ciardi}, {Ford},
  {Gautier}, {Geary}, {Holman}, {Jenkins}, {Li}, {Morehead}, {Morris},
  {Shporer}, {Smith}, {Still}, \& {Van Cleve}}]{fab}
{Fabrycky}, D.~C., {Lissauer}, J.~J., {Ragozzine}, D., {et~al.} 2014, \apj,
  790, 146

\bibitem[{{Fern{\'a}ndez Fern{\'a}ndez} {et~al.}(2024){Fern{\'a}ndez
  Fern{\'a}ndez}, {Wheatley}, {King}, \& {Jenkins}}]{Fernandez24}
{Fern{\'a}ndez Fern{\'a}ndez}, J., {Wheatley}, P.~J., {King}, G.~W., \&
  {Jenkins}, J.~S. 2024, \mnras, 527, 911

\bibitem[{{Feroz} {et~al.}(2019){Feroz}, {Hobson}, {Cameron}, \&
  {Pettitt}}]{Feroz2019}
{Feroz}, F., {Hobson}, M.~P., {Cameron}, E., \& {Pettitt}, A.~N. 2019, The Open
  Journal of Astrophysics, 2, 10

\bibitem[{{Filomeno} {et~al.}(2024){Filomeno}, {Biazzo}, {Baratella},
  {Benatti}, {D'Orazi}, {Desidera}, {Mancini}, \& {Messina}}]{filomeno_subm}
{Filomeno}, S., {Biazzo}, K., {Baratella}, M., {et~al.} 2024, \aap, submitted

\bibitem[{{Ford} \& {Rasio}(2008)}]{2008ApJ...686..621F}
{Ford}, E.~B. \& {Rasio}, F.~A. 2008, \apj, 686, 621

\bibitem[{{Fortney} {et~al.}(2007){Fortney}, {Marley}, \&
  {Barnes}}]{Fortney2007}
{Fortney}, J.~J., {Marley}, M.~S., \& {Barnes}, J.~W. 2007, \apj, 659, 1661

\bibitem[{Frazier {et~al.}(2023)Frazier, Stefánsson, Mahadevan, Yee, Cañas,
  Winn, Luhn, Dai, Doyle, Cegla, Kanodia, Robertson, Wisniewski, Bender, Dong,
  Gupta, Halverson, Hawley, Hebb, Holcomb, Kowalski, Libby-Roberts, Lin,
  McElwain, Ninan, Petrovich, Roy, Schwab, Terrien, \& Wright}]{Frazier_2023}
Frazier, R.~C., Stefánsson, G., Mahadevan, S., {et~al.} 2023, The
  Astrophysical Journal Letters, 944, L41

\bibitem[{{Gaia Coll.} {et~al.}(2021){Gaia Coll.}, {Brown}, {Vallenari},
  {Prusti}, {de Bruijne}, {Babusiaux}, {Biermann}, {Creevey}, {Evans}, {Eyer},
  {Hutton}, {Jansen}, {Jordi}, {Klioner}, {Lammers}, {Lindegren}, {Luri},
  {Mignard}, {Panem}, {Pourbaix}, {Randich}, {Sartoretti}, {Soubiran},
  {Walton}, {Arenou}, {Bailer-Jones}, {Bastian}, {Cropper}, {Drimmel}, {Katz},
  {Lattanzi}, {van Leeuwen}, {Bakker}, {Cacciari}, {Casta{\~n}eda}, {De
  Angeli}, {Ducourant}, {Fabricius}, {Fouesneau}, {Fr{\'e}mat}, {Guerra},
  {Guerrier}, {Guiraud}, {Jean-Antoine Piccolo}, {Masana}, {Messineo},
  {Mowlavi}, {Nicolas}, {Nienartowicz}, {Pailler}, {Panuzzo}, {Riclet}, {Roux},
  {Seabroke}, {Sordo}, {Tanga}, {Th{\'e}venin}, {Gracia-Abril}, {Portell},
  {Teyssier}, {Altmann}, {Andrae}, {Bellas-Velidis}, {Benson}, {Berthier},
  {Blomme}, {Brugaletta}, {Burgess}, {Busso}, {Carry}, {Cellino}, {Cheek},
  {Clementini}, {Damerdji}, {Davidson}, {Delchambre}, {Dell'Oro},
  {Fern{\'a}ndez-Hern{\'a}ndez}, {Galluccio}, {Garc{\'\i}a-Lario},
  {Garcia-Reinaldos}, {Gonz{\'a}lez-N{\'u}{\~n}ez}, {Gosset}, {Haigron},
  {Halbwachs}, {Hambly}, {Harrison}, {Hatzidimitriou}, {Heiter},
  {Hern{\'a}ndez}, {Hestroffer}, {Hodgkin}, {Holl}, {Jan{\ss}en}, {Jevardat de
  Fombelle}, {Jordan}, {Krone-Martins}, {Lanzafame}, {L{\"o}ffler}, {Lorca},
  {Manteiga}, {Marchal}, {Marrese}, {Moitinho}, {Mora}, {Muinonen}, {Osborne},
  {Pancino}, {Pauwels}, {Petit}, {Recio-Blanco}, {Richards}, {Riello},
  {Rimoldini}, {Robin}, {Roegiers}, {Rybizki}, {Sarro}, {Siopis}, {Smith},
  {Sozzetti}, {Ulla}, {Utrilla}, {van Leeuwen}, {van Reeven}, {Abbas}, {Abreu
  Aramburu}, {Accart}, {Aerts}, {Aguado}, {Ajaj}, {Altavilla}, {{\'A}lvarez},
  {{\'A}lvarez Cid-Fuentes}, {Alves}, {Anderson}, {Anglada Varela}, {Antoja},
  {Audard}, {Baines}, {Baker}, {Balaguer-N{\'u}{\~n}ez}, {Balbinot}, {Balog},
  {Barache}, {Barbato}, {Barros}, {Barstow}, {Bartolom{\'e}}, {Bassilana},
  {Bauchet}, {Baudesson-Stella}, {Becciani}, {Bellazzini}, {Bernet}, {Bertone},
  {Bianchi}, {Blanco-Cuaresma}, {Boch}, {Bombrun}, {Bossini}, {Bouquillon},
  {Bragaglia}, {Bramante}, {Breedt}, {Bressan}, {Brouillet}, {Bucciarelli},
  {Burlacu}, {Busonero}, {Butkevich}, {Buzzi}, {Caffau}, {Cancelliere},
  {C{\'a}novas}, {Cantat-Gaudin}, {Carballo}, {Carlucci}, {Carnerero},
  {Carrasco}, {Casamiquela}, {Castellani}, {Castro-Ginard}, {Castro Sampol},
  {Chaoul}, {Charlot}, {Chemin}, {Chiavassa}, {Cioni}, {Comoretto}, {Cooper},
  {Cornez}, {Cowell}, {Crifo}, {Crosta}, {Crowley}, {Dafonte}, {Dapergolas},
  {David}, {David}, {de Laverny}, {De Luise}, {De March}, {De Ridder}, {de
  Souza}, {de Teodoro}, {de Torres}, {del Peloso}, {del Pozo}, {Delbo},
  {Delgado}, {Delgado}, {Delisle}, {Di Matteo}, {Diakite}, {Diener},
  {Distefano}, {Dolding}, {Eappachen}, {Edvardsson}, {Enke}, {Esquej}, {Fabre},
  {Fabrizio}, {Faigler}, {Fedorets}, {Fernique}, {Fienga}, {Figueras},
  {Fouron}, {Fragkoudi}, {Fraile}, {Franke}, {Gai}, {Garabato},
  {Garcia-Gutierrez}, {Garc{\'\i}a-Torres}, {Garofalo}, {Gavras}, {Gerlach},
  {Geyer}, {Giacobbe}, {Gilmore}, {Girona}, {Giuffrida}, {Gomel}, {Gomez},
  {Gonzalez-Santamaria}, {Gonz{\'a}lez-Vidal}, {Granvik},
  {Guti{\'e}rrez-S{\'a}nchez}, {Guy}, {Hauser}, {Haywood}, {Helmi}, {Hidalgo},
  {Hilger}, {H{\l}adczuk}, {Hobbs}, {Holland}, {Huckle}, {Jasniewicz},
  {Jonker}, {Juaristi Campillo}, {Julbe}, {Karbevska}, {Kervella}, {Khanna},
  {Kochoska}, {Kontizas}, {Kordopatis}, {Korn}, {Kostrzewa-Rutkowska},
  {Kruszy{\'n}ska}, {Lambert}, {Lanza}, {Lasne}, {Le Campion}, {Le Fustec},
  {Lebreton}, {Lebzelter}, {Leccia}, {Leclerc}, {Lecoeur-Taibi}, {Liao},
  {Licata}, {Lindstr{\o}m}, {Lister}, {Livanou}, {Lobel}, {Madrero Pardo},
  {Managau}, {Mann}, {Marchant}, {Marconi}, {Marcos Santos}, {Marinoni},
  {Marocco}, {Marshall}, {Martin Polo}, {Mart{\'\i}n-Fleitas}, {Masip},
  {Massari}, {Mastrobuono-Battisti}, {Mazeh}, {McMillan}, {Messina},
  {Michalik}, {Millar}, {Mints}, {Molina}, {Molinaro}, {Moln{\'a}r},
  {Montegriffo}, {Mor}, {Morbidelli}, {Morel}, {Morris}, {Mulone}, {Munoz},
  {Muraveva}, {Murphy}, {Musella}, {Noval}, {Ord{\'e}novic}, {Orr{\`u}},
  {Osinde}, {Pagani}, {Pagano}, {Palaversa}, {Palicio}, {Panahi}, {Pawlak},
  {Pe{\~n}alosa Esteller}, {Penttil{\"a}}, {Piersimoni}, {Pineau}, {Plachy},
  {Plum}, {Poggio}, {Poretti}, {Poujoulet}, {Pr{\v{s}}a}, {Pulone}, {Racero},
  {Ragaini}, {Rainer}, {Raiteri}, {Rambaux}, {Ramos}, {Ramos-Lerate}, {Re
  Fiorentin}, {Regibo}, {Reyl{\'e}}, {Ripepi}, {Riva}, {Rixon}, {Robichon},
  {Robin}, {Roelens}, {Rohrbasser}, {Romero-G{\'o}mez}, {Rowell}, {Royer},
  {Rybicki}, {Sadowski}, {Sagrist{\`a} Sell{\'e}s}, {Sahlmann}, {Salgado},
  {Salguero}, {Samaras}, {Sanchez Gimenez}, {Sanna}, {Santove{\~n}a},
  {Sarasso}, {Schultheis}, {Sciacca}, {Segol}, {Segovia}, {S{\'e}gransan},
  {Semeux}, {Shahaf}, {Siddiqui}, {Siebert}, {Siltala}, {Slezak}, {Smart},
  {Solano}, {Solitro}, {Souami}, {Souchay}, {Spagna}, {Spoto}, {Steele},
  {Steidelm{\"u}ller}, {Stephenson}, {S{\"u}veges}, {Szabados}, {Szegedi-Elek},
  {Taris}, {Tauran}, {Taylor}, {Teixeira}, {Thuillot}, {Tonello}, {Torra},
  {Torra}, {Turon}, {Unger}, {Vaillant}, {van Dillen}, {Vanel}, {Vecchiato},
  {Viala}, {Vicente}, {Voutsinas}, {Weiler}, {Wevers}, {Wyrzykowski}, {Yoldas},
  {Yvard}, {Zhao}, {Zorec}, {Zucker}, {Zurbach}, \& {Zwitter}}]{gaia2021}
{Gaia Coll.}, {Brown}, A.~G.~A., {Vallenari}, A., {et~al.} 2021, \aap, 649, A1

\bibitem[{{Gaia Coll.} {et~al.}(2016){Gaia Coll.}, {Prusti}, {de Bruijne},
  {Brown}, {Vallenari}, {Babusiaux}, {Bailer-Jones}, {Bastian}, {Biermann},
  {Evans}, {Eyer}, {Jansen}, {Jordi}, {Klioner}, {Lammers}, {Lindegren},
  {Luri}, {Mignard}, {Milligan}, {Panem}, {Poinsignon}, {Pourbaix}, {Randich},
  {Sarri}, {Sartoretti}, {Siddiqui}, {Soubiran}, {Valette}, {van Leeuwen},
  {Walton}, {Aerts}, {Arenou}, {Cropper}, {Drimmel}, {H{\o}g}, {Katz},
  {Lattanzi}, {O'Mullane}, {Grebel}, {Holland}, {Huc}, {Passot}, {Bramante},
  {Cacciari}, {Casta{\~n}eda}, {Chaoul}, {Cheek}, {De Angeli}, {Fabricius},
  {Guerra}, {Hern{\'a}ndez}, {Jean-Antoine-Piccolo}, {Masana}, {Messineo},
  {Mowlavi}, {Nienartowicz}, {Ord{\'o}{\~n}ez-Blanco}, {Panuzzo}, {Portell},
  {Richards}, {Riello}, {Seabroke}, {Tanga}, {Th{\'e}venin}, {Torra}, {Els},
  {Gracia-Abril}, {Comoretto}, {Garcia-Reinaldos}, {Lock}, {Mercier},
  {Altmann}, {Andrae}, {Astraatmadja}, {Bellas-Velidis}, {Benson}, {Berthier},
  {Blomme}, {Busso}, {Carry}, {Cellino}, {Clementini}, {Cowell}, {Creevey},
  {Cuypers}, {Davidson}, {De Ridder}, {de Torres}, {Delchambre}, {Dell'Oro},
  {Ducourant}, {Fr{\'e}mat}, {Garc{\'\i}a-Torres}, {Gosset}, {Halbwachs},
  {Hambly}, {Harrison}, {Hauser}, {Hestroffer}, {Hodgkin}, {Huckle}, {Hutton},
  {Jasniewicz}, {Jordan}, {Kontizas}, {Korn}, {Lanzafame}, {Manteiga},
  {Moitinho}, {Muinonen}, {Osinde}, {Pancino}, {Pauwels}, {Petit},
  {Recio-Blanco}, {Robin}, {Sarro}, {Siopis}, {Smith}, {Smith}, {Sozzetti},
  {Thuillot}, {van Reeven}, {Viala}, {Abbas}, {Abreu Aramburu}, {Accart},
  {Aguado}, {Allan}, {Allasia}, {Altavilla}, {{\'A}lvarez}, {Alves},
  {Anderson}, {Andrei}, {Anglada Varela}, {Antiche}, {Antoja}, {Ant{\'o}n},
  {Arcay}, {Atzei}, {Ayache}, {Bach}, {Baker}, {Balaguer-N{\'u}{\~n}ez},
  {Barache}, {Barata}, {Barbier}, {Barblan}, {Baroni}, {Barrado y
  Navascu{\'e}s}, {Barros}, {Barstow}, {Becciani}, {Bellazzini}, {Bellei},
  {Bello Garc{\'\i}a}, {Belokurov}, {Bendjoya}, {Berihuete}, {Bianchi},
  {Bienaym{\'e}}, {Billebaud}, {Blagorodnova}, {Blanco-Cuaresma}, {Boch},
  {Bombrun}, {Borrachero}, {Bouquillon}, {Bourda}, {Bouy}, {Bragaglia},
  {Breddels}, {Brouillet}, {Br{\"u}semeister}, {Bucciarelli}, {Budnik},
  {Burgess}, {Burgon}, {Burlacu}, {Busonero}, {Buzzi}, {Caffau}, {Cambras},
  {Campbell}, {Cancelliere}, {Cantat-Gaudin}, {Carlucci}, {Carrasco},
  {Castellani}, {Charlot}, {Charnas}, {Charvet}, {Chassat}, {Chiavassa},
  {Clotet}, {Cocozza}, {Collins}, {Collins}, {Costigan}, {Crifo}, {Cross},
  {Crosta}, {Crowley}, {Dafonte}, {Damerdji}, {Dapergolas}, {David}, {David},
  {De Cat}, {de Felice}, {de Laverny}, {De Luise}, {De March}, {de Martino},
  {de Souza}, {Debosscher}, {del Pozo}, {Delbo}, {Delgado}, {Delgado}, {di
  Marco}, {Di Matteo}, {Diakite}, {Distefano}, {Dolding}, {Dos Anjos},
  {Drazinos}, {Dur{\'a}n}, {Dzigan}, {Ecale}, {Edvardsson}, {Enke}, {Erdmann},
  {Escolar}, {Espina}, {Evans}, {Eynard Bontemps}, {Fabre}, {Fabrizio},
  {Faigler}, {Falc{\~a}o}, {Farr{\`a}s Casas}, {Faye}, {Federici}, {Fedorets},
  {Fern{\'a}ndez-Hern{\'a}ndez}, {Fernique}, {Fienga}, {Figueras}, {Filippi},
  {Findeisen}, {Fonti}, {Fouesneau}, {Fraile}, {Fraser}, {Fuchs}, {Furnell},
  {Gai}, {Galleti}, {Galluccio}, {Garabato}, {Garc{\'\i}a-Sedano}, {Gar{\'e}},
  {Garofalo}, {Garralda}, {Gavras}, {Gerssen}, {Geyer}, {Gilmore}, {Girona},
  {Giuffrida}, {Gomes}, {Gonz{\'a}lez-Marcos}, {Gonz{\'a}lez-N{\'u}{\~n}ez},
  {Gonz{\'a}lez-Vidal}, {Granvik}, {Guerrier}, {Guillout}, {Guiraud},
  {G{\'u}rpide}, {Guti{\'e}rrez-S{\'a}nchez}, {Guy}, {Haigron},
  {Hatzidimitriou}, {Haywood}, {Heiter}, {Helmi}, {Hobbs}, {Hofmann}, {Holl},
  {Holland}, {Hunt}, {Hypki}, {Icardi}, {Irwin}, {Jevardat de Fombelle},
  {Jofr{\'e}}, {Jonker}, {Jorissen}, {Julbe}, {Karampelas}, {Kochoska},
  {Kohley}, {Kolenberg}, {Kontizas}, {Koposov}, {Kordopatis}, {Koubsky},
  {Kowalczyk}, {Krone-Martins}, {Kudryashova}, {Kull}, {Bachchan},
  {Lacoste-Seris}, {Lanza}, {Lavigne}, {Le Poncin-Lafitte}, {Lebreton},
  {Lebzelter}, {Leccia}, {Leclerc}, {Lecoeur-Taibi}, {Lemaitre}, {Lenhardt},
  {Leroux}, {Liao}, {Licata}, {Lindstr{\o}m}, {Lister}, {Livanou}, {Lobel},
  {L{\"o}ffler}, {L{\'o}pez}, {Lopez-Lozano}, {Lorenz}, {Loureiro},
  {MacDonald}, {Magalh{\~a}es Fernandes}, {Managau}, {Mann}, {Mantelet},
  {Marchal}, {Marchant}, {Marconi}, {Marie}, {Marinoni}, {Marrese},
  {Marschalk{\'o}}, {Marshall}, {Mart{\'\i}n-Fleitas}, {Martino}, {Mary},
  {Matijevi{\v{c}}}, {Mazeh}, {McMillan}, {Messina}, {Mestre}, {Michalik},
  {Millar}, {Miranda}, {Molina}, {Molinaro}, {Molinaro}, {Moln{\'a}r},
  {Moniez}, {Montegriffo}, {Monteiro}, {Mor}, {Mora}, {Morbidelli}, {Morel},
  {Morgenthaler}, {Morley}, {Morris}, {Mulone}, {Muraveva}, {Musella},
  {Narbonne}, {Nelemans}, {Nicastro}, {Noval}, {Ord{\'e}novic},
  {Ordieres-Mer{\'e}}, {Osborne}, {Pagani}, {Pagano}, {Pailler}, {Palacin},
  {Palaversa}, {Parsons}, {Paulsen}, {Pecoraro}, {Pedrosa}, {Pentik{\"a}inen},
  {Pereira}, {Pichon}, {Piersimoni}, {Pineau}, {Plachy}, {Plum}, {Poujoulet},
  {Pr{\v{s}}a}, {Pulone}, {Ragaini}, {Rago}, {Rambaux}, {Ramos-Lerate},
  {Ranalli}, {Rauw}, {Read}, {Regibo}, {Renk}, {Reyl{\'e}}, {Ribeiro},
  {Rimoldini}, {Ripepi}, {Riva}, {Rixon}, {Roelens}, {Romero-G{\'o}mez},
  {Rowell}, {Royer}, {Rudolph}, {Ruiz-Dern}, {Sadowski}, {Sagrist{\`a}
  Sell{\'e}s}, {Sahlmann}, {Salgado}, {Salguero}, {Sarasso}, {Savietto},
  {Schnorhk}, {Schultheis}, {Sciacca}, {Segol}, {Segovia}, {Segransan},
  {Serpell}, {Shih}, {Smareglia}, {Smart}, {Smith}, {Solano}, {Solitro},
  {Sordo}, {Soria Nieto}, {Souchay}, {Spagna}, {Spoto}, {Stampa}, {Steele},
  {Steidelm{\"u}ller}, {Stephenson}, {Stoev}, {Suess}, {S{\"u}veges}, {Surdej},
  {Szabados}, {Szegedi-Elek}, {Tapiador}, {Taris}, {Tauran}, {Taylor},
  {Teixeira}, {Terrett}, {Tingley}, {Trager}, {Turon}, {Ulla}, {Utrilla},
  {Valentini}, {van Elteren}, {Van Hemelryck}, {van Leeuwen}, {Varadi},
  {Vecchiato}, {Veljanoski}, {Via}, {Vicente}, {Vogt}, {Voss}, {Votruba},
  {Voutsinas}, {Walmsley}, {Weiler}, {Weingrill}, {Werner}, {Wevers},
  {Whitehead}, {Wyrzykowski}, {Yoldas}, {{\v{Z}}erjal}, {Zucker}, {Zurbach},
  {Zwitter}, {Alecu}, {Allen}, {Allende Prieto}, {Amorim},
  {Anglada-Escud{\'e}}, {Arsenijevic}, {Azaz}, {Balm}, {Beck}, {Bernstein},
  {Bigot}, {Bijaoui}, {Blasco}, {Bonfigli}, {Bono}, {Boudreault}, {Bressan},
  {Brown}, {Brunet}, {Bunclark}, {Buonanno}, {Butkevich}, {Carret}, {Carrion},
  {Chemin}, {Ch{\'e}reau}, {Corcione}, {Darmigny}, {de Boer}, {de Teodoro}, {de
  Zeeuw}, {Delle Luche}, {Domingues}, {Dubath}, {Fodor}, {Fr{\'e}zouls},
  {Fries}, {Fustes}, {Fyfe}, {Gallardo}, {Gallegos}, {Gardiol}, {Gebran},
  {Gomboc}, {G{\'o}mez}, {Grux}, {Gueguen}, {Heyrovsky}, {Hoar}, {Iannicola},
  {Isasi Parache}, {Janotto}, {Joliet}, {Jonckheere}, {Keil}, {Kim},
  {Klagyivik}, {Klar}, {Knude}, {Kochukhov}, {Kolka}, {Kos}, {Kutka}, {Lainey},
  {LeBouquin}, {Liu}, {Loreggia}, {Makarov}, {Marseille}, {Martayan},
  {Martinez-Rubi}, {Massart}, {Meynadier}, {Mignot}, {Munari}, {Nguyen},
  {Nordlander}, {Ocvirk}, {O'Flaherty}, {Olias Sanz}, {Ortiz}, {Osorio},
  {Oszkiewicz}, {Ouzounis}, {Palmer}, {Park}, {Pasquato}, {Peltzer}, {Peralta},
  {P{\'e}turaud}, {Pieniluoma}, {Pigozzi}, {Poels}, {Prat}, {Prod'homme},
  {Raison}, {Rebordao}, {Risquez}, {Rocca-Volmerange}, {Rosen}, {Ruiz-Fuertes},
  {Russo}, {Sembay}, {Serraller Vizcaino}, {Short}, {Siebert}, {Silva},
  {Sinachopoulos}, {Slezak}, {Soffel}, {Sosnowska}, {Strai{\v{z}}ys}, {ter
  Linden}, {Terrell}, {Theil}, {Tiede}, {Troisi}, {Tsalmantza}, {Tur},
  {Vaccari}, {Vachier}, {Valles}, {Van Hamme}, {Veltz}, {Virtanen}, {Wallut},
  {Wichmann}, {Wilkinson}, {Ziaeepour}, \& {Zschocke}}]{gaia2016}
{Gaia Coll.}, {Prusti}, T., {de Bruijne}, J.~H.~J., {et~al.} 2016, \aap, 595,
  A1

\bibitem[{{Gaia Coll.} {et~al.}(2023){Gaia Coll.}, {Vallenari}, {Brown},
  {Prusti}, {de Bruijne}, {Arenou}, {Babusiaux}, {Biermann}, {Creevey},
  {Ducourant}, {Evans}, {Eyer}, {Guerra}, {Hutton}, {Jordi}, {Klioner},
  {Lammers}, {Lindegren}, {Luri}, {Mignard}, {Panem}, {Pourbaix}, {Randich},
  {Sartoretti}, {Soubiran}, {Tanga}, {Walton}, {Bailer-Jones}, {Bastian},
  {Drimmel}, {Jansen}, {Katz}, {Lattanzi}, {van Leeuwen}, {Bakker}, {Cacciari},
  {Casta{\~n}eda}, {De Angeli}, {Fabricius}, {Fouesneau}, {Fr{\'e}mat},
  {Galluccio}, {Guerrier}, {Heiter}, {Masana}, {Messineo}, {Mowlavi},
  {Nicolas}, {Nienartowicz}, {Pailler}, {Panuzzo}, {Riclet}, {Roux},
  {Seabroke}, {Sordo}, {Th{\'e}venin}, {Gracia-Abril}, {Portell}, {Teyssier},
  {Altmann}, {Andrae}, {Audard}, {Bellas-Velidis}, {Benson}, {Berthier},
  {Blomme}, {Burgess}, {Busonero}, {Busso}, {C{\'a}novas}, {Carry}, {Cellino},
  {Cheek}, {Clementini}, {Damerdji}, {Davidson}, {de Teodoro}, {Nu{\~n}ez
  Campos}, {Delchambre}, {Dell'Oro}, {Esquej}, {Fern{\'a}ndez-Hern{\'a}ndez},
  {Fraile}, {Garabato}, {Garc{\'\i}a-Lario}, {Gosset}, {Haigron}, {Halbwachs},
  {Hambly}, {Harrison}, {Hern{\'a}ndez}, {Hestroffer}, {Hodgkin}, {Holl},
  {Jan{\ss}en}, {Jevardat de Fombelle}, {Jordan}, {Krone-Martins}, {Lanzafame},
  {L{\"o}ffler}, {Marchal}, {Marrese}, {Moitinho}, {Muinonen}, {Osborne},
  {Pancino}, {Pauwels}, {Recio-Blanco}, {Reyl{\'e}}, {Riello}, {Rimoldini},
  {Roegiers}, {Rybizki}, {Sarro}, {Siopis}, {Smith}, {Sozzetti}, {Utrilla},
  {van Leeuwen}, {Abbas}, {{\'A}brah{\'a}m}, {Abreu Aramburu}, {Aerts},
  {Aguado}, {Ajaj}, {Aldea-Montero}, {Altavilla}, {{\'A}lvarez}, {Alves},
  {Anders}, {Anderson}, {Anglada Varela}, {Antoja}, {Baines}, {Baker},
  {Balaguer-N{\'u}{\~n}ez}, {Balbinot}, {Balog}, {Barache}, {Barbato},
  {Barros}, {Barstow}, {Bartolom{\'e}}, {Bassilana}, {Bauchet}, {Becciani},
  {Bellazzini}, {Berihuete}, {Bernet}, {Bertone}, {Bianchi}, {Binnenfeld},
  {Blanco-Cuaresma}, {Blazere}, {Boch}, {Bombrun}, {Bossini}, {Bouquillon},
  {Bragaglia}, {Bramante}, {Breedt}, {Bressan}, {Brouillet}, {Brugaletta},
  {Bucciarelli}, {Burlacu}, {Butkevich}, {Buzzi}, {Caffau}, {Cancelliere},
  {Cantat-Gaudin}, {Carballo}, {Carlucci}, {Carnerero}, {Carrasco},
  {Casamiquela}, {Castellani}, {Castro-Ginard}, {Chaoul}, {Charlot}, {Chemin},
  {Chiaramida}, {Chiavassa}, {Chornay}, {Comoretto}, {Contursi}, {Cooper},
  {Cornez}, {Cowell}, {Crifo}, {Cropper}, {Crosta}, {Crowley}, {Dafonte},
  {Dapergolas}, {David}, {David}, {de Laverny}, {De Luise}, {De March}, {De
  Ridder}, {de Souza}, {de Torres}, {del Peloso}, {del Pozo}, {Delbo},
  {Delgado}, {Delisle}, {Demouchy}, {Dharmawardena}, {Di Matteo}, {Diakite},
  {Diener}, {Distefano}, {Dolding}, {Edvardsson}, {Enke}, {Fabre}, {Fabrizio},
  {Faigler}, {Fedorets}, {Fernique}, {Fienga}, {Figueras}, {Fournier},
  {Fouron}, {Fragkoudi}, {Gai}, {Garcia-Gutierrez}, {Garcia-Reinaldos},
  {Garc{\'\i}a-Torres}, {Garofalo}, {Gavel}, {Gavras}, {Gerlach}, {Geyer},
  {Giacobbe}, {Gilmore}, {Girona}, {Giuffrida}, {Gomel}, {Gomez},
  {Gonz{\'a}lez-N{\'u}{\~n}ez}, {Gonz{\'a}lez-Santamar{\'\i}a},
  {Gonz{\'a}lez-Vidal}, {Granvik}, {Guillout}, {Guiraud},
  {Guti{\'e}rrez-S{\'a}nchez}, {Guy}, {Hatzidimitriou}, {Hauser}, {Haywood},
  {Helmer}, {Helmi}, {Sarmiento}, {Hidalgo}, {Hilger}, {H{\l}adczuk}, {Hobbs},
  {Holland}, {Huckle}, {Jardine}, {Jasniewicz}, {Jean-Antoine Piccolo},
  {Jim{\'e}nez-Arranz}, {Jorissen}, {Juaristi Campillo}, {Julbe}, {Karbevska},
  {Kervella}, {Khanna}, {Kontizas}, {Kordopatis}, {Korn}, {K{\'o}sp{\'a}l},
  {Kostrzewa-Rutkowska}, {Kruszy{\'n}ska}, {Kun}, {Laizeau}, {Lambert},
  {Lanza}, {Lasne}, {Le Campion}, {Lebreton}, {Lebzelter}, {Leccia}, {Leclerc},
  {Lecoeur-Taibi}, {Liao}, {Licata}, {Lindstr{\o}m}, {Lister}, {Livanou},
  {Lobel}, {Lorca}, {Loup}, {Madrero Pardo}, {Magdaleno Romeo}, {Managau},
  {Mann}, {Manteiga}, {Marchant}, {Marconi}, {Marcos}, {Marcos Santos},
  {Mar{\'\i}n Pina}, {Marinoni}, {Marocco}, {Marshall}, {Martin Polo},
  {Mart{\'\i}n-Fleitas}, {Marton}, {Mary}, {Masip}, {Massari},
  {Mastrobuono-Battisti}, {Mazeh}, {McMillan}, {Messina}, {Michalik}, {Millar},
  {Mints}, {Molina}, {Molinaro}, {Moln{\'a}r}, {Monari}, {Mongui{\'o}},
  {Montegriffo}, {Montero}, {Mor}, {Mora}, {Morbidelli}, {Morel}, {Morris},
  {Muraveva}, {Murphy}, {Musella}, {Nagy}, {Noval}, {Oca{\~n}a}, {Ogden},
  {Ordenovic}, {Osinde}, {Pagani}, {Pagano}, {Palaversa}, {Palicio},
  {Pallas-Quintela}, {Panahi}, {Payne-Wardenaar}, {Pe{\~n}alosa Esteller},
  {Penttil{\"a}}, {Pichon}, {Piersimoni}, {Pineau}, {Plachy}, {Plum}, {Poggio},
  {Pr{\v{s}}a}, {Pulone}, {Racero}, {Ragaini}, {Rainer}, {Raiteri}, {Rambaux},
  {Ramos}, {Ramos-Lerate}, {Re Fiorentin}, {Regibo}, {Richards}, {Rios Diaz},
  {Ripepi}, {Riva}, {Rix}, {Rixon}, {Robichon}, {Robin}, {Robin}, {Roelens},
  {Rogues}, {Rohrbasser}, {Romero-G{\'o}mez}, {Rowell}, {Royer}, {Ruz Mieres},
  {Rybicki}, {Sadowski}, {S{\'a}ez N{\'u}{\~n}ez}, {Sagrist{\`a} Sell{\'e}s},
  {Sahlmann}, {Salguero}, {Samaras}, {Sanchez Gimenez}, {Sanna},
  {Santove{\~n}a}, {Sarasso}, {Schultheis}, {Sciacca}, {Segol}, {Segovia},
  {S{\'e}gransan}, {Semeux}, {Shahaf}, {Siddiqui}, {Siebert}, {Siltala},
  {Silvelo}, {Slezak}, {Slezak}, {Smart}, {Snaith}, {Solano}, {Solitro},
  {Souami}, {Souchay}, {Spagna}, {Spina}, {Spoto}, {Steele},
  {Steidelm{\"u}ller}, {Stephenson}, {S{\"u}veges}, {Surdej}, {Szabados},
  {Szegedi-Elek}, {Taris}, {Taylor}, {Teixeira}, {Tolomei}, {Tonello}, {Torra},
  {Torra}, {Torralba Elipe}, {Trabucchi}, {Tsounis}, {Turon}, {Ulla}, {Unger},
  {Vaillant}, {van Dillen}, {van Reeven}, {Vanel}, {Vecchiato}, {Viala},
  {Vicente}, {Voutsinas}, {Weiler}, {Wevers}, {Wyrzykowski}, {Yoldas}, {Yvard},
  {Zhao}, {Zorec}, {Zucker}, \& {Zwitter}}]{gaia2023}
{Gaia Coll.}, {Vallenari}, A., {Brown}, A.~G.~A., {et~al.} 2023, \aap, 674, A1

\bibitem[{{Gaidos} {et~al.}(2023){Gaidos}, {Hirano}, {Lee}, {Harakawa},
  {Hodapp}, {Jacobson}, {Kotani}, {Kudo}, {Kurokawa}, {Kuzuhara}, {Nishikawa},
  {Omiya}, {Serizawa}, {Tamura}, {Ueda}, \&
  {Vievard}}]{gaidos2023MNRAS.518.3777G}
{Gaidos}, E., {Hirano}, T., {Lee}, R.~A., {et~al.} 2023, \mnras, 518, 3777

\bibitem[{Gilbertson {et~al.}(2020)Gilbertson, Ford, Jones, \&
  Stenning}]{Gilbertson_2020}
Gilbertson, C., Ford, E.~B., Jones, D.~E., \& Stenning, D.~C. 2020, The
  Astrophysical Journal, 905, 155

\bibitem[{{Goldberg} \& {Batygin}(2022)}]{goldberg}
{Goldberg}, M. \& {Batygin}, K. 2022, \aj, 163, 201

\bibitem[{{Goldberg} \& {Batygin}(2023)}]{goldberg2}
{Goldberg}, M. \& {Batygin}, K. 2023, \apj, 948, 12

\bibitem[{{Goyal} {et~al.}(2023){Goyal}, {Dai}, \& {Wang}}]{goyal2}
{Goyal}, A.~V., {Dai}, F., \& {Wang}, S. 2023, \apj, 955, 118

\bibitem[{{Goyal} \& {Wang}(2022)}]{goyal}
{Goyal}, A.~V. \& {Wang}, S. 2022, \apj, 933, 162

\bibitem[{Haywood {et~al.}(2014)Haywood, Collier~Cameron, Queloz, Barros,
  Deleuil, Fares, Gillon, Lanza, Lovis, Moutou, Pepe, Pollacco, Santerne,
  Ségransan, \& Unruh}]{haywood2014}
Haywood, R.~D., Collier~Cameron, A., Queloz, D., {et~al.} 2014, \mnras, 443,
  2517

\bibitem[{{Hedges} {et~al.}(2021){Hedges}, {Hughes}, {Zhou}, {David}, {Becker},
  {Giacalone}, {Vanderburg}, {Rodriguez}, {Bieryla}, {Wirth}, {Atherton},
  {Fetherolf}, {Collins}, {Price-Whelan}, {Bedell}, {Quinn}, {Gan}, {Ricker},
  {Latham}, {Vanderspek}, {Seager}, {Winn}, {Jenkins}, {Kielkopf}, {Schwarz},
  {Dressing}, {Gonzales}, {Crossfield}, {Matthews}, {Jensen}, {Furlan},
  {Gnilka}, {Howell}, {Lester}, {Scott}, {Feliz}, {Lund}, {Siverd}, {Stevens},
  {Narita}, {Fukui}, {Murgas}, {Palle}, {Sutton}, {Stassun}, {Bouma}, {Vezie},
  {Villase{\~n}or}, {Quintana}, \& {Smith}}]{hedges2021}
{Hedges}, C., {Hughes}, A., {Zhou}, G., {et~al.} 2021, \aj, 162, 54

\bibitem[{{Hippke} {et~al.}(2019){Hippke}, {David}, {Mulders}, \&
  {Heller}}]{wotan2019AJ....158..143H}
{Hippke}, M., {David}, T.~J., {Mulders}, G.~D., \& {Heller}, R. 2019, \aj, 158,
  143

\bibitem[{{H{\o}g} {et~al.}(2000){H{\o}g}, {Fabricius}, {Makarov}, {Urban},
  {Corbin}, {Wycoff}, {Bastian}, {Schwekendiek}, \& {Wicenec}}]{2000}
{H{\o}g}, E., {Fabricius}, C., {Makarov}, V.~V., {et~al.} 2000, \aap, 355, L27

\bibitem[{{Hunter} {et~al.}(2012){Hunter}, {Macgregor}, {Szabo}, {Wellington},
  \& {Bellgard}}]{YABI}
{Hunter}, A., {Macgregor}, A.~B., {Szabo}, T., {Wellington}, C., \& {Bellgard},
  M.~I. 2012, Source Code for Biology and Medicine, 7, 1

\bibitem[{{Izidoro} {et~al.}(2017){Izidoro}, {Ogihara}, {Raymond},
  {Morbidelli}, {Pierens}, {Bitsch}, {Cossou}, \& {Hersant}}]{izidoro}
{Izidoro}, A., {Ogihara}, M., {Raymond}, S.~N., {et~al.} 2017, \mnras, 470,
  1750

\bibitem[{{Jackson} {et~al.}(2008){Jackson}, {Greenberg}, \&
  {Barnes}}]{2008ApJ...678.1396J}
{Jackson}, B., {Greenberg}, R., \& {Barnes}, R. 2008, \apj, 678, 1396

\bibitem[{{Johnstone} {et~al.}(2021){Johnstone}, {Bartel}, \&
  {G{\"u}del}}]{Johnstone+2021}
{Johnstone}, C.~P., {Bartel}, M., \& {G{\"u}del}, M. 2021, \aap, 649, A96

\bibitem[{{Kaminski} {et~al.}(2018){Kaminski}, {Trifonov}, {Caballero},
  {Quirrenbach}, {Ribas}, {Reiners}, {Amado}, {Zechmeister}, {Dreizler},
  {Perger}, {Tal-Or}, {Bonfils}, {Mayor}, {Astudillo-Defru}, {Bauer},
  {B{\'e}jar}, {Cifuentes}, {Colom{\'e}}, {Cort{\'e}s-Contreras}, {Delfosse},
  {D{\'\i}ez-Alonso}, {Forveille}, {Guenther}, {Hatzes}, {Henning}, {Jeffers},
  {K{\"u}rster}, {Lafarga}, {Luque}, {Mandel}, {Montes}, {Morales},
  {Passegger}, {Pedraz}, {Reffert}, {Sadegi}, {Schweitzer}, {Seifert}, {Stahl},
  \& {Udry}}]{2018A&A...618A.115K}
{Kaminski}, A., {Trifonov}, T., {Caballero}, J.~A., {et~al.} 2018, \aap, 618,
  A115

\bibitem[{{Kaplan} {et~al.}(2019){Kaplan}, {Bender}, {Terrien}, {Ninan}, {Roy},
  \& {Mahadevan}}]{kaplan2019ASPC..523..567K}
{Kaplan}, K.~F., {Bender}, C.~F., {Terrien}, R.~C., {et~al.} 2019, in
  Astronomical Society of the Pacific Conference Series, Vol. 523, Astronomical
  Data Analysis Software and Systems XXVII, ed. P.~J. {Teuben}, M.~W. {Pound},
  B.~A. {Thomas}, \& E.~M. {Warner}, 567

\bibitem[{Kipping(2010)}]{kipping2010}
Kipping, D.~M. 2010, Monthly Notices of the Royal Astronomical Society, 408,
  1758

\bibitem[{{Kochanek} {et~al.}(2017){Kochanek}, {Shappee}, {Stanek}, {Holoien},
  {Thompson}, {Prieto}, {Dong}, {Shields}, {Will}, {Britt}, {Perzanowski}, \&
  {Pojma{\'n}ski}}]{2017PASP..129j4502K}
{Kochanek}, C.~S., {Shappee}, B.~J., {Stanek}, K.~Z., {et~al.} 2017, \pasp,
  129, 104502

\bibitem[{{Kraus} {et~al.}(2015){Kraus}, {Cody}, {Covey}, {Rizzuto}, {Mann}, \&
  {Ireland}}]{2015ApJ...807....3K}
{Kraus}, A.~L., {Cody}, A.~M., {Covey}, K.~R., {et~al.} 2015, \apj, 807, 3

\bibitem[{{Kreidberg}(2015)}]{Kreidberg2015}
{Kreidberg}, L. 2015, \pasp, 127, 1161

\bibitem[{{Kubyshkina} {et~al.}(2018{\natexlab{a}}){Kubyshkina}, {Fossati},
  {Erkaev}, {Cubillos}, {Johnstone}, {Kislyakova}, {Lammer}, {Lendl}, \&
  {Odert}}]{kuby+2018a}
{Kubyshkina}, D., {Fossati}, L., {Erkaev}, N.~V., {et~al.} 2018{\natexlab{a}},
  \apjl, 866, L18

\bibitem[{{Kubyshkina} {et~al.}(2018{\natexlab{b}}){Kubyshkina}, {Fossati},
  {Erkaev}, {Johnstone}, {Cubillos}, {Kislyakova}, {Lammer}, {Lendl}, \&
  {Odert}}]{kuby+2018b}
{Kubyshkina}, D., {Fossati}, L., {Erkaev}, N.~V., {et~al.} 2018{\natexlab{b}},
  \aap, 619, A151

\bibitem[{{Lammers} {et~al.}(2023){Lammers}, {Hadden}, \& {Murray}}]{lammers}
{Lammers}, C., {Hadden}, S., \& {Murray}, N. 2023, \mnras, 525, L66

\bibitem[{{Lillo-Box} {et~al.}(2020){Lillo-Box}, {Lopez}, {Santerne},
  {Nielsen}, {Barros}, {Deleuil}, {Acu{\~n}a}, {Mousis}, {Sousa}, {Adibekyan},
  {Armstrong}, {Barrado}, {Bayliss}, {Brown}, {Demangeon}, {Dumusque},
  {Figueira}, {Hojjatpanah}, {Osborn}, {Santos}, \&
  {Udry}}]{2020A&A...640A..48L}
{Lillo-Box}, J., {Lopez}, T.~A., {Santerne}, A., {et~al.} 2020, \aap, 640, A48

\bibitem[{{Lind} {et~al.}(2009){Lind}, {Asplund}, \& {Barklem}}]{NLTE_li}
{Lind}, K., {Asplund}, M., \& {Barklem}, P.~S. 2009, \aap, 503, 541

\bibitem[{{Locci} {et~al.}(2019){Locci}, {Cecchi-Pestellini}, \&
  {Micela}}]{Locci19}
{Locci}, D., {Cecchi-Pestellini}, C., \& {Micela}, G. 2019, \aap, 624, A101

\bibitem[{{Lopez} \& {Fortney}(2014)}]{LopFor14}
{Lopez}, E.~D. \& {Fortney}, J.~J. 2014, \apj, 792, 1

\bibitem[{{Maggio} {et~al.}(2022){Maggio}, {Locci}, {Pillitteri}, {Benatti},
  {Claudi}, {Desidera}, {Micela}, {Damasso}, {Sozzetti}, \& {Suarez
  Mascare{\~n}o}}]{2022ApJ...925..172M}
{Maggio}, A., {Locci}, D., {Pillitteri}, I., {et~al.} 2022, \apj, 925, 172

\bibitem[{Mahmud {et~al.}(2011)Mahmud, Crockett, Johns-Krull, Prato, Hartigan,
  Jaffe, \& Beichman}]{Mahmud_2011}
Mahmud, N.~I., Crockett, C.~J., Johns-Krull, C.~M., {et~al.} 2011, The
  Astrophysical Journal, 736, 123

\bibitem[{{Mallorqu{\'\i}n} {et~al.}(2023){Mallorqu{\'\i}n}, {B{\'e}jar},
  {Lodieu}, {Zapatero Osorio}, {Tabernero}, {Su{\'a}rez Mascare{\~n}o},
  {Zechmeister}, {Luque}, {Pall{\'e}}, \& {Montes}}]{2023A&A...671A.163M}
{Mallorqu{\'\i}n}, M., {B{\'e}jar}, V.~J.~S., {Lodieu}, N., {et~al.} 2023,
  \aap, 671, A163

\bibitem[{{Mamajek} \& {Hillenbrand}(2008)}]{Mama+Hille2008}
{Mamajek}, E.~E. \& {Hillenbrand}, L.~A. 2008, \apj, 687, 1264

\bibitem[{{Mann} {et~al.}(2016{\natexlab{a}}){Mann}, {Gaidos}, {Mace},
  {Johnson}, {Bowler}, {LaCourse}, {Jacobs}, {Vanderburg}, {Kraus}, {Kaplan},
  \& {Jaffe}}]{2016ApJ...818...46M}
{Mann}, A.~W., {Gaidos}, E., {Mace}, G.~N., {et~al.} 2016{\natexlab{a}}, \apj,
  818, 46

\bibitem[{{Mann} {et~al.}(2016{\natexlab{b}}){Mann}, {Newton}, {Rizzuto},
  {Irwin}, {Feiden}, {Gaidos}, {Mace}, {Kraus}, {James}, {Ansdell},
  {Charbonneau}, {Covey}, {Ireland}, {Jaffe}, {Johnson}, {Kidder}, \&
  {Vanderburg}}]{2016AJ....152...61M}
{Mann}, A.~W., {Newton}, E.~R., {Rizzuto}, A.~C., {et~al.} 2016{\natexlab{b}},
  \aj, 152, 61

\bibitem[{{Mantovan} {et~al.}(2024){Mantovan}, {Malavolta}, {Desidera},
  {Zingales}, {Borsato}, {Piotto}, {Maggio}, {Locci}, {Polychroni}, {Turrini},
  {Baratella}, {Biazzo}, {Nardiello}, {Stassun}, {Nascimbeni}, {Benatti},
  {John}, {Watkins}, {Bieryla}, {Lissauer}, {Twicken}, {Lanza}, {Winn},
  {Messina}, {Montalto}, {Sozzetti}, {Boffin}, {Cheryasov}, {Strakhov},
  {Murgas}, {D'Arpa}, {Barkaoui}, {Benni}, {Bignamini}, {Bonomo}, {Borsa},
  {Cabona}, {Cameron}, {Claudi}, {Cochran}, {Collins}, {Damasso}, {Dong},
  {Endl}, {Fukui}, {F{\H{u}}r{\'e}sz}, {Gandolfi}, {Ghedina}, {Jenkins},
  {Kab{\'a}th}, {Latham}, {Lorenzi}, {Luque}, {Maldonado}, {McLeod},
  {Molinaro}, {Narita}, {Nowak}, {Orell-Miquel}, {Pall{\'e}}, {Parviainen},
  {Pedani}, {Quinn}, {Relles}, {Rowden}, {Scandariato}, {Schwarz}, {Seager},
  {Shporer}, {Vanderburg}, \& {Wilson}}]{2024A&A...682A.129M}
{Mantovan}, G., {Malavolta}, L., {Desidera}, S., {et~al.} 2024, \aap, 682, A129

\bibitem[{{Mantovan} {et~al.}(2022){Mantovan}, {Montalto}, {Piotto}, {Wilson},
  {Collier Cameron}, {Majidi}, {Borsato}, {Granata}, \&
  {Nascimbeni}}]{2022MNRAS.516.4432M}
{Mantovan}, G., {Montalto}, M., {Piotto}, G., {et~al.} 2022, \mnras, 516, 4432

\bibitem[{{Millholland} {et~al.}(2017){Millholland}, {Wang}, \&
  {Laughlin}}]{millholland}
{Millholland}, S., {Wang}, S., \& {Laughlin}, G. 2017, \apjl, 849, L33

\bibitem[{{Montalto} {et~al.}(2021){Montalto}, {Piotto}, {Marrese},
  {Nascimbeni}, {Prisinzano}, {Granata}, {Marinoni}, {Desidera}, {Ortolani},
  {Aerts}, {Alei}, {Altavilla}, {Benatti}, {B{\"o}rner}, {Cabrera}, {Claudi},
  {Deleuil}, {Fabrizio}, {Gizon}, {Goupil}, {Heras}, {Magrin}, {Malavolta},
  {Mas-Hesse}, {Pagano}, {Paproth}, {Pertenais}, {Pollacco}, {Ragazzoni},
  {Ramsay}, {Rauer}, \& {Udry}}]{pic}
{Montalto}, M., {Piotto}, G., {Marrese}, P.~M., {et~al.} 2021, \aap, 653, A98

\bibitem[{{Morrison} {et~al.}(2020){Morrison}, {Dawson}, \&
  {MacDonald}}]{morrison}
{Morrison}, S.~J., {Dawson}, R.~I., \& {MacDonald}, M. 2020, \apj, 904, 157

\bibitem[{{Nardiello} {et~al.}(2021){Nardiello}, {Deleuil}, {Mantovan},
  {Malavolta}, {Lacedelli}, {Libralato}, {Bedin}, {Borsato}, {Granata}, \&
  {Piotto}}]{2021MNRAS.505.3767N}
{Nardiello}, D., {Deleuil}, M., {Mantovan}, G., {et~al.} 2021, \mnras, 505,
  3767

\bibitem[{{Nardiello} {et~al.}(2022){Nardiello}, {Malavolta}, {Desidera},
  {Baratella}, {D'Orazi}, {Messina}, {Biazzo}, {Benatti}, {Damasso}, {Rajpaul},
  {Bonomo}, {Dolcetta}, {Mallonn}, {Cale}, {Plavchan}, {El Mufti}, {Bignamini},
  {Borsa}, {Carleo}, {Claudi}, {Covino}, {Lanza}, {Maldonado}, {Mancini},
  {Micela}, {Molinari}, {Pinamonti}, {Piotto}, {Poretti}, {Scandariato},
  {Sozzetti}, {Andreuzzi}, {Boschin}, {Cosentino}, {Fiorenzano}, {Harutyunyan},
  {Knapic}, {Pedani}, {Affer}, {Maggio}, \& {Rainer}}]{nardiello2022}
{Nardiello}, D., {Malavolta}, L., {Desidera}, S., {et~al.} 2022, \aap, 664,
  A163

\bibitem[{{Newton} {et~al.}(2019){Newton}, {Mann}, {Tofflemire}, {Pearce},
  {Rizzuto}, {Vanderburg}, {Martinez}, {Wang}, {Ruffio}, {Kraus}, {Johnson},
  {Thao}, {Wood}, {Rampalli}, {Nielsen}, {Collins}, {Dragomir}, {Hellier},
  {Anderson}, {Barclay}, {Brown}, {Feiden}, {Hart}, {Isopi}, {Kielkopf},
  {Mallia}, {Nelson}, {Rodriguez}, {Stockdale}, {Waite}, {Wright}, {Lissauer},
  {Ricker}, {Vanderspek}, {Latham}, {Seager}, {Winn}, {Jenkins}, {Bouma},
  {Burke}, {Davies}, {Fausnaugh}, {Li}, {Morris}, {Mukai}, {Villase{\~n}or},
  {Villeneuva}, {De Rosa}, {Macintosh}, {Mengel}, {Okumura}, \&
  {Wittenmyer}}]{2019ApJ...880L..17N}
{Newton}, E.~R., {Mann}, A.~W., {Tofflemire}, B.~M., {et~al.} 2019, \apjl, 880,
  L17

\bibitem[{{Osborn} {et~al.}(2022){Osborn}, {Bonfanti}, {Gandolfi}, {Hedges},
  {Leleu}, {Fortier}, {Futyan}, {Gutermann}, {Maxted}, {Borsato}, {Collins},
  {Gomes da Silva}, {G{\'o}mez Maqueo Chew}, {Hooton}, {Lendl}, {Parviainen},
  {Salmon}, {Schanche}, {Serrano}, {Sousa}, {Tuson}, {Ulmer-Moll}, {Van
  Grootel}, {Wells}, {Wilson}, {Alibert}, {Alonso}, {Anglada}, {Asquier},
  {Barrado y Navascues}, {Baumjohann}, {Beck}, {Benz}, {Biondi}, {Bonfils},
  {Bouchy}, {Brandeker}, {Broeg}, {B{\'a}rczy}, {Barros}, {Cabrera}, {Charnoz},
  {Collier Cameron}, {Csizmadia}, {Davies}, {Deleuil}, {Delrez}, {Demory},
  {Ehrenreich}, {Erikson}, {Fossati}, {Fridlund}, {Gillon}, {G{\"o}mez-Munoz},
  {G{\"u}del}, {Heng}, {Hoyer}, {Isaak}, {Kiss}, {Laskar}, {Lecavelier des
  Etangs}, {Lovis}, {Magrin}, {Malavolta}, {McCormac}, {Nascimbeni},
  {Olofsson}, {Ottensamer}, {Pagano}, {Pall{\'e}}, {Peter}, {Piazza}, {Piotto},
  {Pollacco}, {Queloz}, {Ragazzoni}, {Rando}, {Rauer}, {Reimers}, {Ribas},
  {Demangeon}, {Smith}, {Sabin}, {Santos}, {Scandariato}, {Schroffenegger},
  {Schwarz}, {Shporer}, {Simon}, {Steller}, {Szab{\'o}}, {S{\'e}gransan},
  {Thomas}, {Udry}, {Walter}, \& {Walton}}]{osborn2022}
{Osborn}, H.~P., {Bonfanti}, A., {Gandolfi}, D., {et~al.} 2022, \aap, 664, A156

\bibitem[{{Owen} \& {Wu}(2013)}]{2013ApJ...775..105O}
{Owen}, J.~E. \& {Wu}, Y. 2013, \apj, 775, 105

\bibitem[{{Penz} {et~al.}(2008){Penz}, {Micela}, \& {Lammer}}]{Penz08a}
{Penz}, T., {Micela}, G., \& {Lammer}, H. 2008, \aap, 477, 309

\bibitem[{{Pizzolato} {et~al.}(2003){Pizzolato}, {Maggio}, {Micela},
  {Sciortino}, \& {Ventura}}]{Pizzo+2003}
{Pizzolato}, N., {Maggio}, A., {Micela}, G., {Sciortino}, S., \& {Ventura}, P.
  2003, \aap, 397, 147

\bibitem[{{Pollacco} {et~al.}(2006){Pollacco}, {Skillen}, {Collier Cameron},
  {Christian}, {Hellier}, {Irwin}, {Lister}, {Street}, {West}, {Anderson},
  {Clarkson}, {Deeg}, {Enoch}, {Evans}, {Fitzsimmons}, {Haswell}, {Hodgkin},
  {Horne}, {Kane}, {Keenan}, {Maxted}, {Norton}, {Osborne}, {Parley}, {Ryans},
  {Smalley}, {Wheatley}, \& {Wilson}}]{2006PASP..118.1407P}
{Pollacco}, D.~L., {Skillen}, I., {Collier Cameron}, A., {et~al.} 2006, \pasp,
  118, 1407

\bibitem[{Prato {et~al.}(2008)Prato, Huerta, Johns-Krull, Mahmud, Jaffe, \&
  Hartigan}]{Prato_2008}
Prato, L., Huerta, M., Johns-Krull, C.~M., {et~al.} 2008, The Astrophysical
  Journal, 687, L103

\bibitem[{{Quirrenbach} {et~al.}(2014){Quirrenbach}, {Amado}, {Caballero},
  {Mundt}, {Reiners}, {Ribas}, {Seifert}, {Abril}, {Aceituno},
  {Alonso-Floriano}, {Ammler-von Eiff}, {Antona Jim{\'e}nez},
  {Anwand-Heerwart}, {Azzaro}, {Bauer}, {Barrado}, {Becerril}, {B{\'e}jar},
  {Ben{\'\i}tez}, {Berdi{\~n}as}, {C{\'a}rdenas}, {Casal}, {Claret},
  {Colom{\'e}}, {Cort{\'e}s-Contreras}, {Czesla}, {Doellinger}, {Dreizler},
  {Feiz}, {Fern{\'a}ndez}, {Galad{\'\i}}, {G{\'a}lvez-Ortiz},
  {Garc{\'\i}a-Piquer}, {Garc{\'\i}a-Vargas}, {Garrido}, {Gesa}, {G{\'o}mez
  Galera}, {Gonz{\'a}lez {\'A}lvarez}, {Gonz{\'a}lez Hern{\'a}ndez},
  {Gr{\"o}zinger}, {Gu{\`a}rdia}, {Guenther}, {de Guindos},
  {Guti{\'e}rrez-Soto}, {Hagen}, {Hatzes}, {Hauschildt}, {Helmling}, {Henning},
  {Hermann}, {Hern{\'a}ndez Casta{\~n}o}, {Herrero}, {Hidalgo}, {Holgado},
  {Huber}, {Huber}, {Jeffers}, {Joergens}, {de Juan}, {Kehr}, {Klein},
  {K{\"u}rster}, {Lamert}, {Lalitha}, {Laun}, {Lemke}, {Lenzen}, {L{\'o}pez del
  Fresno}, {L{\'o}pez Mart{\'\i}}, {L{\'o}pez-Santiago}, {Mall}, {Mandel},
  {Mart{\'\i}n}, {Mart{\'\i}n-Ruiz}, {Mart{\'\i}nez-Rodr{\'\i}guez}, {Marvin},
  {Mathar}, {Mirabet}, {Montes}, {Morales Mu{\~n}oz}, {Moya}, {Naranjo},
  {Ofir}, {Oreiro}, {Pall{\'e}}, {Panduro}, {Passegger}, {P{\'e}rez-Calpena},
  {P{\'e}rez Medialdea}, {Perger}, {Pluto}, {Ram{\'o}n}, {Rebolo}, {Redondo},
  {Reffert}, {Reinhardt}, {Rhode}, {Rix}, {Rodler}, {Rodr{\'\i}guez},
  {Rodr{\'\i}guez-L{\'o}pez}, {Rodr{\'\i}guez-P{\'e}rez}, {Rohloff}, {Rosich},
  {S{\'a}nchez-Blanco}, {S{\'a}nchez Carrasco}, {Sanz-Forcada}, {Sarmiento},
  {Sch{\"a}fer}, {Schiller}, {Schmidt}, {Schmitt}, {Solano}, {Stahl}, {Storz},
  {St{\"u}rmer}, {Su{\'a}rez}, {Ulbrich}, {Veredas}, {Wagner}, {Winkler},
  {Zapatero Osorio}, {Zechmeister}, {Abell{\'a}n de Paco},
  {Anglada-Escud{\'e}}, {del Burgo}, {Klutsch}, {Lizon}, {L{\'o}pez-Morales},
  {Morales}, {Perryman}, {Tulloch}, \& {Xu}}]{2014SPIE.9147E..1FQ}
{Quirrenbach}, A., {Amado}, P.~J., {Caballero}, J.~A., {et~al.} 2014, in
  Society of Photo-Optical Instrumentation Engineers (SPIE) Conference Series,
  Vol. 9147, Ground-based and Airborne Instrumentation for Astronomy V, ed.
  S.~K. {Ramsay}, I.~S. {McLean}, \& H.~{Takami}, 91471F

\bibitem[{{Quirrenbach} {et~al.}(2018){Quirrenbach}, {Amado}, {Ribas},
  {Reiners}, {Caballero}, {Seifert}, {Aceituno}, {Azzaro}, {Baroch}, {Barrado},
  {Bauer}, {Becerril}, {B{\`e}jar}, {Ben{\'\i}tez}, {Brinkm{\"o}ller}, {Cardona
  Guill{\'e}n}, {Cifuentes}, {Colom{\'e}}, {Cort{\'e}s-Contreras}, {Czesla},
  {Dreizler}, {Fr{\"o}lich}, {Fuhrmeister}, {Galad{\'\i}-Enr{\'\i}quez},
  {Gonz{\'a}lez Hern{\'a}ndez}, {Gonz{\'a}lez Peinado}, {Guenther}, {de
  Guindos}, {Hagen}, {Hatzes}, {Hauschildt}, {Helmling}, {Henning}, {Herbort},
  {Hern{\'a}ndez Casta{\~n}o}, {Herrero}, {Hintz}, {Jeffers}, {Johnson}, {de
  Juan}, {Kaminski}, {Klahr}, {K{\"u}rster}, {Lafarga}, {Sairam}, {Lamp{\'o}n},
  {Lara}, {Launhardt}, {L{\'o}pez del Fresno}, {L{\'o}pez-Puertas}, {Luque},
  {Mandel}, {Marfil}, {Mart{\'\i}n}, {Mart{\'\i}n-Ruiz}, {Mathar}, {Montes},
  {Morales}, {Nagel}, {Nortmann}, {Nowak}, {Pall{\'e}}, {Passegger}, {Pavlov},
  {Pedraz}, {P{\'e}rez-Medialdea}, {Perger}, {Rebolo}, {Reffert},
  {Rodr{\'\i}guez}, {Rodr{\'\i}guez L{\'o}pez}, {Rosich}, {Sabotta}, {Sadegi},
  {Salz}, {S{\'a}nchez-L{\'o}pez}, {Sanz-Forcada}, {Sarkis}, {Sch{\"a}fer},
  {Schiller}, {Schmitt}, {Sch{\"o}fer}, {Schweitzer}, {Shulyak}, {Solano},
  {Stahl}, {Tala Pinto}, {Trifonov}, {Zapatero Osorio}, {Yan}, {Zechmeister},
  {Abell{\'a}n}, {Abril}, {Alonso-Floriano}, {Ammler-von Eiff},
  {Anglada-Escud{\'e}}, {Anwand-Heerwart}, {Arroyo-Torres}, {Berdi{\~n}as},
  {Bergondy}, {Bl{\"u}mcke}, {del Burgo}, {Cano}, {Carro}, {C{\'a}rdenas},
  {Casal}, {Claret}, {D{\'\i}ez-Alonso}, {Doellinger}, {Dorda}, {Feiz},
  {Fern{\'a}ndez}, {Ferro}, {Gaisn{\'e}}, {Gallardo}, {G{\'a}lvez-Ortiz},
  {Garc{\'\i}a-Piquer}, {Garc{\'\i}a-Vargas}, {Garrido}, {Gesa}, {G{\'o}mez
  Galera}, {Gonz{\'a}lez-{\'A}lvarez}, {Gonz{\'a}lez-Cuesta}, {Grohnert},
  {Gr{\"o}zinger}, {Gu{\`a}rdia}, {Guijarro}, {Hedrosa}, {Hermann}, {Hermelo},
  {Hern{\'a}ndez Arab{\'\i}}, {Hern{\'a}ndez Hernando}, {Hidalgo}, {Holgado},
  {Huber}, {Huber}, {Huke}, {Kehr}, {Kim}, {Klein}, {Kl{\"u}ter}, {Klutsch},
  {Labarga}, {Labiche}, {Lamert}, {Laun}, {L{\'a}zaro}, {Lemke}, {Lenzen},
  {Llamas}, {Lizon}, {Lodieu}, {L{\'o}pez Gonz{\'a}lez}, {L{\'o}pez-Morales},
  {L{\'o}pez Salas}, {L{\'o}pez-Santiago}, {Mag{\'a}n Madinabeitia}, {Mall},
  {Mancini}, {Mar{\'\i}n Molina}, {Mart{\'\i}nez-Rodr{\'\i}guez}, {Maroto
  Fern{\'a}ndez}, {Marvin}, {Mirabet}, {Moreno-Raya}, {Moya}, {Mundt},
  {Naranjo}, {Panduro}, {Pascual}, {P{\'e}rez-Calpena}, {Perryman}, {Pluto},
  {Ram{\'o}n}, {Redondo}, {Reinhart}, {Rhode}, {Rix}, {Rodler}, {Rohloff},
  {S{\'a}nchez-Blanco}, {S{\'a}nchez Carrasco}, {Sarmiento}, {Schmidt},
  {Storz}, {Strachan}, {St{\"u}rmer}, {Su{\'a}rez}, {Tabernero}, {Tal-Or},
  {Tulloch}, {Ulbrich}, {Veredas}, {Vico Linares}, {Vidal-Dasilva},
  {Vilardell}, {Wagner}, {Winkler}, {Wolthoff}, {Xu}, \&
  {Zhao}}]{2018SPIE10702E..0WQ}
{Quirrenbach}, A., {Amado}, P.~J., {Ribas}, I., {et~al.} 2018, in Society of
  Photo-Optical Instrumentation Engineers (SPIE) Conference Series, Vol. 10702,
  Ground-based and Airborne Instrumentation for Astronomy VII, ed. C.~J.
  {Evans}, L.~{Simard}, \& H.~{Takami}, 107020W

\bibitem[{{Rainer} {et~al.}(2023){Rainer}, {Desidera}, {Borsa}, {Barbato},
  {Biazzo}, {Bonomo}, {Gratton}, {Messina}, {Scandariato}, {Affer}, {Benatti},
  {Carleo}, {Cabona}, {Covino}, {Lanza}, {Ligi}, {Maldonado}, {Mancini},
  {Nardiello}, {Sicilia}, {Sozzetti}, {Bignamini}, {Cosentino}, {Knapic},
  {Mart{\'\i}nez Fiorenzano}, {Molinari}, {Pedani}, \&
  {Poretti}}]{Raineretal2023}
{Rainer}, M., {Desidera}, S., {Borsa}, F., {et~al.} 2023, \aap, 676, A90

\bibitem[{Rajpaul {et~al.}(2015)Rajpaul, Aigrain, Osborne, Reece, \&
  Roberts}]{rajpaul_10.1093/mnras/stv1428}
Rajpaul, V., Aigrain, S., Osborne, M.~A., Reece, S., \& Roberts, S. 2015,
  Monthly Notices of the Royal Astronomical Society, 452, 2269

\bibitem[{Reiners \& Zechmeister(2020)}]{Reiners_2020}
Reiners, A. \& Zechmeister, M. 2020, The Astrophysical Journal Supplement
  Series, 247, 11

\bibitem[{{Ricker} {et~al.}(2015){Ricker}, {Winn}, {Vanderspek}, {Latham},
  {Bakos}, {Bean}, {Berta-Thompson}, {Brown}, {Buchhave}, {Butler}, {Butler},
  {Chaplin}, {Charbonneau}, {Christensen-Dalsgaard}, {Clampin}, {Deming},
  {Doty}, {De Lee}, {Dressing}, {Dunham}, {Endl}, {Fressin}, {Ge}, {Henning},
  {Holman}, {Howard}, {Ida}, {Jenkins}, {Jernigan}, {Johnson}, {Kaltenegger},
  {Kawai}, {Kjeldsen}, {Laughlin}, {Levine}, {Lin}, {Lissauer}, {MacQueen},
  {Marcy}, {McCullough}, {Morton}, {Narita}, {Paegert}, {Palle}, {Pepe},
  {Pepper}, {Quirrenbach}, {Rinehart}, {Sasselov}, {Sato}, {Seager},
  {Sozzetti}, {Stassun}, {Sullivan}, {Szentgyorgyi}, {Torres}, {Udry}, \&
  {Villasenor}}]{ricker2015JATIS...1a4003R}
{Ricker}, G.~R., {Winn}, J.~N., {Vanderspek}, R., {et~al.} 2015, Journal of
  Astronomical Telescopes, Instruments, and Systems, 1, 014003

\bibitem[{{Roberts} {et~al.}(1987){Roberts}, {Lehar}, \& {Dreher}}]{Roberts87}
{Roberts}, D.~H., {Lehar}, J., \& {Dreher}, J.~W. 1987, \aj, 93, 968

\bibitem[{{Sanz-Forcada} {et~al.}(2022){Sanz-Forcada}, {L{\'o}pez-Puertas},
  {Nortmann}, \& {Lamp{\'o}n}}]{SF22}
{Sanz-Forcada}, J., {L{\'o}pez-Puertas}, M., {Nortmann}, L., \& {Lamp{\'o}n},
  M. 2022, 21st Cambridge Workshop on Cool Stars, Stellar Systems, and the Sun

\bibitem[{{Sanz-Forcada} {et~al.}(2011){Sanz-Forcada}, {Micela}, {Ribas},
  {Pollock}, {Eiroa}, {Velasco}, {Solano}, \& {Garc{\'\i}a-{\'A}lvarez}}]{SF11}
{Sanz-Forcada}, J., {Micela}, G., {Ribas}, I., {et~al.} 2011, \aap, 532, A6

\bibitem[{{Schwab} {et~al.}(2016){Schwab}, {Rakich}, {Gong}, {Mahadevan},
  {Halverson}, {Roy}, {Terrien}, {Robertson}, {Hearty}, {Levi}, {Monson},
  {Wright}, {McElwain}, {Bender}, {Blake}, {St{\"u}rmer}, {Gurevich},
  {Chakraborty}, \& {Ramsey}}]{schwab2016SPIE.9908E..7HS}
{Schwab}, C., {Rakich}, A., {Gong}, Q., {et~al.} 2016, in Society of
  Photo-Optical Instrumentation Engineers (SPIE) Conference Series, Vol. 9908,
  Ground-based and Airborne Instrumentation for Astronomy VI, ed. C.~J.
  {Evans}, L.~{Simard}, \& H.~{Takami}, 99087H

\bibitem[{{Shappee} {et~al.}(2014){Shappee}, {Prieto}, {Grupe}, {Kochanek},
  {Stanek}, {De Rosa}, {Mathur}, {Zu}, {Peterson}, {Pogge}, {Komossa}, {Im},
  {Jencson}, {Holoien}, {Basu}, {Beacom}, {Szczygie{\l}}, {Brimacombe},
  {Adams}, {Campillay}, {Choi}, {Contreras}, {Dietrich}, {Dubberley},
  {Elphick}, {Foale}, {Giustini}, {Gonzalez}, {Hawkins}, {Howell}, {Hsiao},
  {Koss}, {Leighly}, {Morrell}, {Mudd}, {Mullins}, {Nugent}, {Parrent},
  {Phillips}, {Pojmanski}, {Rosing}, {Ross}, {Sand}, {Terndrup}, {Valenti},
  {Walker}, \& {Yoon}}]{2014ApJ...788...48S}
{Shappee}, B.~J., {Prieto}, J.~L., {Grupe}, D., {et~al.} 2014, \apj, 788, 48

\bibitem[{{Smith} {et~al.}(2012){Smith}, {Stumpe}, {Van Cleve}, {Jenkins},
  {Barclay}, {Fanelli}, {Girouard}, {Kolodziejczak}, {McCauliff}, {Morris}, \&
  {Twicken}}]{2012PASP..124.1000S}
{Smith}, J.~C., {Stumpe}, M.~C., {Van Cleve}, J.~E., {et~al.} 2012, \pasp, 124,
  1000

\bibitem[{{Sneden}(1973)}]{1973sneden}
{Sneden}, C.~A. 1973, PhD thesis, University of Texas, Austin

\bibitem[{{Sousa} {et~al.}(2015){Sousa}, {Santos}, {Adibekyan}, {Delgado-Mena},
  \& {Israelian}}]{2015sousa}
{Sousa}, S.~G., {Santos}, N.~C., {Adibekyan}, V., {Delgado-Mena}, E., \&
  {Israelian}, G. 2015, \aap, 577, A67

\bibitem[{{Sozzetti} {et~al.}(2007){Sozzetti}, {Torres}, {Charbonneau},
  {Latham}, {Holman}, {Winn}, {Laird}, \& {O'Donovan}}]{2007ApJ...664.1190S}
{Sozzetti}, A., {Torres}, G., {Charbonneau}, D., {et~al.} 2007, \apj, 664, 1190

\bibitem[{{Spina} {et~al.}(2020){Spina}, {Nordlander}, {Casey}, {Bedell},
  {D'Orazi}, {Mel{\'e}ndez}, {Karakas}, {Desidera}, {Baratella}, {Yana
  Galarza}, \& {Casali}}]{2020spina}
{Spina}, L., {Nordlander}, T., {Casey}, A.~R., {et~al.} 2020, \apj, 895, 52

\bibitem[{{Stumpe} {et~al.}(2014){Stumpe}, {Smith}, {Catanzarite}, {Van Cleve},
  {Jenkins}, {Twicken}, \& {Girouard}}]{2014PASP..126..100S}
{Stumpe}, M.~C., {Smith}, J.~C., {Catanzarite}, J.~H., {et~al.} 2014, \pasp,
  126, 100

\bibitem[{{Stumpe} {et~al.}(2012){Stumpe}, {Smith}, {Van Cleve}, {Twicken},
  {Barclay}, {Fanelli}, {Girouard}, {Jenkins}, {Kolodziejczak}, {McCauliff}, \&
  {Morris}}]{2012PASP..124..985S}
{Stumpe}, M.~C., {Smith}, J.~C., {Van Cleve}, J.~E., {et~al.} 2012, \pasp, 124,
  985

\bibitem[{{Terquem} \& {Papaloizou}(2007)}]{terquem}
{Terquem}, C. \& {Papaloizou}, J. C.~B. 2007, \apj, 654, 1110

\bibitem[{{Terquem} \& {Papaloizou}(2019)}]{terquem2}
{Terquem}, C. \& {Papaloizou}, J. C.~B. 2019, \mnras, 482, 530

\bibitem[{{Trifonov} {et~al.}(2018){Trifonov}, {K{\"u}rster}, {Zechmeister},
  {Tal-Or}, {Caballero}, {Quirrenbach}, {Amado}, {Ribas}, {Reiners}, {Reffert},
  {Dreizler}, {Hatzes}, {Kaminski}, {Launhardt}, {Henning}, {Montes},
  {B{\'e}jar}, {Mundt}, {Pavlov}, {Schmitt}, {Seifert}, {Morales}, {Nowak},
  {Jeffers}, {Rodr{\'\i}guez-L{\'o}pez}, {del Burgo}, {Anglada-Escud{\'e}},
  {L{\'o}pez-Santiago}, {Mathar}, {Ammler-von Eiff}, {Guenther}, {Barrado},
  {Gonz{\'a}lez Hern{\'a}ndez}, {Mancini}, {St{\"u}rmer}, {Abril}, {Aceituno},
  {Alonso-Floriano}, {Antona}, {Anwand-Heerwart}, {Arroyo-Torres}, {Azzaro},
  {Baroch}, {Bauer}, {Becerril}, {Ben{\'\i}tez}, {Berdi{\~n}as}, {Bergond},
  {Bl{\"u}mcke}, {Brinkm{\"o}ller}, {Cano}, {C{\'a}rdenas V{\'a}zquez},
  {Casal}, {Cifuentes}, {Claret}, {Colom{\'e}}, {Cort{\'e}s-Contreras},
  {Czesla}, {D{\'\i}ez-Alonso}, {Feiz}, {Fern{\'a}ndez}, {Ferro},
  {Fuhrmeister}, {Galad{\'\i}-Enr{\'\i}quez}, {Garcia-Piquer}, {Garc{\'\i}a
  Vargas}, {Gesa}, {G{\'o}mez Galera}, {Gonz{\'a}lez-Peinado}, {Gr{\"o}zinger},
  {Grohnert}, {Gu{\`a}rdia}, {Guijarro}, {de Guindos}, {Guti{\'e}rrez-Soto},
  {Hagen}, {Hauschildt}, {Hedrosa}, {Helmling}, {Hermelo}, {Hern{\'a}ndez
  Arab{\'\i}}, {Hern{\'a}ndez Casta{\~n}o}, {Hern{\'a}ndez Hernando},
  {Herrero}, {Huber}, {Huke}, {Johnson}, {de Juan}, {Kim}, {Klein},
  {Kl{\"u}ter}, {Klutsch}, {Lafarga}, {Lamp{\'o}n}, {Lara}, {Laun}, {Lemke},
  {Lenzen}, {L{\'o}pez del Fresno}, {L{\'o}pez-Gonz{\'a}lez},
  {L{\'o}pez-Puertas}, {L{\'o}pez Salas}, {Luque}, {Mag{\'a}n Madinabeitia},
  {Mall}, {Mandel}, {Marfil}, {Mar{\'\i}n Molina}, {Maroto Fern{\'a}ndez},
  {Mart{\'\i}n}, {Mart{\'\i}n-Ruiz}, {Marvin}, {Mirabet}, {Moya},
  {Moreno-Raya}, {Nagel}, {Naranjo}, {Nortmann}, {Ofir}, {Oreiro}, {Pall{\'e}},
  {Panduro}, {Pascual}, {Passegger}, {Pedraz}, {P{\'e}rez-Calpena}, {P{\'e}rez
  Medialdea}, {Perger}, {Perryman}, {Pluto}, {Rabaza}, {Ram{\'o}n}, {Rebolo},
  {Redondo}, {Reinhardt}, {Rhode}, {Rix}, {Rodler}, {Rodr{\'\i}guez},
  {Rodr{\'\i}guez Trinidad}, {Rohloff}, {Rosich}, {Sadegi},
  {S{\'a}nchez-Blanco}, {S{\'a}nchez Carrasco}, {S{\'a}nchez-L{\'o}pez},
  {Sanz-Forcada}, {Sarkis}, {Sarmiento}, {Sch{\"a}fer}, {Schiller},
  {Sch{\"o}fer}, {Schweitzer}, {Solano}, {Stahl}, {Strachan}, {Su{\'a}rez},
  {Tabernero}, {Tala}, {Tulloch}, {Veredas}, {Vico Linares}, {Vilardell},
  {Wagner}, {Winkler}, {Wolthoff}, {Xu}, {Yan}, \& {Zapatero
  Osorio}}]{2018A&A...609A.117T}
{Trifonov}, T., {K{\"u}rster}, M., {Zechmeister}, M., {et~al.} 2018, \aap, 609,
  A117

\bibitem[{{Turrini} {et~al.}(2023){Turrini}, {Marzari}, {Polychroni}, {Claudi},
  {Desidera}, {Mesa}, {Pinamonti}, {Sozzetti}, {Su{\'a}rez Mascare{\~n}o},
  {Damasso}, {Benatti}, {Malavolta}, {Micela}, {Zinzi}, {B{\'e}jar}, {Biazzo},
  {Bignamini}, {Bonavita}, {Borsa}, {del Burgo}, {Chauvin}, {Delorme},
  {Gonz{\'a}lez Hern{\'a}ndez}, {Gratton}, {Hagelberg}, {Janson}, {Langlois},
  {Lanza}, {Lazzoni}, {Lodieu}, {Maggio}, {Mancini}, {Molinari}, {Molinaro},
  {Murgas}, \& {Nardiello}}]{2023A&A...679A..55T}
{Turrini}, D., {Marzari}, F., {Polychroni}, D., {et~al.} 2023, \aap, 679, A55

\bibitem[{{Van Eylen} {et~al.}(2019){Van Eylen}, {Albrecht}, {Huang},
  {MacDonald}, {Dawson}, {Cai}, {Foreman-Mackey}, {Lundkvist}, {Silva Aguirre},
  {Snellen}, \& {Winn}}]{vaneylen2019AJ....157...61V}
{Van Eylen}, V., {Albrecht}, S., {Huang}, X., {et~al.} 2019, \aj, 157, 61

\bibitem[{Wang(2017)}]{wang_2017}
Wang, S. 2017, Research Notes of the AAS, 1, 26

\bibitem[{{Weiss} {et~al.}(2018){Weiss}, {Marcy}, {Petigura}, {Fulton},
  {Howard}, {Winn}, {Isaacson}, {Morton}, {Hirsch}, {Sinukoff}, {Cumming},
  {Hebb}, \& {Cargile}}]{weiss_rad}
{Weiss}, L.~M., {Marcy}, G.~W., {Petigura}, E.~A., {et~al.} 2018, \aj, 155, 48

\bibitem[{Xu \& Wang(2024)}]{xu_bump}
Xu, W. \& Wang, S. 2024, The Astrophysical Journal Letters, 962, L4

\bibitem[{{Yana Galarza} {et~al.}(2019){Yana Galarza}, {Mel{\'e}ndez},
  {Lorenzo-Oliveira}, {Valio}, {Reggiani}, {Carlos}, {Ponte}, {Spina},
  {Haywood}, \& {Gandolfi}}]{2019galarza}
{Yana Galarza}, J., {Mel{\'e}ndez}, J., {Lorenzo-Oliveira}, D., {et~al.} 2019,
  \mnras, 490, L86

\bibitem[{{Zakhozhay} {et~al.}(2022){Zakhozhay}, {Launhardt}, {M{\"u}ller},
  {Brems}, {Eigenthaler}, {Gennaro}, {Hempel}, {Hempel}, {Henning}, {Kennedy},
  {Kim}, {K{\"u}rster}, {Lachaume}, {Manerikar}, {Patel}, {Pavlov}, {Reffert},
  \& {Trifonov}}]{2022A&A...667A..63Z}
{Zakhozhay}, O.~V., {Launhardt}, R., {M{\"u}ller}, A., {et~al.} 2022, \aap,
  667, A63

\bibitem[{{Zechmeister} \& {K{\"u}rster}(2009)}]{2009A&A...496..577Z}
{Zechmeister}, M. \& {K{\"u}rster}, M. 2009, \aap, 496, 577

\bibitem[{{Zechmeister} {et~al.}(2018){Zechmeister}, {Reiners}, {Amado},
  {Azzaro}, {Bauer}, {B{\'e}jar}, {Caballero}, {Guenther}, {Hagen}, {Jeffers},
  {Kaminski}, {K{\"u}rster}, {Launhardt}, {Montes}, {Morales}, {Quirrenbach},
  {Reffert}, {Ribas}, {Seifert}, {Tal-Or}, \& {Wolthoff}}]{2018A&A...609A..12Z}
{Zechmeister}, M., {Reiners}, A., {Amado}, P.~J., {et~al.} 2018, \aap, 609, A12

\bibitem[{Zhang {et~al.}(2023)Zhang, Knutson, Dai, Wang, Ricker, Schwarz, Mann,
  \& Collins}]{Zhang_2023}
Zhang, M., Knutson, H.~A., Dai, F., {et~al.} 2023, The Astronomical Journal,
  165, 62

\end{thebibliography}
\begin{appendix}
\section{Additional spectroscopic data} \label{app:data}    
\subsection{CARMENES spectroscopic observations}
Between 6 April and 07 October 2021 we collected 22 spectra with the CARMENES spectrograph mounted on the 3.5\,m telescope Calar Alto Observatory, Almer\'{i}a, Spain, under the observing programs S21-3.5-006 and F21-3.5-005 (PI Pall\'{e}). The exposure time was set to 600s leading to a S/N per pixel of 47--133 at 7370\,\AA. The CARMENES spectrograph has two arms \citep{2014SPIE.9147E..1FQ,2018SPIE10702E..0WQ}, the visible (VIS) arm covering the spectral range 5200--9600\,$\AA$ and a near-infrared (NIR) arm covering the spectral range 9600--17100\,$\AA$. CARMENES performance, data reduction and wavelength calibration are described in \citet{2018A&A...609A.117T} and \citet{2018A&A...618A.115K}. Relative radial velocity values, chromatic index (CRX), differential line width (dLW), and H$\alpha$ index values were obtained using \texttt{serval}\citep{2018A&A...609A..12Z}. All RV measurements were corrected for barycentric motion, secular acceleration and nightly zero-points. In this work, we analysed relative RVs measured from VIS spectra. Their uncertainties are in the range 3.2--10.2\,\ms with a mean value of 5.7\,\ms.

\subsection{NEID spectroscopic observations}
NEID is a fiber-fed, red-optical (3800–9300\,\AA), environmentally stabilized echelle spectrograph \citep{schwab2016SPIE.9908E..7HS} installed on the WIYN 3.5 m telescope at Kitt Peak National Observatory. In our study of TOI-2076, we obtained 15 high-resolution observations (resolving power of R $\sim$113,000) using NEID, spanning from May 14 to June 27. Each observation was conducted with an exposure time of 1200 seconds. The data were processed with the updated version 1.2.1 of the NEID Data Reduction Pipeline (DRP; \citealt{kaplan2019ASPC..523..567K}), utilising the CCF \citep{baranne1996A&AS..119..373B} method to calculate radial velocities (RVs). The NEID CCF RVs from this dataset demonstrated a median single measurement precision of 1.7 \ms, reflecting a level of accuracy similar to that of HARPS-N.\\
\section{Description of the test models for RVs and spectroscopic activity diagnostic time series} \label{app:rvmodeldetails}

All our test models to fit the RVs included a component to correct for stellar activity. The signal due to the stellar activity has been fitted through a GP regression adopting a QP kernel, which has been widely and effectively used to mitigate the activity term in the RV time series. We consider this GP kernel an optimal choice in our case, based on the results of the frequency content analysis discussed in Sect. \ref{sec:activity}. A generic element of the QP covariance matrix (e.g. \citealt{haywood2014}) is defined as follows:
		\begin{gather} 
			\label{eq:eqgpqpkernel}
			k_{QP}(t, t^{\prime}) = h^2\cdot\exp\Bigg[-\frac{(t-t^{\prime})^2}{2\lambda^2} - \frac{\sin^{2}\Bigg(\pi(t-t^{\prime})/\theta\Bigg)}{2w^2}\Bigg] + \nonumber \\
			+\, (\sigma^{2}_{\rm RV}(t)+\sigma^{2}_{\rm jit})\cdot\delta_{t, t^{\prime}}
		\end{gather}
Here, $t$ and $t^{\prime}$ represent two different epochs of observations, $\sigma_{\rm RV}$ is the radial velocity uncertainty of a specific instrument, and $\delta_{t, t^{\prime}}$ is the Kronecker delta. We take other sources of uncorrelated noise -- instrumental and/or astrophysical -- into account by adding a constant jitter term $\sigma_{\rm jit}$ for each spectrograph in quadrature to the formal uncertainties $\sigma_{\rm RV}$. The GP hyper-parameters are $h$, which denotes the scale amplitude of the correlated signal (specific for each instrument); $\theta$, which represents the periodic timescale of the correlated signal, and corresponds to the stellar rotation period; $w$, which describes the "weight" of the rotation period harmonic content within a complete stellar rotation (i.e. a low value of $w$ indicates that the periodic variations contain a significant contribution from the harmonics of the rotation periods); and $\lambda$, that represents the decay timescale of the correlations, and is related to the temporal evolution of the magnetically active regions responsible for the correlated signal observed in the RVs. 

Hereafter we provide a description of the different GP-based models used in our work. The priors of the corresponding free (hyper-)parameters are listed in Table \ref{tab:rvpriors}.

\subsection{GP regression not trained on activity indicators}
We applied the GP regression to all the available RVs time series (HARPS-N, CARMENES, and NEID). In a first attempt, we used a single QP kernel to model the complete time series, i.e. one $w$ and one $\lambda$ free hyper-parameter was adopted over the whole time span of the data, with one scale amplitude $h$ for each instrument. In a second test, the analysis was performed on a seasonal basis, i.e. we divided the dataset into three chunks of time (or ``seasons'': BJD 2459068--2459479; 2459571--2459837; 2459950--2460187), and used a triplet of $h$, $w$ and $\lambda$ free hyper-parameters to model each chunk of data. We selected this test model taking into account the possibility that the properties of the RV pattern as seen in the stellar magnetic activity could change from one season to another.

\subsection{GP regression trained on the chromospheric activity diagnostic \logrhk}

We used the stellar activity proxy \logrhk to help constraining the hyper-parameters of the QP activity component in the full RV time series of HARPS-N. To that purpose, we modelled the \logrhk time series derived from HARPS-N spectra with a model including a sinusoid to fit the long-term modulation (see Sect. \ref{sec:freqcontentanalysis}), and a QP kernel sharing the $w$, $\theta$ and $\lambda$ hyper-parameters with the QP kernel used for the GP regression of the HARPS-N RVs (i.e. those GP hyper-parameters are ``trained'' on the activity diagnostic time series).  

\subsection{Multidimensional GP regression}
A novel framework to model stellar activity in RV time series simultaneously with activity diagnostics, defined as ``multidimensional GP regression'', was first proposed by \cite{rajpaul_10.1093/mnras/stv1428} and later further developed by, e.g., \cite{Gilbertson_2020} and \cite{pyaneti2}. Within this framework, a set of simultaneous spectroscopic observables derived from the same spectra, containing signals induced by stellar activity, are jointly modelled assuming that they are described by the same underlying function and its derivatives, which are generated by adopting a specific GP kernel. Recently, the multidimensional GP regression has been frequently tested as a potentially effective method to measure the masses of planets orbiting active stars, with an age and activity-induced RV scatter comparable to those of TOI-2076 (e.g. \citealt{2022MNRAS.514.1606B,nardiello2022,barragan_10.1093/mnras/stad1139,damasso_2023A&A...672A.126D}) 
In our study, we adopted the framework detailed in \cite{pyaneti2} (e.g. see Section 3.1 therein for the extensive description of the physical basis of the method and theoretical approach), which makes use of the first time derivative of a GP-based function. For our analyses, we included in our \texttt{MultiNest}-based framework the \texttt{Python} modules specifically developed to perform a multidimensional GP regression which are part of the \texttt{pyaneti} package\footnote{\url{https://github.com/oscaribv/pyaneti}} \citep{pyaneti,pyaneti2}. We have performed a multidimensional GP regression analysis using the HARPS-N RVs and the activity diagnostics \logrhk, BIS, and FWHM time series derived from the same spectra. We have simultaneously fit the RV dataset with one of the activity indicators at a time (i.e. we performed a two-dimensional GP regression), resulting in three alternative multidimensional GP test models. We adopted the QP covariance matrix described in Eq. \ref{eq:eqgpqpkernel} as the GP kernel. When \logrhk or FWHM was used, we simultaneously fit a sinusoid to model the long-term modulation observed for the two activity indicators (see Sect. \ref{sec:freqcontentanalysis}), and we did not consider the first-time derivative of the GP-based function to model the activity diagnostic time series, following the discussion in Sect. 3 of \cite{rajpaul_10.1093/mnras/stv1428}. 
In our study, we label with ($A_{\rm RV}$, $B_{\rm RV}$), ($A_{\rm BIS}$, $B_{\rm BIS}$), $A_{\rm log\,R^{\prime}_\mathrm{HK}}$, and $A_{\rm FWHM}$ the coefficients that relate the individual time series to the underlying GP-based function (coefficients \textit{A}) and its first-time derivative (coefficients \textit{B}). These coefficients are treated as free parameters in our test models. The coefficients of the first-time derivative, $B_{\rm log\,R^{\prime}_\mathrm{HK}}$ and $B_{\rm FWHM}$, are fixed to zero. 

\begin{table*}[htbp]
		\caption{Prior used to model RVs and spectroscopic activity diagnostics.}
		\label{tab:rvpriors}
		\small
\begin{threeparttable}
			\begin{tabular}{ll}
				\hline
				\textbf{Parameter}   & \textbf{Prior}\\ 
				\hline
				\noalign{\smallskip} 
				\textit{GP hyper-parameters:} & \\
				\noalign{\smallskip}
				$h_{\rm HARPS-N}$ $[\ms]$ & $\mathcal{U}$(0,100) \\
                    \noalign{\smallskip}
				$h_{\rm CARMENES-VIS}$ $[\ms]$ & $\mathcal{U}$(0,100)\\
                    \noalign{\smallskip}
                    $h_{\rm NEID}$ $[\ms]$ & $\mathcal{U}$(0,100)\\
                    \noalign{\smallskip}
				$\ln \lambda$ [days] & $\mathcal{U}$(0,10) \\
				\noalign{\smallskip}
				$w$ & $\mathcal{U}$(0,2) \\ 
				\noalign{\smallskip}
                    $\theta$ [days] & $\mathcal{U}$(0,10)  \\
				\noalign{\smallskip}
				\textit{Planet-related parameters:} &  \\
				\noalign{\smallskip}  
				$K_{\rm b}$ $[\ms]$ & $\mathcal{U}$(0,20) \\
				\noalign{\smallskip} 
				orbital period, $P_{\rm b}$ [days] & $\mathcal{N}$(10.35523,0.00001)  \\
				\noalign{\smallskip}
				T$_{\rm conj,\,b}$ [BJD-2450000] & $\mathcal{N}$(8950.8289,0.0005) \\
                    \noalign{\smallskip}  
				$K_{\rm c}$ $[\ms]$ & $\mathcal{U}$(0,20) \\
				\noalign{\smallskip} 
				orbital period, $P_{\rm c}$ [days] & $\mathcal{N}$(21.01549,0.00003)  \\
				\noalign{\smallskip}
				T$_{\rm conj,\,c}$ [BJD-2450000] & $\mathcal{N}$(8937.8283,0.0007) \\
                    \noalign{\smallskip}  
				$K_{\rm d}$ $[\ms]$ & $\mathcal{U}$(0,20) \\
				\noalign{\smallskip} 
				orbital period, $P_{\rm d}$ [days] & $\mathcal{N}$(35.125514,0.00007)  \\
				\noalign{\smallskip}
				T$_{\rm conj,\,d}$ [BJD-2450000] & $\mathcal{N}$(8938.2965,0.001) \\
                    \noalign{\smallskip}
                    acceleration, $\dot{\gamma}$ [$\ms d^{-1}$] & $\mathcal{U}$(-1,1) \\
				\noalign{\smallskip}
                    \textit{activity diagnostic-related parameters:} &  \\
				\noalign{\smallskip} 
                    semi-amplitude of a long-term activity signal (\logrhk) [dex] & $\mathcal{U}$(0,1) \\
                    \noalign{\smallskip} 
                    semi-amplitude of a long-term activity signal (FWHM) [m/s] & $\mathcal{U}$(0,100) \\
                    \noalign{\smallskip} 
                    period of a long-term activity signal (\logrhk and FWHM) [days] & $\mathcal{U}$(0,2000) \\
                    \noalign{\smallskip} 
                    Reference time of a long-term activity signal (\logrhk and FWHM) [BJD-2450000] & $\mathcal{U}$(9000,11000) \\
                    \noalign{\smallskip} 
                    $\sigma_{\rm jit,\: \log\,R^{\prime}_\mathrm{HK}}$ [dex] & $\mathcal{U}$(0,0.1) \\
                    \noalign{\smallskip} 
                    offset, $\gamma_{\rm \log\,R^{\prime}_\mathrm{HK}}$ [dex] & $\mathcal{U}$(-4.6,-4) \\
                    \noalign{\smallskip} 
                    $\sigma_{\rm jit,\: FWHM}$ [m/s] & $\mathcal{U}$(0,50) \\
                    \noalign{\smallskip} 
                    offset, $\gamma_{\rm FWHM}$ [m/s] & $\mathcal{U}$(8000,9500) \\
                    \noalign{\smallskip} 
                    \textit{Multidimensional GP-related parameters:} & \\
				  \noalign{\smallskip}
                    $A_{\rm RV}$, $B_{\rm RV}$ [m/s] & $\mathcal{U}$(-100,100) \\
                    \noalign{\smallskip}
                    $A_{\rm BIS}$, $B_{\rm BIS}$ [m/s] & $\mathcal{U}$(-100,100) \\
                    \noalign{\smallskip}
                    $A_{\rm FWHM}$ [m/s] & $\mathcal{U}$(-100,100) \\
                    \noalign{\smallskip}
                    $A_{\rm log\,R^{\prime}_\mathrm{HK}}$ [dex] & $\mathcal{U}$(-0.1,0.1) \\
                    \noalign{\smallskip}
                    \textit{Instrument-related parameters:} & \\
				  \noalign{\smallskip}
				$\sigma_{\rm jit,\: HARPS-N}$ $[\ms]$ & $\mathcal{U}$(0,50) (all RV extractions) \\ 
                    \noalign{\smallskip}
                    offset, $\gamma_{\rm HARPS-N}$ $[\ms]$ & $\mathcal{U}$(-13000,-12500) (DRS) \\ 
                    \noalign{\smallskip}
				offset, $\gamma_{\rm HARPS-N}$ $[\ms]$ & $\mathcal{U}$(-100,100) (SERVAL) \\ 
                    \noalign{\smallskip}
				offset, $\gamma_{\rm HARPS-N}$ $[\ms]$ & $\mathcal{U}$(-12700,-12300) (LBL) \\ 
				\noalign{\smallskip}
				$\sigma_{\rm jit,\: CARMENES-VIS}$ $[\ms]$ & $\mathcal{U}$(0,50) \\ 
				\noalign{\smallskip}
				$\gamma_{\rm CARMENES-VIS}$ $[\ms]$ & $\mathcal{U}$(-100,100) \\
                    \noalign{\smallskip}          
				$\sigma_{\rm jit,\: NEID}$ $[\ms]$ & $\mathcal{U}$(0,50) \\ 
				\noalign{\smallskip}
				$\gamma_{\rm NEID}$ $[\ms]$ & $\mathcal{U}$(-12800,-12600) \\
                    \noalign{\smallskip}
                    \textit{\texttt{TRADES} specific parameters:} & \\
				  \noalign{\smallskip}
                    planet mass [M$_{\odot}$] & $\mathcal{U}$(0,0.0003)\tnote{a}\\
                    \noalign{\smallskip}
                    planet eccentricity & $\mathcal{N}$(0,0.1)\tnote{b}\\
                    \noalign{\smallskip}
                    argument of periastron [deg] & $\mathcal{U}$(0,360)\tnote{a}\\
                    \noalign{\smallskip}
                    mean anomaly [deg] & $\mathcal{U}$(0,360)\tnote{a}\\
                    \noalign{\smallskip}
                    stellar mass [M$_{\odot}$] & 0.849 (fixed) \\
                    \noalign{\smallskip}
                    longitude of ascending node [deg] & 180 (fixed) \\ 
                    \noalign{\smallskip}
				\hline
        		\end{tabular}
    \begin{tablenotes}
     \item[a] Same prior adopted for all the transiting planets in the system.
     \item[b] Following the result of \cite{vaneylen2019AJ....157...61V} for multi-planet systems. Only positive values were sampled. The same prior has been adopted for all three transiting planets in the system.
    \end{tablenotes}
   \end{threeparttable}
	\end{table*}

\section{Additional plots and tables} \label{app:additionalplots}

\begin{figure*}
    \centering
    \includegraphics[width=0.6\linewidth, angle=180]{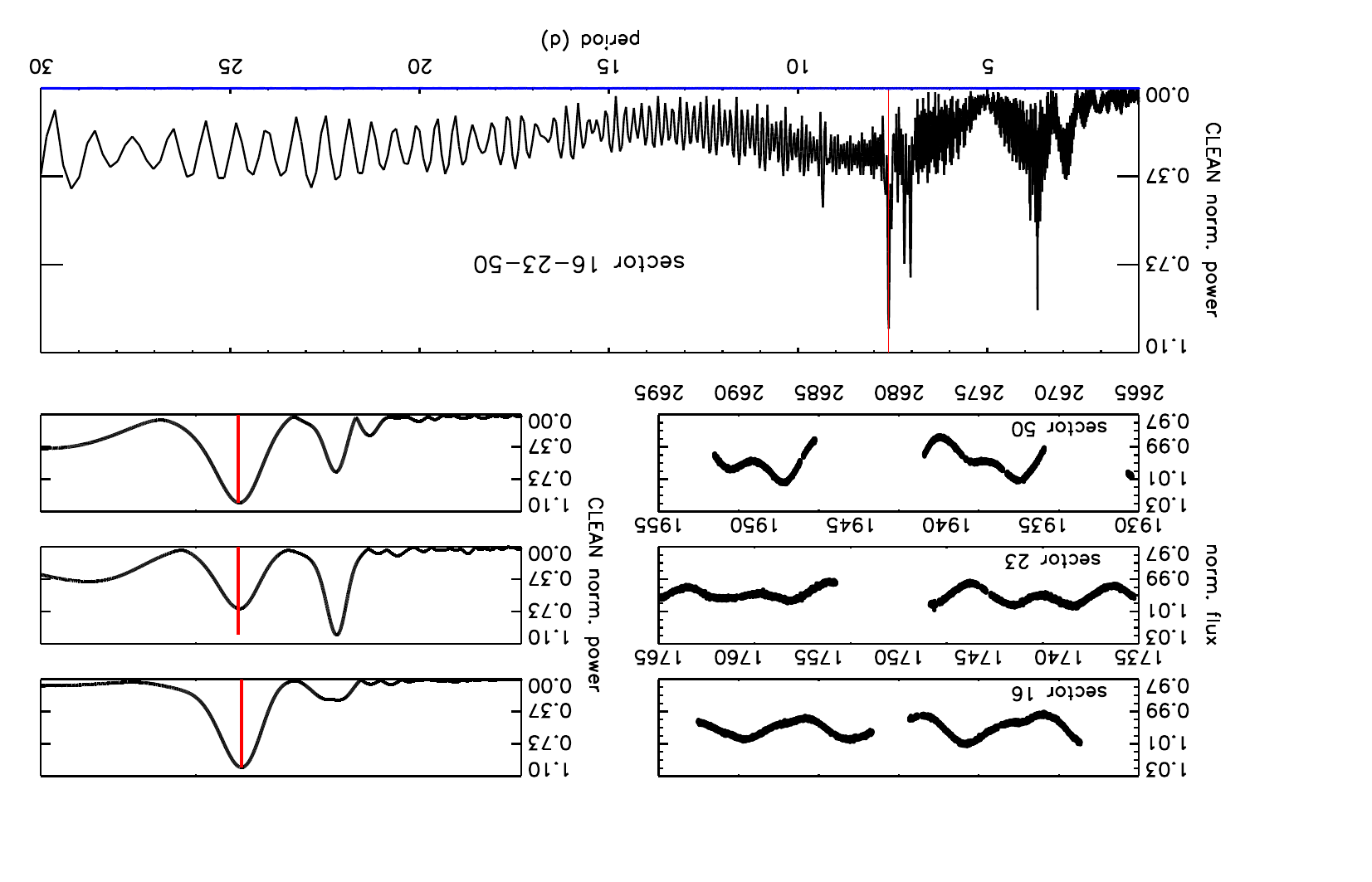}
    \caption{Periodograms of the TESS light curve calculated with the \texttt{CLEAN} algorithm.}
    \label{fig:lcperiod}
\end{figure*}

\begin{figure*}
    \centering
    \includegraphics[width=1\linewidth]{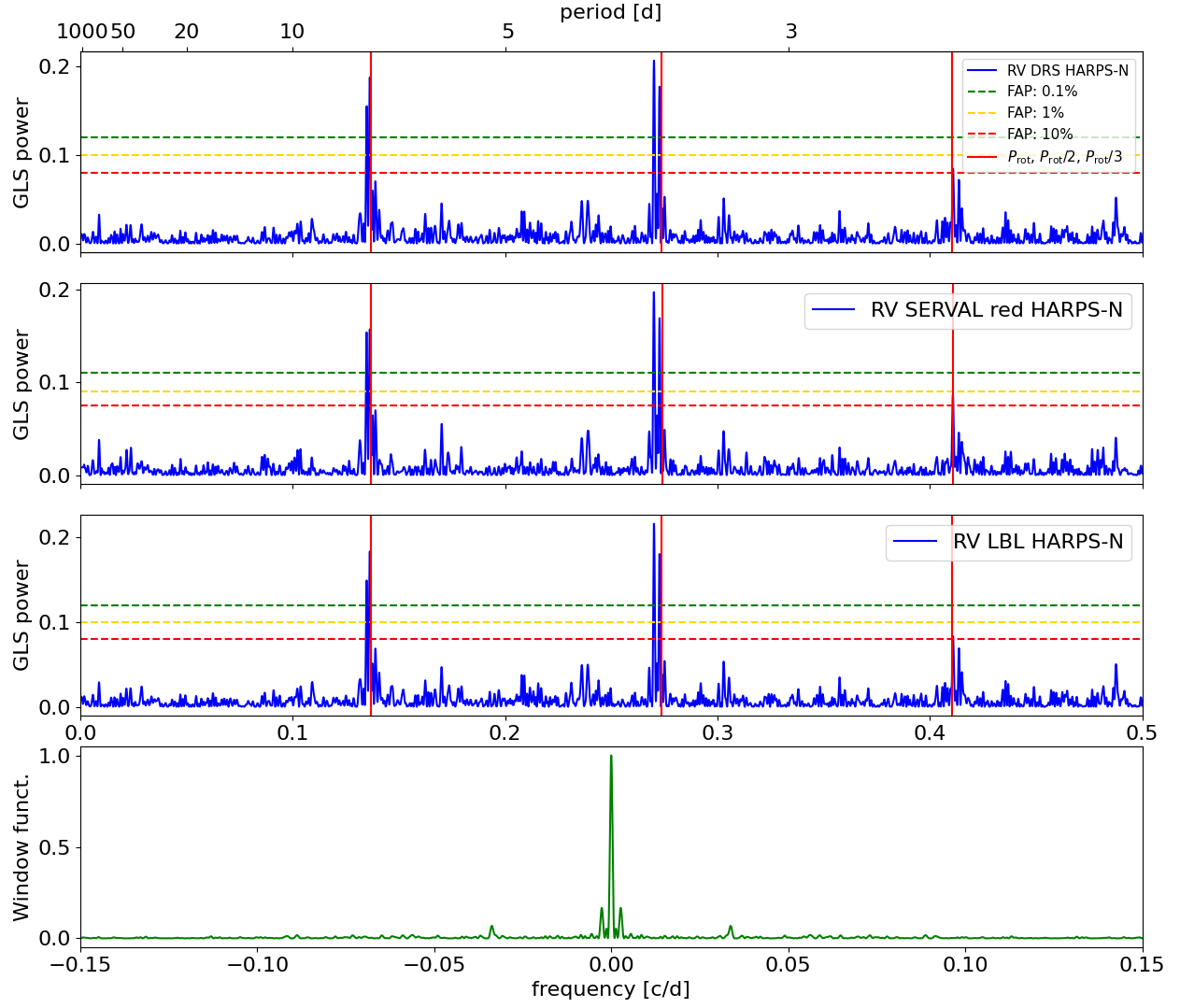}
    \caption{GLS periodograms of the RV dataset extracted from HARPS-N spectra using different algorithms, except that shown in Fig. \ref{fig:gls_diagnostics_harpn} (see Table \ref{tab:rvsummary}).}
    \label{fig:gls_rv_harpn}
\end{figure*}





\begin{figure*}[ht]
    \centering
    \subfigure[]{\includegraphics[width=0.45\textwidth]{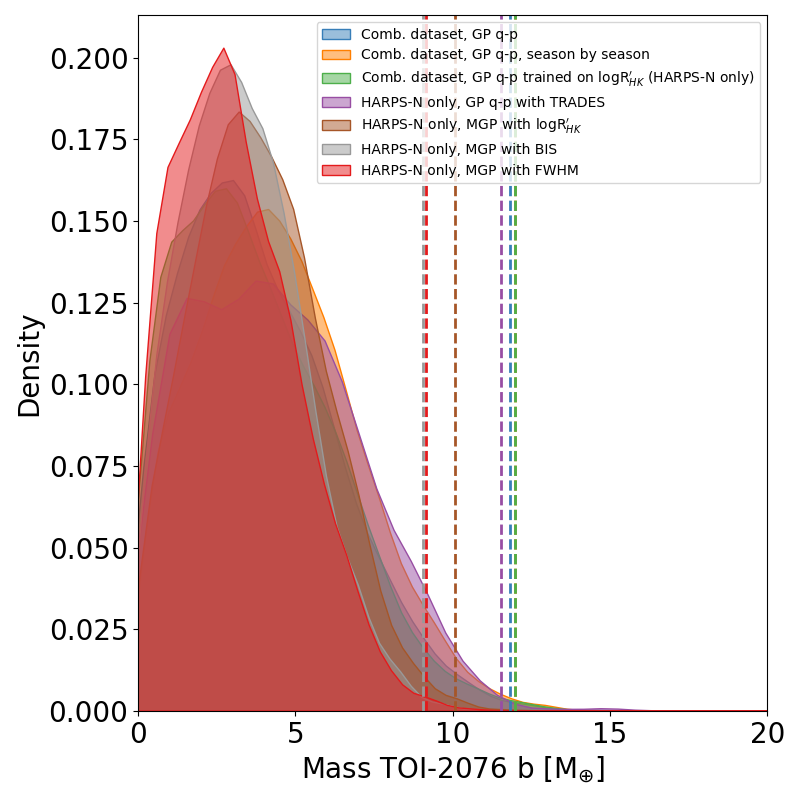}}
    \subfigure[]{\includegraphics[width=0.45\textwidth]{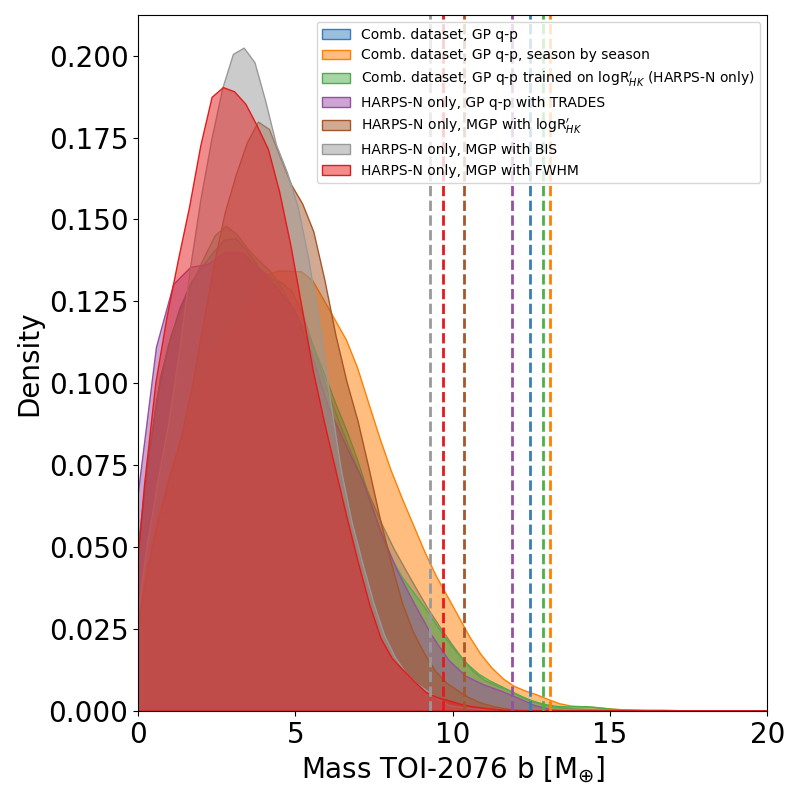}}
    \subfigure[]{\includegraphics[width=0.45\textwidth]{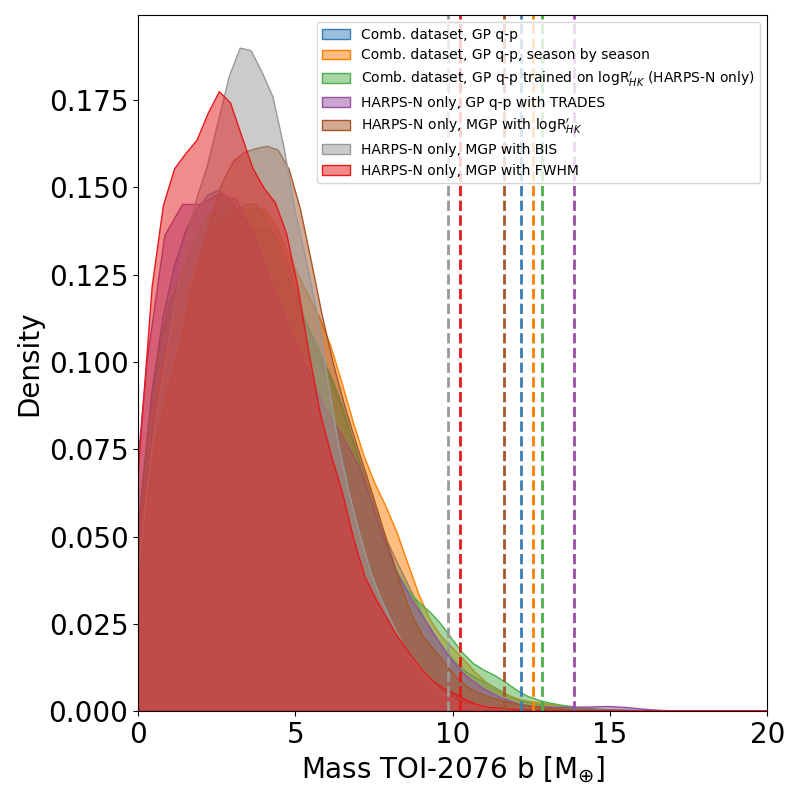}}
    \caption{Posterior mass distributions for the planet BD+40 2790/TOI-2076~b. The dashed vertical lines indicate the 3$\sigma$ upper limits for each distribution. Each panel corresponds to a HARPS-N RV dataset extracted with a different method (see Table \ref{tab:rvsummary}): (a) DRS; (b) SERVAL; (c) LBL }
    \label{fig:mass_planet_b_2}
\end{figure*}

\newpage

\begin{figure*}[ht]
    \centering
    \subfigure[]{\includegraphics[width=0.45\textwidth]{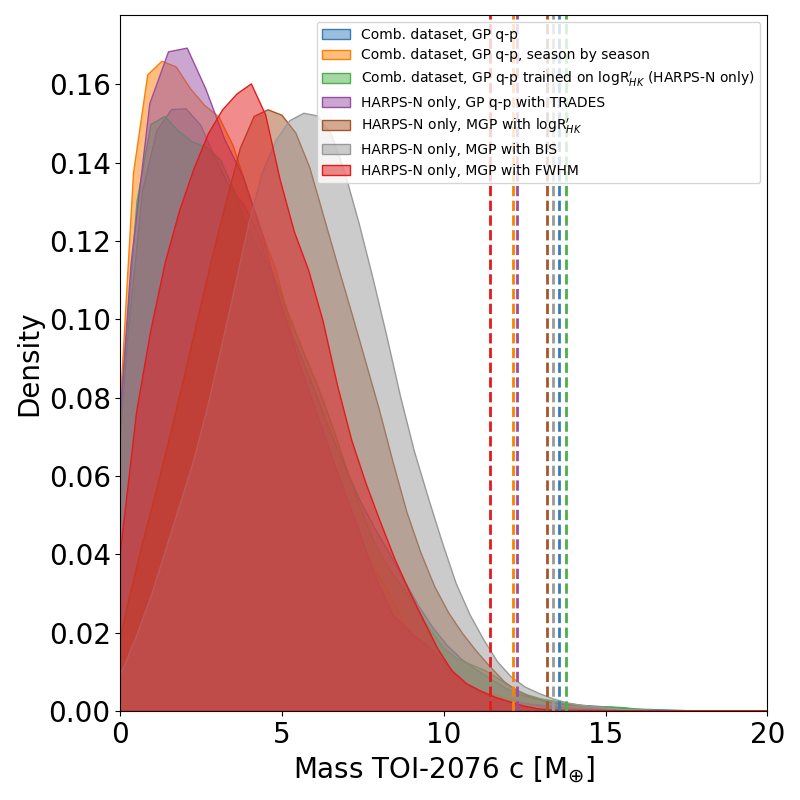}}
    \subfigure[]{\includegraphics[width=0.45\textwidth]{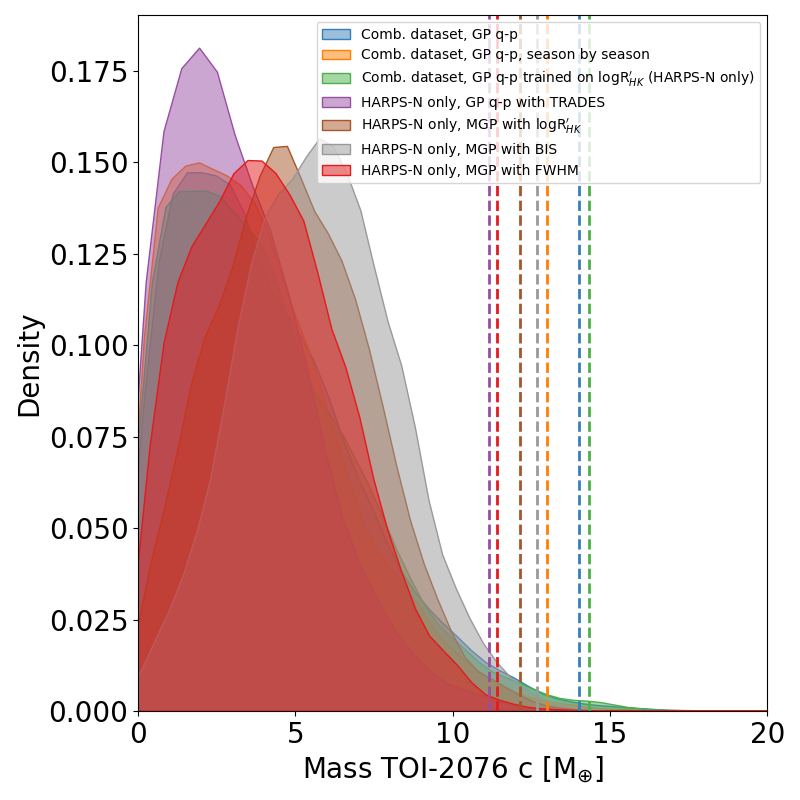}}
    \subfigure[]{\includegraphics[width=0.45\textwidth]{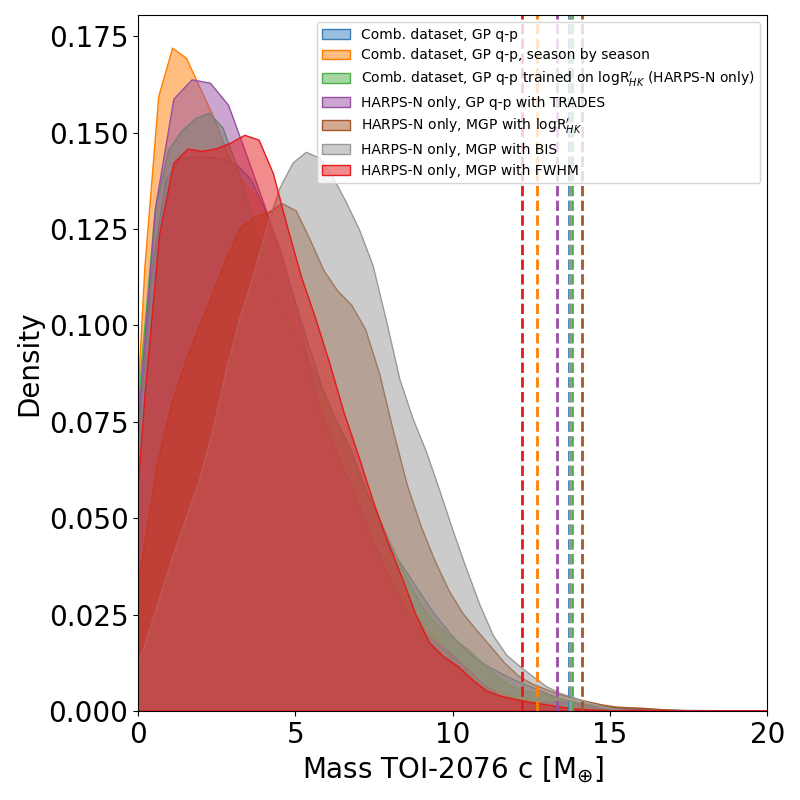}}
    \caption{Same as in Fig. \ref{fig:mass_planet_b_2} but for planet BD+40 2790/TOI-2076~c.}
    \label{fig:mass_planet_c_2}
\end{figure*}

\newpage

\begin{figure*}[ht]
    \centering
    \subfigure[]{\includegraphics[width=0.45\textwidth]{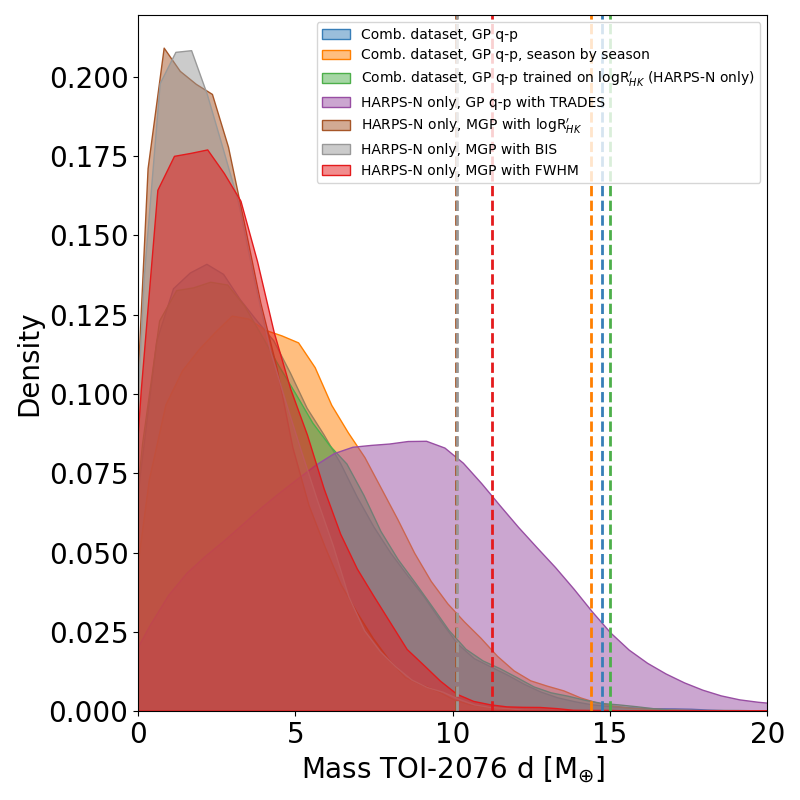}}
    \subfigure[]{\includegraphics[width=0.45\textwidth]{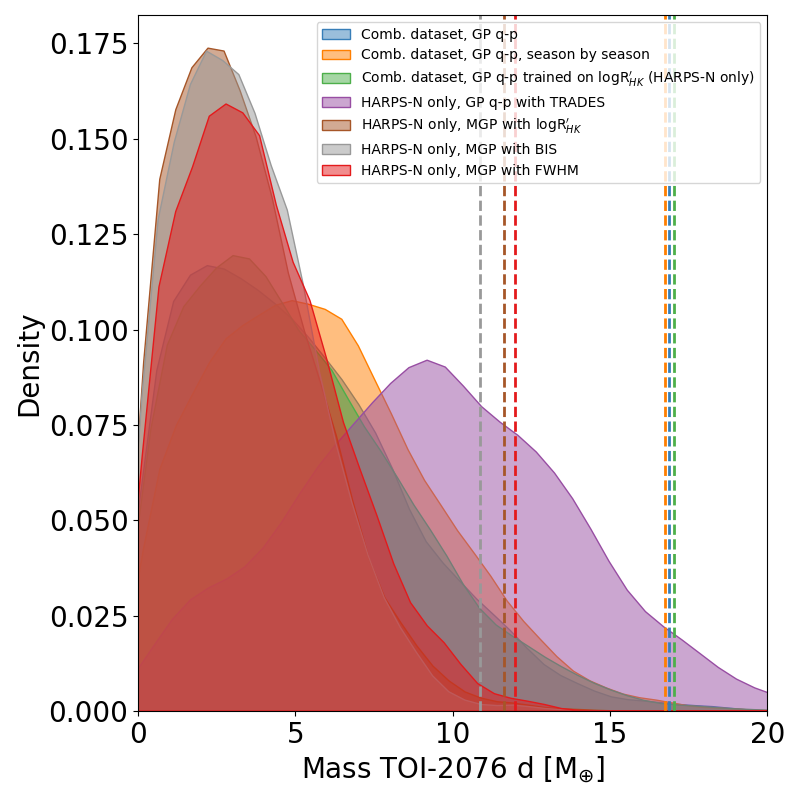}}
    \subfigure[]{\includegraphics[width=0.45\textwidth]{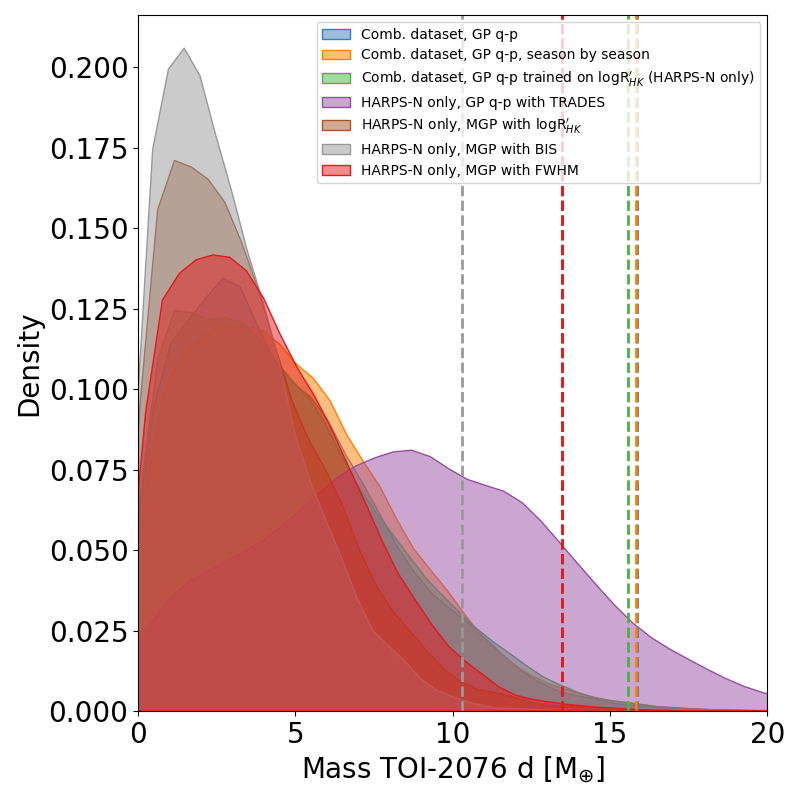}}
    \caption{Same as in Fig. \ref{fig:mass_planet_b_2} but for planet BD+40 2790/TOI-2076~d.}
    \label{fig:mass_planet_d_2}
\end{figure*}

\begin{figure}
    \centering
    \subfigure[]{\includegraphics[width=0.35\textwidth]{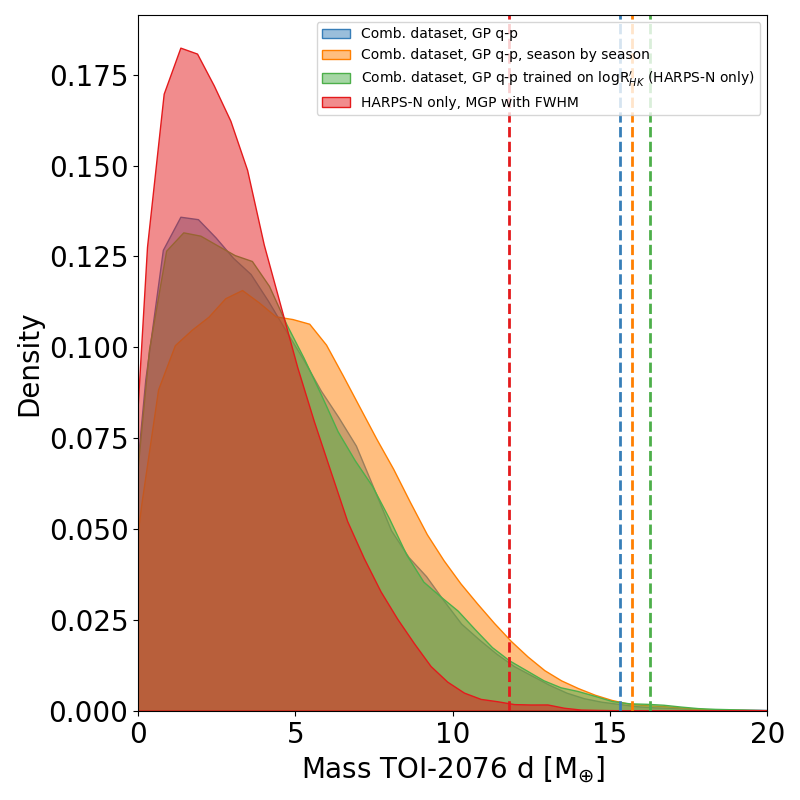}}
    \subfigure[]{\includegraphics[width=0.35\textwidth]{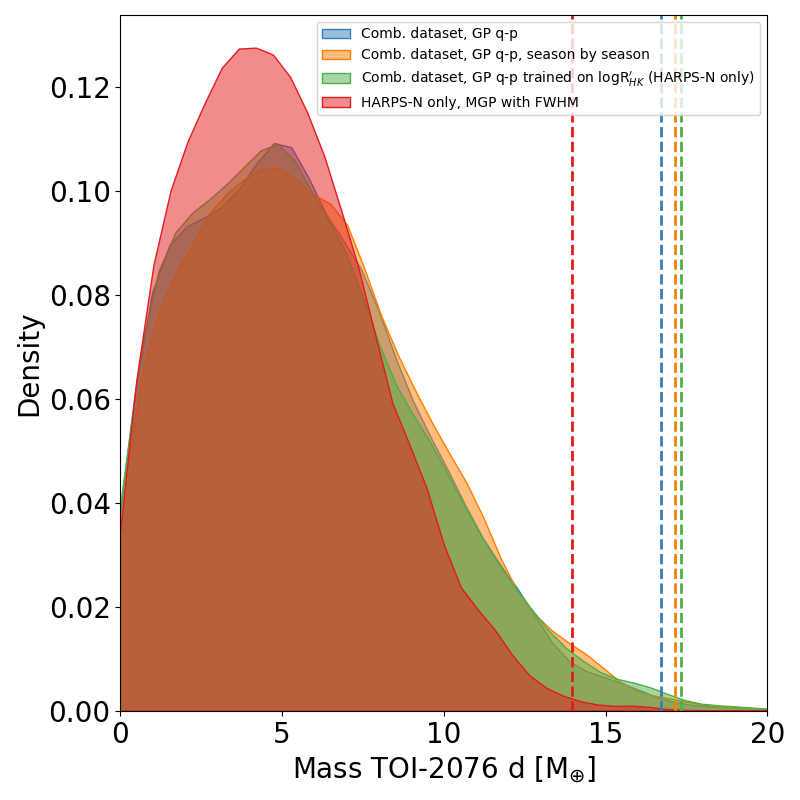}}
    \subfigure[]{\includegraphics[width=0.35\textwidth]{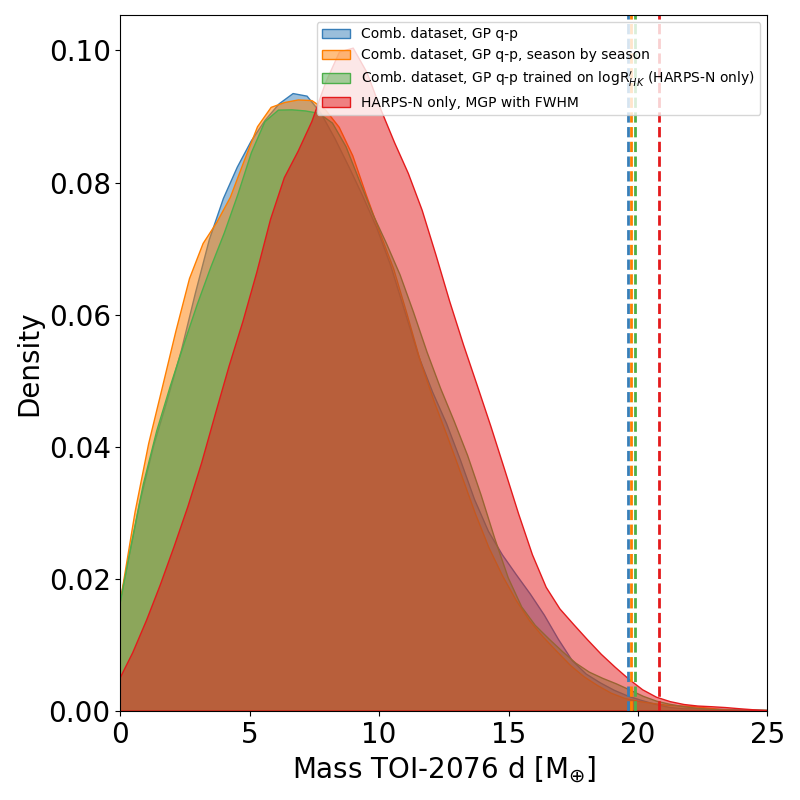}}
    \caption{Posterior mass distributions for the planet BD+40 2790/TOI-2076~d. The dashed vertical lines indicate the 3$\sigma$ upper limits for each distribution. Each panel corresponds to a HARPS-N RV dataset extracted with a different method (Table \ref{tab:rvsummary}): (a) SERVAL, wavelength range nr. 1; (b) SERVAL, wavelength range nr. 2; (c) SERVAL, wavelength range nr. 4. }
    \label{fig:additional_mass_planet_d}
\end{figure}



\begin{figure}
    \includegraphics[width=0.42\textwidth]{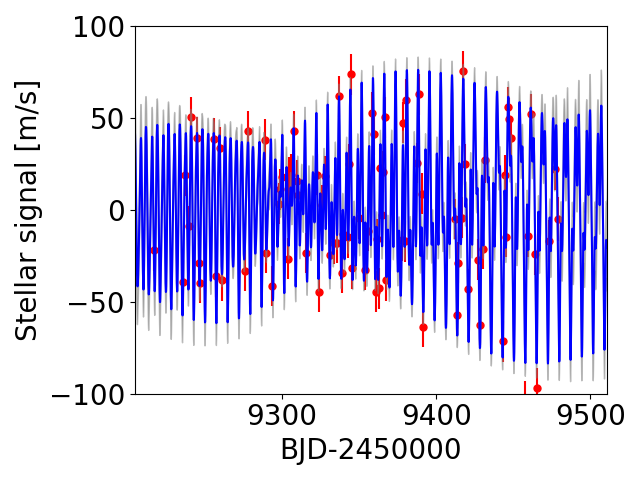}
    \includegraphics[width=0.45\textwidth]{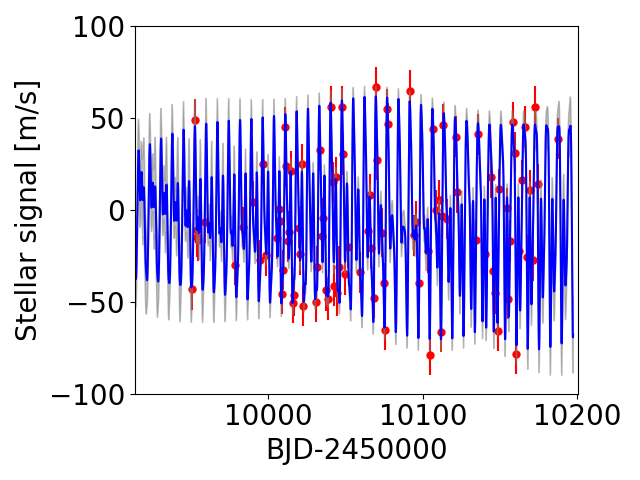}
    \caption{Examples of the stellar activity component in the HARPS-N RV time series (SERVAL, all echelle orders) as fitted through a GP regression using a quasi-periodic kernel (the first and third observing season are shown). The curve in blue represents the GP best-fit model; the grey area is 1$\sigma$ confidence interval. The RV error bars include a jitter term (11 \ms) added in quadrature to the measurement uncertainties. }
    \label{fig:gpactivity1}
\end{figure}

\begin{figure}
    \includegraphics[width=0.42\textwidth]{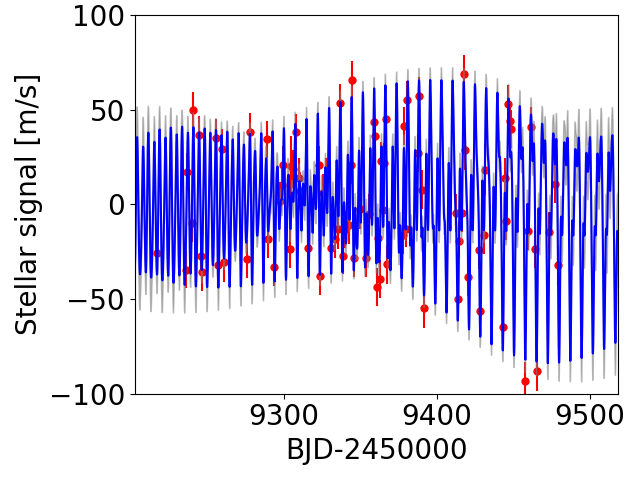}
    \includegraphics[width=0.45\textwidth]{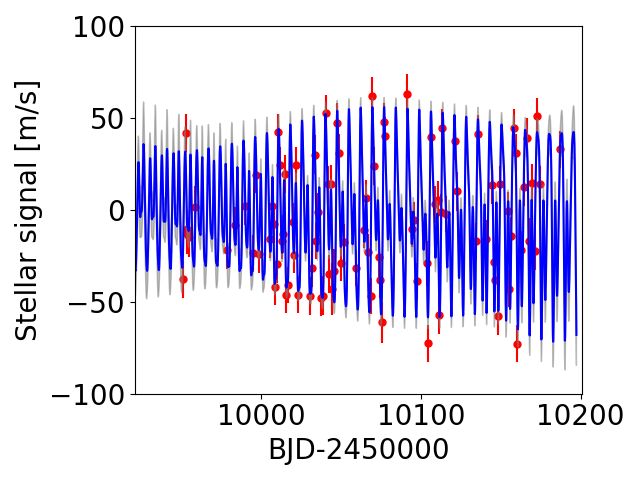}
    \caption{Same as in Fig. \ref{fig:gpactivity1}, but for the HARPS-N RV time series extracted with SERVAL in the wavelength range nr. 3 (Table \ref{tab:rvsummary}). The RV error bars include a jitter term (9.7 \ms) added in quadrature to the measurement uncertainties. }
        \label{fig:gpactivity2}
\end{figure}

\begin{sidewaystable*}
    \caption{Results of the photo-evaporation modeling, assuming two different X-rays and EUV stellar flux evolution models.}
    \label{table:p08sf22results}     
    \centering              
    \begin{tabular}{c c|c c c c|c c c c|c c c c}         
	\hline    
    \noalign{\smallskip}
    \noalign{\smallskip}
    \multicolumn{14}{l}{X-rays and EUV radiation evolution model: \cite{Penz08a} and \cite{SF22}}\\
    \noalign{\smallskip}
    \hline
    \textbf{Core Radius} &\textbf{ Core Mass} & \textbf{Mass }& \textbf{Radius} & \textbf{$f_{\rm atm}$} & \textbf{mass loss rate} & \textbf{Mass} & \textbf{Radius} & \textbf{$f_{\rm atm}$} &\textbf{ mass loss rate} & \textbf{Mass} & \textbf{Radius} & \textbf{$f_{\rm atm}$} &\textbf{ mass loss rate} \\     
    (\rearth) & (\mearth) & ($M_\oplus$) & ($R_\oplus$) & (\%) & (g/s)  & ($M_\oplus$) & ($R_\oplus$) & (\%) & (g/s) & ($M_\oplus$) & ($R_\oplus$) & (\%) & (g/s) \\    
    \hline
    \multicolumn{2}{c}{Planet b} & \multicolumn{4}{c}{current age} & \multicolumn{4}{c}{at 10\,Myr}& \multicolumn{4}{c}{at 1\,Gyr} \\
     1.9&11.9 & 12.0 & 2.54 & 0.8 & $2.8 \times 10^{10}$ & 12.10 & 3.2 & 1.4 & $1.4 \times 10^{11}$ & 11.9 & 2.2 & 0.3 &  $7.6 \times 10^{9}$ \\
     1.8 & 9.9 & 10.0 & 2.54 & 0.9 & $3.3 \times 10^{10}$ & 10.10& 3.4 & 1.8 & $1.9 \times 10^{11}$ & 9.9 & 2.1 & 0.2 &  $7.9 \times 10^{9}$  \\
    1.7& 7.9  & 8.0 & 2.54 & 1.0 & $4.1 \times 10^{10}$ & 8.10 & 3.8 & 2.6 & $2.8 \times 10^{11}$  & 7.9  & 1.9 & $8.4 \times 10^{-2}$ &  $7.3 \times 10^{9}$  \\
    1.4 & 3.9  & 4.0 & 2.54 & 1.2 & $7.5 \times 10^{10}$ & 4.60 & 7.7 & 13.6 & $2.1 \times 10^{12}$ & 3.95 & 1.4 & 0.0 &  0.0 \\
    \hline                       
    \multicolumn{2}{c}{Planet c} & \multicolumn{4}{c}{current age} & \multicolumn{4}{c}{at 10\,Myr}& \multicolumn{4}{c}{at 1\,Gyr}\\
     1.9& 12.6  & 13.0 & 3.35 & 3.3 & $2.4 \times 10^{10}$  & 13.10& 4.2 & 3.8 & $1.1 \times 10^{11}$  & 12.9 & 3.1 & 2.9 & $7.6 \times 10^{9}$  \\
     1.8& 9.7  & 10.0 & 3.35 & 3.4 & $3.0 \times 10^{10}$   & 10.10 & 4.3 & 4.2 & $1.5 \times 10^{11}$  & 9.9 & 3.0 & 2.8 & $9.2 \times 10^{9}$  \\
     1.6& 6.8  & 7.0 & 3.35 & 3.5 & $4.2 \times 10^{10}$   & 7.10 & 4.7 & 5.1 & $2.5 \times 10^{11}$ & 6.9 & 2.8 & 2.3 & $1.1 \times 10^{10}$  \\
     1.4& 3.9  & 4.0 & 3.35 & 3.5 & $6.7 \times 10^{10}$   & 4.26 & 6.4 & 9.5 & $7.1 \times 10^{11}$  & 3.9 & 2.2 & 0.9 & $9.1 \times 10^{9}$  \\
   \hline                       
    \multicolumn{2}{c}{Planet d} & \multicolumn{4}{c}{current age} & \multicolumn{4}{c}{at 10\,Myr} & \multicolumn{4}{c}{at 1\,Gyr}\\
     2.0 & 18.5   & 19.0 & 3.29 & 3.0 & $0.8 \times 10^{10}$  & 19.02 & 3.8 & 3.1 & $3.8 \times 10^{10}$  & 19.0 & 3.1 & 2.9 &  $2.6 \times 10^{9}$  \\
      1.9&14.5   & 15.0 & 3.29 & 3.2 & $1.1 \times 10^{10}$  & 15.03 & 3.9 & 3.4 & $5.0 \times 10^{10}$ & 15.0 & 3.1 & 3.0 &  $3.4 \times 10^{9}$  \\
       1.8  & 9.7   & 10.0 & 3.29 & 3.4 & $1.2\times 10^{10}$  & 10.04 & 4.1 & 3.8 & $7.7 \times 10^{10}$  & 10.0 & 3.0 & 3.0 &  $4.9 \times 10^{9}$ \\
       1.4 & 3.9   & 4.0 & 3.29 & 3.4 & $3.5\times 10^{10}$  & 4.10 & 5.2 & 6.1 & $2.7 \times 10^{11}$ & 3.9 & 2.6 & 1.8 &  $7.4 \times 10^{9}$ \\
   	\hline    
    \noalign{\smallskip}
    \noalign{\smallskip}
    \multicolumn{10}{l}{X-rays and EUV radiation evolution model: \cite{Johnstone+2021} }\\
    \noalign{\smallskip}
    \hline
    \textbf{Core Radius} &\textbf{ Core Mass} & \textbf{Mass }& \textbf{Radius} & \textbf{$f_{\rm atm}$} & \textbf{mass loss rate} & \textbf{Mass} & \textbf{Radius} & \textbf{$f_{\rm atm}$} &\textbf{ mass loss rate}  & \textbf{Mass} & \textbf{Radius} & \textbf{$f_{\rm atm}$} &\textbf{ mass loss rate} \\   
    (\rearth) & (\mearth) & ($M_\oplus$) & ($R_\oplus$) & (\%) & (g/s)  & ($M_\oplus$) & ($R_\oplus$) & (\%) & (g/s)  & ($M_\oplus$) & ($R_\oplus$) & (\%) & (g/s) \\    
    \hline
        \multicolumn{2}{c}{Planet b} & \multicolumn{4}{c}{current age} & \multicolumn{4}{c}{at 10\,Myr}& \multicolumn{4}{c}{at 1\,Gyr}\\
     1.9&11.9 & 12.0 & 2.54 & 0.8 & $2.8 \times 10^{10}$ & 12.10 & 3.2 & 1.4 & $1.4 \times 10^{11}$ & 12.0 & 2.3 & 0.5 & $6.9 \times 10^{9}$  \\
     1.8 & 9.9 & 10.0 & 2.54 & 0.9 & $2.0 \times 10^{10}$ & 10.0& 3.2 & 1.4 & $9.8 \times 10^{10}$ & 10.0 & 2.2 & 0.4 & $7.6 \times 10^{9}$  \\
     1.7& 7.9  & 8.0 & 2.54 & 1.0 & $2.4 \times 10^{10}$ & 8.10 & 3.4 & 1.8 & $1.4 \times 10^{11}$ & 7.9 & 2.1 & 0.4 & $8.q \times 10^{9}$  \\
     1.4 & 3.9  & 4.0 & 2.54 & 1.2 & $4.5 \times 10^{10}$ & 4.20 & 5.4 & 6.2 & $7.1 \times 10^{11}$ & 3.95 & 1.44 & 0.0 & 0.0  \\
    \hline                       
    \multicolumn{2}{c}{Planet c} & \multicolumn{4}{c}{current age} & \multicolumn{4}{c}{at 10\,Myr} & \multicolumn{4}{c}{at 1\,Gyr}\\
     1.9& 12.6  & 13.0 & 3.35 & 3.3 & $1.4 \times 10^{10}$  & 13.0& 4.1 & 3.6 & $6.7 \times 10^{10}$ & 13.0 & 3.1 & 3.1 & $6.3 \times 10^{9}$  \\
     1.8& 9.7  & 10.0 & 3.35 & 3.4 & $1.8 \times 10^{10}$   & 10.0 & 4.2 & 3.8 & $8.9 \times 10^{10}$ & 10.0 & 3.1 & 3,0 & $7.8 \times 10^{9}$  \\
     1.6& 6.8  & 7.0 & 3.35 & 3.5 & $2.5 \times 10^{10}$   & 7.10 & 4.4 & 4.3 & $1.4 \times 10^{11}$ & 6.9 & 2.9 & 2.7 & $9.8 \times 10^{9}$  \\
     1.4& 3.9  & 4.0 & 3.35 & 3.5 & $4.1 \times 10^{10}$   & 4.13 & 5.4 & 6.4 & $3.4 \times 10^{11}$ & 3.9 & 2.5 & 1.6 & $1.1 \times 10^{10}$  \\
   \hline                       
    \multicolumn{2}{c}{Planet d} & \multicolumn{4}{c}{current age} & \multicolumn{4}{c}{at 10\,Myr}& \multicolumn{4}{c}{at 1\,Gyr}\\
     2.0 & 18.5   & 19.0 & 3.29 & 3.0 & $4.5 \times 10^{9}$  & 19.0 & 3.8 & 3.1 & $2.2 \times 10^{10}$ & 19.0 & 3.1 & 3.0 & $2.1 \times 10^{9}$  \\
      1.9&14.5   & 15.0 & 3.29 & 3.2 & $6.0 \times 10^{9}$  & 15.0 & 3.9 & 3.3 & $2.9 \times 10^{10}$ & 15.0 & 3.1 & 3.1 & $2.7 \times 10^{9}$  \\
      1.8  & 9.7   & 10.0 & 3.29 & 3.4 & $8.9\times 10^{9}$  & 10.0 & 4.0 & 3.7 & $4.5 \times 10^{10}$ & 10.0 & 3.1 & 3.2 & $4.0 \times 10^{9}$ \\  
       1.4 & 3.9   & 4.0 & 3.29 & 3.4 & $2.1\times 10^{10}$  & 4.05 & 4.7 & 6.1 & $1.4 \times 10^{11}$ & 4.0 & 2.7 & 2.4 & $7.2 \times 10^{9}$ \\  
    \hline   
    \end{tabular}
\end{sidewaystable*}

\end{appendix}

\end{document}